\documentclass[a4paper,twocolumn,superscriptaddress,11pt,longbibliography,accepted=2020-23-08]{quantumarticle}
\pdfoutput=1

\usepackage[utf8]{inputenc}
\usepackage[english]{babel}
\usepackage[T1]{fontenc}
\usepackage{amsmath}
\usepackage{amsfonts}
\usepackage{amssymb}
\usepackage{amsthm}
\usepackage{hyperref}
\usepackage[numbers,sort&compress]{natbib}
\usepackage{tikz}
\usepackage{lipsum}
\usepackage[normalem]{ulem}
\usepackage{exscale}
\usepackage{bbm}
\usepackage{graphicx}
\usepackage{latexsym}
\usepackage{enumerate}
\usepackage{bbold}
\usepackage{color}
\usepackage{nicefrac}
\usepackage{mathtools}
\usepackage{chngcntr}
\usepackage{tcolorbox}
\usepackage{wrapfig}
\usetikzlibrary{decorations.pathreplacing}
\usetikzlibrary{graphs}
\usepackage{tikz}
\usepackage{pgfplots}
\usepackage{array, threeparttable, booktabs}

\newcommand{\proj}[1]{\ketbra{#1}{#1}}
\newcommand{\ket}[1]{\left|#1\right\rangle}
\newcommand{\bra}[1]{\langle#1|}

\newcommand{\ketbra}[2]{|#1\rangle\langle#2|}

\newcommand{\La}{\langle}
\newcommand{\Ra}{\rangle}
\newcommand{\refstate}{\ket{\psi'}}

\newcommand{\refstateproj}{\proj{\psi'}}
\newcommand{\physstate}{\ket{\psi}}
\newcommand{\physstatedm}{\rho}
\newcommand{\physstatebra}{\bra{\psi}}
\newcommand{\iso}{\Phi}
\newcommand{\junk}{\ket{\xi}}
\newcommand{\junkdm}{\sigma}
\newcommand{\sx}{\sigma_{\textrm{x}}}
\newcommand{\sz}{\sigma_{\textrm{z}}}
\newcommand{\sy}{\sigma_{\textrm{y}}}
\newcommand{\st}{\cos\theta\sx + \sin\theta\sz}
\newcommand{\tp}{\otimes}
\newcommand{\tr}{\text{tr}}
\newcommand{\maxent}{\ket{\phi^{\text{+}}}}
\newcommand{\tx}[1]{\textrm{#1}}

\let\arrowedvec=\vec
\let\vec\mathbf

\newcommand{\rA}{\text{A}}
\newcommand{\rB}{\text{B}}
\newcommand{\rP}{\text{P}}

\newcommand{\M}{\mathsf{M}}
\newcommand{\N}{\mathsf{N}}
\newcommand{\Bb}{\mathcal{B}}
\newcommand{\Ss}{\mathsf{S}}
\newcommand{\T}{\mathsf{T}}

\newcommand{\A}{\mathsf{A}}
\newcommand{\B}{\mathsf{B}}
\newcommand{\Xx}{\mathsf{X}}
\newcommand{\Zz}{\mathsf{Z}}
\newcommand{\Yy}{\mathsf{Y}}

\theoremstyle{definition}
\newtheorem{definition}{Definition}
\theoremstyle{remark}

\definecolor{quantumviolet}{HTML}{53257F} %Quantum violet
\definecolor{quantumgray}{HTML}{555555} %Quantum gray
\definecolor{mygray}{gray}{0.95} %Quantum gray

\newtcolorbox[auto counter,number within=section]{boxfigure}[2][]{%
colback=mygray,colframe=quantumviolet,fonttitle=\bfseries,width=\textwidth,float*=ht,lower separated=false, halign=justify,title=Box~\thetcbcounter: #2,#1}

\begin{document}

\title{Self-testing of quantum systems: a review}% Force line breaks with \\

\author{Ivan \v{S}upi\'{c}}
\affiliation{D{\'{e}}partement de Physique Appliqu\'{e}e, Universit\'{e} de Gen\`{e}ve, 1211 Gen\`{e}ve, Switzerland}
\author{Joseph Bowles}

%\email{bowles.physics@gmail.com}
\affiliation{ICFO-Institut de Ciencies Fotoniques,  The Barcelona Institute of Science and Technology,  08860 Castelldefels (Barcelona),  Spain}

\maketitle

\onecolumn
\begin{abstract}
Self-testing is a method to infer the underlying physics of a quantum experiment in a black box scenario. As such it represents the strongest form of certification for quantum systems. In recent years a considerable self-testing literature has developed, leading to progress in related device-independent quantum information protocols and deepening our understanding of quantum correlations. In this work we give a thorough and self-contained introduction and review of self-testing and its application to other areas of quantum information.
\end{abstract}\\
\twocolumn

\maketitle
\section{Introduction}
In contrast to classical theories, states in quantum physics can be entangled and sets of measurements can be incompatible. As shown by Bell in 1964 \cite{bell}, these features imply striking observable phenomena. In particular, the outcomes of incompatible measurements made on the local subsystems of an entangled quantum state can exhibit correlations that are provably stronger than any resulting from a classical theory, a phenomenon known as Bell nonlocality. %As a result, the observation of Bell nonlocality in the outcome statistics of an experiment implies the existence of both entanglement and incompatibility in the underlying physics. 
The field of Bell nonlocality has since grown considerably (see \cite{bellreview} for a recent review article), and the existence of Bell nonlocal correlations in nature is now a well established fact \cite{hensen,loophole2,loophole3}.

As more was understood about Bell nonlocality, a number of works \citep{Summers1987,Popescu1992,BMR,Tsirelson1993} eventually pointed out that there exist Bell nonlocal correlations that---as well as requiring entanglement and incompatibility---can only be produced by making \emph{particular} sets of incompatible measurements on \emph{particular} entangled states. These works have since given birth to the field of self-testing, which broadly speaking aims to understand the structure of the set of quantum correlations and identify those correlations that admit a unique physical realisation. 

An important milestone in the development of self-testing was the 2004 work of Mayers and Yao \cite{Mayers2004}. This work set the terminology and formalism that was to be adopted by later works, and includes the first usage of the term `self-testing' in this context. A similar idea was already present in \cite{Mayers98} in a cryptographic context, using the term `self-checking' instead of `self-testing'. These early works also introduced the paradigm of \emph{device-independence}, to which self-testing is intimately related. In particular, a self-testing protocol can be seen as a device-independent---or black box---certification of a quantum system, assuming that the system can be prepared many times in an independent, identically distributed manner. Self-testing is consequently relevant to many device-independent quantum information protocols and has led to related progress in this area. More recently, self-testing has become synonymous with any protocol for certifying any type of quantum system under a small set of assumptions. %\textcolor{red}{A part of the community objects equating self-testing with device-independent certification, as several steps or assumptions have to be made to reach the second from the first. However, a general consensus is that self-testing is an indispensable tool in studying the set of quantum correlations and its relation to the set of quantum strategies used to obtain the correlations.}

In this work we give a up-to-date review of the field of self-testing. We hope that it will be of use to people both unfamiliar with the field, as well as serving as a reference for those within it. The review is organised as follows. In section \ref{di} we give a gentle introduction to device-independence and its connection to self-testing. We then formally introduce self-testing, giving the mathematical definitions in section \ref{defs} and a simple example in section \ref{sec:example} that illustrates many important concepts. Sections \ref{sec:bipartite} to \ref{sec:measurements} are a thorough literature review of state and measurement self-testing, explaining the tools and techniques that are commonly used along the way. In section \ref{sec:extensions} we review extensions of self-testing to other scenarios, and in section \ref{sec:applications} the application of self-testing to other fields in quantum information theory. In section \ref{sec:experiments} we cover experimental realisations of self-testing protocols. Finally, in section \ref{sec:openquestions} we discuss some possible future directions for the field and a number of open problems. 

We point the reader to the related review articles \cite{Montanaro_Wolf} and \cite{scarani} where discussions about self-testing can also be found. \cite{Montanaro_Wolf} deals with the classical and quantum certification of both classical and quantum properties of an object, with self-testing being identified as classical certification of quantum properties. \cite{scarani} provides a pedagogical review of the device-independent approach to quantum physics. Self-testing is discussed as one of device-independent protocols.  We also recommend \cite{McKague_thesis,Jed1,Jed2} as valuable texts for first time readers.

\tableofcontents

\begin{figure*}[t]
    \centering
    \includegraphics[scale=1.05]{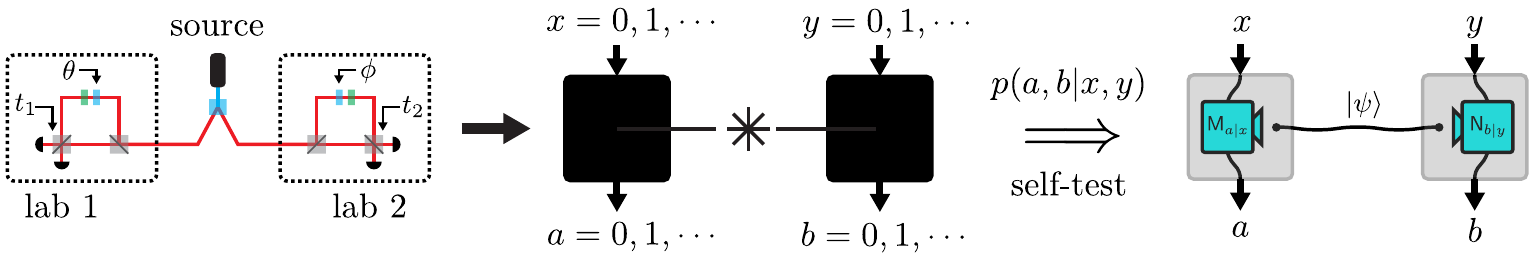}
    \caption{(left) A source produces systems and distributes them between two laboratories that can perform different local experiments by varying the settings of their equipment. (centre) In the device-independent approach, each local laboratory is treated as a black box that takes as input a label that corresponds to a particular choice of settings for the experiment, and outputs a label that denotes the corresponding result. After repeating the experiments sufficiently many times, the probabilities $p(a,b\vert x,y)$ can be estimated. (right) When self-testing a state and measurements, one aims to infer the form of the state of the source and measurement operators describing the laboratories from knowledge of the probabilities alone, i.e.\ in a black box scenario.}
    \label{fig_scenario}
\end{figure*}

\section{Self-testing as a device-independent protocol}\label{di}
The treatment of complex systems as black boxes is a powerful tool in many scientific domains, providing a minimalist level of abstraction that allows one to focus on \emph{what} a device or system does without the need to model precisely \emph{how} this is achieved. In quantum information theory, this approach is known as the \emph{device-independent} (DI) approach.

In order to explain the idea of the device-independent approach we imagine the following scenario. Consider two laboratories, run by two experimenters called Carmela and Deng. In their laboratories (let's imagine they are quantum optics laboratories) both Carmela and Deng have access to some equipment (e.g.\ lasers, beamsplitters, waveplates, photon detectors,...) which they can use to perform different experiments. A given experiment consists of a choice of \emph{settings} (e.g.\ laser intensity, angle of the waveplates, type of beamsplitter,...) that after a run of the experiment provides a \emph{result} (e.g.\ photon detection location, time of detection,...). Furthermore, a source is positioned between the laboratories and emits physical systems (e.g.\ photons) that are sent to  Carmela's and Deng's laboratories; see figure \ref{fig_scenario}, left. 

Suppose that Carmela and Deng would like to learn if the source is emitting entangled particles (where the entanglement is with respect to the two laboratories). One way to achieve this is to use their equipment to perform tomography of the state of the source, i.e.\ Carmela and Deng perform a number of experiments each with different settings, collect statistics of the results, and use quantum state tomography to reconstruct the density matrix of the state, which can then be checked to determine if it is entangled (for instance using an entanglement witness). This is indeed what is done in many experiments around the world. 

Imagine now however that two computer scientists---called Alice and Bob---are visiting each of the labs. Despite knowing the mathematical definition of entanglement, they will have problems convincing themselves that the source is producing entanglement. Firstly, they do not understand the experimental setup, so they do not know what the different settings do. Moreover, even if they were told what the settings do, they do not have a good understanding of quantum optics. As a result, they will not be able to reconstruct the state of the source in order to check if it is entangled, as was the case for Carmela and Deng. 

Alice, however, proposes the following: even though they do not understand what the settings do, they can still change them and observe \emph{something}. That is, they can simply model their laboratories as black boxes. Each laboratory is treated as a device (a black box) that takes an input (the settings) and returns and output (the result), but the physical mechanism behind how this occurs is unknown (see figure \ref{fig_scenario}, centre). Similarly, they do not assume anything about the source; all they know is that it is distributing some physical systems that may or may not be entangled. Alice denotes each of her possible settings as $x=0,1,\dots$ and Bob denotes each of his possible settings as $y=0,1,\dots$ . Similarly Alice and Bob denote the possible results of their experiments by $a=0,1,\dots$ and $b=0,1,\dots$.

After trying the different settings sufficiently many times and collecting statistics, Alice and Bob can estimate the probabilities (also called the \emph{correlations})
\begin{align}\label{pab_intro}
    p(a,b\vert x,y),
\end{align}
that is, the probabilities to see the results $a$ and $b$ given that the settings $x$ and $y$ are used. It is important here to stress that although Alice and Bob can estimate these probabilities, they are ignorant about the underlying physics; from their perspective the experiments could have been made on atoms, electrons, neutrinos or any other physical system. This scenario is called the \emph{device-independent} scenario. Remarkably, even with such little knowledge, Alice and Bob can still conclude that the source emits entangled states. 

\begin{figure}
    \centering
    \includegraphics[scale=0.9]{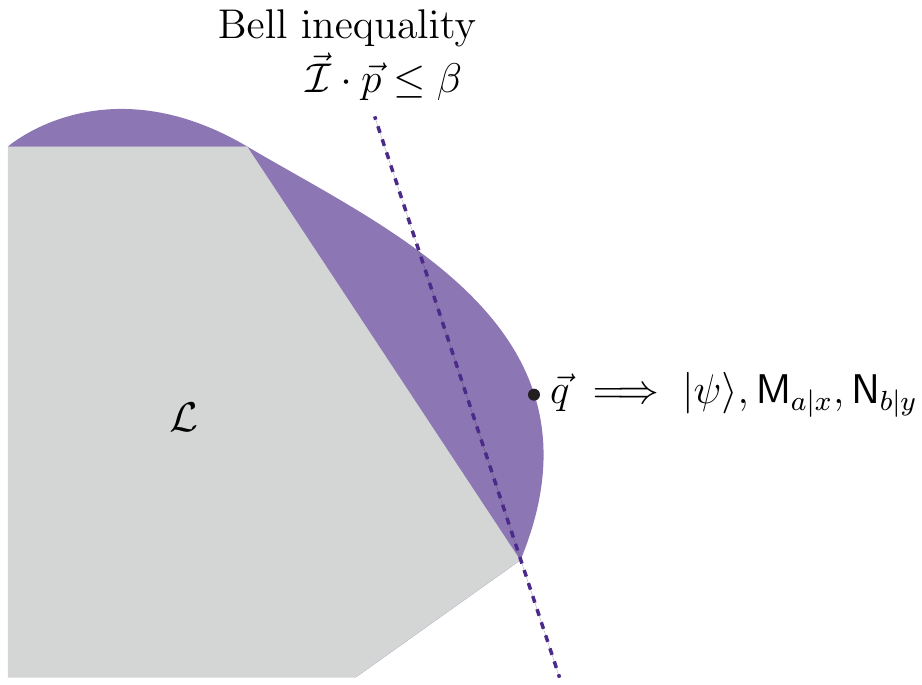}
    \caption{\label{fig:st_fig} Geometric representation of self-testing. The figure is a 2-dimensional representation of the space of probability vectors $\arrowedvec{p}=(p(00\vert 00),p(01\vert 00),\dots)$. The grey area $\mathcal{L}$ is the set of probability vectors that are obtainable with separable states (commonly called the `local set'). A Bell inequality (dotted line) consists of a pair $(\arrowedvec{\mathcal{I}},\beta)$ such that the half space $\arrowedvec{\mathcal{I}}\cdot\arrowedvec{p}\leq\beta$ contains $\mathcal{L}$. Entangled quantum states are capable of producing probability vectors that lie outside of $\mathcal{L}$ (purple area) and thus violate Bell inequalities. Often, extremal points (e.g.\ $\arrowedvec{q}$) of the quantum set, which maximally violate some Bell inequality, admit---up to local transformations---a unique realisation in terms of a particular entangled state and measurements. Self-testing involves identifying such probability distributions and proving their unique realisation.}
\end{figure}
 
The trick to achieving this is to use \emph{Bell nonlocality}, a counter-intuitive phenomenon discovered by John Bell in 1964 \cite{bell} (see also box \ref{box:bell}). At the heart of Bell nonlocality are objects called \emph{Bell inequalities}. A Bell inequality consists of a function $\mathcal{I}$ of the probabilities $\{p(a,b\vert x,y)\}$ such that, for a source producing separable (i.e\ non-entangled) states one has
\begin{align}
    \mathcal{I}(\{p(a,b\vert x,y)\})\leq \beta.
\end{align}
%
%Since separability is a mathematical definition, independent of the type of physical system in question, the bound $\beta$ 
Importantly, the bound $\beta$ holds for \emph{any} source producing \emph{any} kind of physical systems, provided that these systems are not entangled. This is a consequence of the fact that the definition of separability is independent of the physical system in question. Interestingly, Bell inequalities can be violated by entangled sources. That is, for some entangled sources one can achieve $\mathcal{I}(p(a,b\vert x,y))>\beta$. This situation can be understood geometrically in the vector space of probabilities, where Bell inequalities are defined by linear hyperplanes; see Fig. \ref{fig:st_fig}.

Alice and Bob can therefore do the following. They compare their probabilities against as many Bell inequalities as they know. If they see that one is violated, then it must be that the source is entangled! Moreover, they are able to conclude this despite knowing nothing about how the experiment was actually performed; all the information that was needed was the probabilities $\{p(a,b\vert x,y)\}$. Such a procedure is called a \emph{device-independent certification of entanglement}. 

Suppose now that we change the task: instead of only detecting entanglement, Alice and Bob want to know the particular entangled state of the source. Due to the device-independent scenario, they cannot know the type of physical system that the source is producing. However, they may hope to write down the state vector $\ket{\psi}$ of the source, without specifying which types of physical degrees of freedom it describes. This often turns out to be possible (up to some local transformations, see the definitions in the following section), as long as one observes the \emph{maximum} possible violation achievable in quantum theory of a corresponding Bell inequality. Such a procedure is called a \emph{device-independent self-test} or simply a \emph{self-test} of the state. Often, this maximal violation also allows one to self-test the measurements, i.e.\ to determine the form of the measurement operators that describe how the outcomes $a$ and $b$ are produced in the local laboratories. 

Alice and Bob will nevertheless face a problem: they will never see the maximum violation of a Bell inequality due to experimental noise and finite statistics. At best, they will be able to lower bound the violation up to some  statistical confidence. To become a practically relevant protocol, self-testing therefore has to be combined with two other tools: (i) noise-resistant self-testing methods (called \emph{robust self-testing}; see section \ref{robust-intro}) which give distance upper bounds to the desired state and/or measurements %that are self-tested in the noiseless case, 
as a function of experimental noise; and (ii) tools of statistical analysis (such as Chernoff bounds) that allow for statistically valid estimations of probabilities and corresponding confidence levels. Self-testing can thus be seen as a theoretical tool that, when augmented by statistical techniques, can be transformed into a protocol for device-independent state and measurement certification.

% Self-testing can thus be seen as \textcolor{purple}{a restricted}\joe{why restricted?} form of device-independent \textcolor{purple}{certification} of the state and measurements performed in an experiment.  

A number of remarks are in order before proceeding to formal definitions of self-testing in the next section. First, one may wonder whether self-testing is possible using only a single device. That is, given a single device from which one observes the conditional probabilities $p(a\vert x)$, is it possible to determine the quantum state $\ket{\psi}$ inside? Notice that one possibility is that there is a pre-programmed classical computer inside the device that simply simulates the statistics $p(a\vert x)$. In the device-independent scenario one cannot rule out this possibility. Thus, one cannot hope to certify any non-classical properties of $\ket{\psi}$ with only a single device. It is for this reason that it is crucial to move to a multipartite scenario in order to certify non-classical states. Indeed, it is precisely the phenomenon of Bell nonlocality that forbids an explanation using local pre-programmed classical devices.

Second, it is worth mentioning some practical advantages of the device-independent scenario. Suppose Alice and Bob are sold two devices that are claimed to contain a particular entangled state and perform certain measurements on it. Using self-testing, they will be able to conclude that the devices are indeed working correctly without having to understand precisely how they operate. This is clearly desirable from the perspective of Alice and Bob, especially for cryptographic applications where one would prefer not to trust the devices. From a more experimental perspective, the device-independent scenario naturally treats experimental errors at the level of the observed statistics, meaning for example, that a false positive detection of entanglement will never occur. Being the strongest form of device-independent certification, self-testing has proven to be useful for many device-independent protocols; see section \ref{sec:applications} for more information. 

Third, we note that the notion of device-independence referred to by the large majority of self-testing works assumes independent, identically distributed (i.i.d.\ ) rounds of the experiment, i.e.\ the state and the measurements are assumed to be the same in each round and do not depend on past measurement results or choices. Such an assumption allows us to talk about the probability distribution $p(a,b\vert x,y)$ that is valid in every experimental round. This however is not the strongest possible notion of device-independence where one further drops the i.i.d.\ assumption (typically used in the field of quantum cryptography). Self-testing under this more stringent notion of device-independence is little explored (although arguably relevant); see the concluding remarks in section \ref{sec:openquestions} for further discussion.

%\textcolor{red}{Direct usefulness of self-testing for device-independent certification is disputed based on the notion of device-independence itself. The community working on Bell nonlocality usually accepts the assumption that source producing identical and independently distributed (i.i.d.) copies as  device-independent. In this case, after some large number of rounds it is possible to estimate the probability distribution $p(a,b|x,y)$. In a more stringent device-independent scenario the i.i.d. assumption is not tolerated. Such approach, more often pursued by researchers working on quantum cryptography and quantum computing, deals with the situation in which in $n$ different rounds the source might produce $n$ systems in different states, which might even be correlated. There is no access to any probability distribution, but just a sequence of $n$ sets of inputs and $n$ sets of outputs. If self-testing is to be used for certification one has to define what is being certified, as there might not be a well defined state the source produces in all rounds.}

Finally, we note that aside from its motivation for device-independent certification, self-testing can also be viewed as part of the general study of quantum correlations. In particular, quantum correlations are defined via the Born rule, whereby a state and measurement operators are mapped to a probability distribution. Generally, this map does not have an inverse, since many combinations of state and measurements can lead to the same probabilities. Self-testing identifies those points of the set of quantum probability distributions for which such an inverse exists, up to the local transformations defined in the next section. That is, it associates some (extremal) probability distributions with a unique realisation using a particular state and measurements. For this reason self-testing has been used to derive important results about the set of quantum correlations. 

\iffalse
\textcolor{purple}{It is important to bear in mind that we described here a bit idealistic situation in which Alice and Bob have access to $\{p(a,b|x,y)\}$. Such probability distributions are, in principle, never experimentally accessible unless some strong assumptions about the source are made. As one can learn by simply looking at the contents of this review self-testing is primarily a theoretical tool which takes the existence of $\{p(a,b|x,y)\}$ for granted. Relating obtained experimental data to those which would provide certain probability distribution $\{p(a,b|x,y)\}$ remains a challenging task. Thus, in practice self-testing is more an intermediary, but a very important step towards assumption-free device-independent quantum state certification.  
}
\fi

\section{Definitions}\label{defs}
\subsection{Notation}
We first introduce some notation. $\mathcal{L}(\mathcal{H})$ denotes the set of linear operators acting on Hilbert space $\mathcal{H}$.
Uppercase Roman letters denote a local party, most often either Alice, $\rA$, or Bob, $\rB$. Roman letters in either subscript or superscript denote the Hilbert space in which a state lives or on which an operator acts e.g. $\ket{\psi}_{\rA}\in\mathcal{H}_{\rA}$. Consecutive labels denote tensor products of Hilbert spaces, e.g. $\ket{\psi}_{\rA\rB}\in\mathcal{H}_{\rA}\otimes\mathcal{H}_{\rB}$. Labels containing the same letter are implicitly assumed to refer to different local Hilbert spaces of a single subsystem, e.g.\ $\rA$ and $\rA'$ refer to two Hilbert spaces of Alice. 

As a result, self-testing is often a useful tool in a variety of device-independent protocols (see section \ref{sec:applications} for explicit examples). In the following section, we formalise these ideas and define precisely what it means to self-test a state and  measurements.

\subsection{The self-testing scenario}\label{sec:scenario}
The device-independent scenario described in the previous section is commonly called a \emph{Bell test}, and the probabilities $p(a,b \vert x,y)$ are called \emph{correlations}. From quantum theory, we know that there exist measurement operators $\M_{a\vert x}\in \mathcal{L}(\mathcal{H}_\rA)$ acting on Alice's local Hilbert space and satisfying
\begin{align}
    \M_{a|x}\succcurlyeq 0 \quad \forall x,a, \quad \sum_{a}\M_{a\vert x}=\openone_{\rA}\quad\forall x
\end{align}
that describe how the outcomes $a$ are obtained given settings $x$. Similarly there exist measurement operators $\N_{b\vert y}\in\mathcal{L}(\mathcal{H}_\rB)$ for Bob acting on his local Hilbert space. From here on we work with a Naimark dilation of each of the measurements. The measurement operators are therefore projective:
\begin{align}
    \M_{a|x}\M_{a'|x}&=\delta_{a,a'}\M_{a|x} \quad &&\forall x,a,a' \; , \nonumber\\
    \N_{b|y}\N_{b'|y}&=\delta_{b,b'}\N_{b|y} \quad &&\forall y,b,b'. \; %;\\[4pt]
%    \sum_a \M_{a|x}&=\openone_{\rA} \; \forall x \; ; \quad
%    \sum_b \N_{b|y}=\openone_{\rB} \; \forall y
\end{align}
This can be justified if one takes the position that projective measurements and unitary evolution are the only fundamental operations in quantum theory. From this perspective, any measurement necessarily involves a projective measurement over a dilated space, the degrees of freedom of which belong to Alice and Bob's local Hilbert spaces. For further discussion on this assumption, see Appendix \ref{app:assumptions}.

%if one takes the position that projective measurements and unitary evolution are the only fundamental operations in quantum theory. From this perspective, all other operations (CPTP maps, POVM measurements) can be seen as consequences of these fundamental operations. From this perspective, any local measurement necessarily involves a projective measurement over a dilated space that is part of Alice's or Bob's laboratory.}

%Thus, if we take $\mathcal{H}_{\rA}$ and $\mathcal{H}_{\rB}$ to describe all local degrees of freedom of Alice's and Bob's laboratories, then the projection postulate tells us that the measurement operators are projective:
%

Now, from the Born rule, there must exist some quantum state $\rho_{\rA\rB}\in\mathcal{L}(\mathcal{H}_{\rA}\tp\mathcal{H}_{\rB})\succcurlyeq 0$, $\tr\,\rho_{\rA\rB}=1$ such that 
\begin{align}\label{pab}
p(a,b\vert x,y)=  \tr\left[\physstatedm_{\rA\rB} \;\M_{a\vert x}\otimes \N_{b\vert y}\right].
\end{align}
%
%Here $\rho_{\rA\rB}$ describes the quantum state of all the degrees of freedom to which Alice and Bob have access, i.e.\ the state of the source and any additional local degrees of freedom of the laboratories. 
In self-testing, one aims to infer the form of the state and the measurements in \eqref{pab} from knowledge of the correlations $p(a,b\vert x,y)$ alone; i.e.\ in the device-independent scenario. 

In order to write \eqref{pab}, one nevertheless needs to make some basic physical assumptions that constitute the definition of the device-independent scenario in most self-testing works. These are
\begin{enumerate}
    \item The experiment admits a quantum description; i.e.\ there exists a quantum state and measurement operators that lead to the observed outcomes via the Born rule. 
    \item The laboratories of Alice and Bob are located at separate locations in space and there is no communication between the laboratories; e.g.\ Alice cannot send the choice of setting $x$ to Bob or vice-versa. 
    \item The settings $x$ and $y$ are chosen freely and independently of all other systems in the experiment. For example, the physical system used to generate $x$ does not have any correlations (quantum or classical) with the particles or the source or the laboratory of Bob. 
    \item Each round of the experiment is independent from all other rounds and physically equivalent to all others. That is, there exists a single density matrix and measurement operators that are valid in every round. The statistics of all the rounds are thus independent and identically distributed (i.i.d.\ ), so that we may consider $p(a,b\vert x,y)$ only.
\end{enumerate}

%\joe{In writing \eqref{pab}, we are explicitly assuming that Alice's and Bob's measurement operators act on different local Hilbert spaces. This is natural given that the two labs are located at spatially separated locations between which we believe there is no communication. As is the case in Bell nonlocality experiments that close the `locality loophole', this assumption can be justified on relativistic grounds by enforcing space-like separation between the measurement events in each round.}

Following the majority of self-testing works, we will often choose to work with a purification $\physstate_{\rA\rB\rP}$ of $\physstatedm_{\rA\rB}$, where the purification space $\mathcal{H}_{\rP}$ is external to both Alice's and Bob's laboratories. This will be quite convenient mathematically, since working with state vectors over density matrices will significantly shorten the length of some equations. We stress that we do not assume that the state shared between Alice and Bob is pure; indeed it is given by $\tr_\rP[\vert \psi\rangle
\langle\psi\vert_{\rA\rB\rP}]=\rho_{\rA\rB}$. Furthermore, our definitions of self-testing will be such that the purification space is untouched by the self-testing protocol. Hence, working with $\physstate_{\rA\rB\rP}$ and tracing out the purification space is equivalent to working with $\physstatedm_{\rA\rB}$. For this reason, the definitions in the following section can be equivalently phrased using $\physstatedm_{\rA\rB}$ by simply discarding the purification space. Using such a purification, the probabilities \eqref{pab} can be written
\begin{align}\label{pab_pure}
p(a,b\vert x,y)= \physstatebra \M_{a\vert x}\otimes \N_{b\vert y}\tp\openone_\rP \physstate_{\rA\rB\rP},
\end{align}
where $\openone_\rP$ is the identity operator on the purification space.

Now, let's imagine we have in mind a particular pure state\footnote{self-testing of mixed states will not be possible; see section \ref{st_geometry}} $\refstate_{\rA'\rB'}$ and projective measurements $\{\M'_{a\vert x}\},\{\N'_{b\vert y}\}$ that we would like to self-test. We call this state and measurements the \emph{reference state} and \emph{reference measurements}. The state $\physstatedm_{\rA\rB}$ and measurements $\{\M_{a\vert x}\},\{\N_{b\vert y}\}$ that correspond to the actual experiment are called the \emph{physical state} and \emph{physical measurements}. Similarly, the realisation $\{\refstate_{\rA'\rB'},\{\M'_{a\vert x}\},\{\N'_{b\vert y}\}\}$ is called the \emph{reference experiment} and $\{\physstatedm_{\rA\rB},\{\M_{a\vert x}\},\{\N_{b\vert y}\}\}$ the \emph{physical experiment}.

Note that it will not be possible to infer exactly the reference state and measurements from the correlations alone. This is for the following two reasons:
\begin{enumerate}[i]
    \item Due to the unitary invariance of the trace, one can reproduce the statistics of any state $\refstate$ and measurements $\{\M'_{a\vert x}\}$, $\{\N'_{b\vert y}\}$ by instead using the rotated state 
$U\tp V\refstate$ and measurements $\{U\M'_{a\vert x}U^\dagger\}$,  $\{V\N'_{b\vert y}V^\dagger\}$, where $U$ and $V$ are unitary transformations. Hence, one can never conclude that the state is $\refstate$ since it may in fact be $U\otimes
V\refstate$.
\item One cannot rule out additional degrees of freedom on which the measurement operators act trivially. That is, a state $\refstate\tp\junk$ and measurements $\{\M'_{a\vert x}\tp\openone\}$, $\{\N'_{b\vert y}\tp\openone\}$ (where the identity operators act on the local spaces of $\junk$) gives the same correlations as $\refstate$ and $\{\M'_{a\vert x}\}$, $\{\N'_{b\vert y}\}$.
\end{enumerate}
To be able to define what it means to infer a particular state in the device-independent scenario, we thus need to define an equivalence between states that takes into account the above unknowns (that is, local unitary transformations and additional unused degrees of freedom). To do this, we make use of the concept of a \emph{local isometry}. 

\begin{figure}
    \centering
    \includegraphics[scale=1.6]{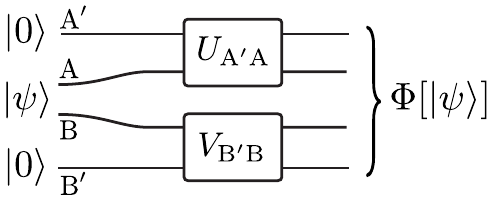}
    \caption{A local isometry applied to a quantum state. Local ancillas are added to the state and unitary transformations are applied locally.}
    \label{iso_fig}
\end{figure}
An isometry $\iso:\mathcal{H}_{\rA_1}\rightarrow\mathcal{H}_{\rA_2}$ is a linear transformation on quantum states that preserves the inner product, and can thus be seen as a unitary operator that can increase the dimension of the space. A general isometry on a state $\ket{\psi}_{\rA_1}$ can be achieved by embedding $\ket{\psi}_{\rA_1}$ in a larger Hilbert space $\mathcal{H}_{\rA_2}$ and performing a unitary transformation; i.e.\ $\iso[\ket{\psi}_{\rA_1}]=U\ket{\psi}_{\rA_2}$. A \emph{local isometry}
\begin{align}
\iso_{\rA_1}\tp\iso_{\rB_1}:\mathcal{H}_{\rA_1}\tp\mathcal{H}_{\rB_1}\rightarrow\mathcal{H}_{\rA_2}\tp\mathcal{H}_{\rB_2}
\end{align}
is an isometry that can be realised with local quantum operations; i.e.\ it is a tensor product of isometries acting locally. One way to implement a local isometry---often used in self-testing works---is to embed the initial state using local ancilla states $\ket{00}_{\rA'\rB'}$, and then perform a local unitary transformation (see figure \ \ref{iso_fig}):
\begin{align}\nonumber
    \iso_\rA\tp\iso_\rB [ \physstate_{\rA\rB}]=U_{\rA'\rA}\otimes V_{\rB'\rB}\left[\ket{00}_{\rA'\rB'}
    \otimes\physstate_{\rA\rB}\right].
\end{align}
%
% With the notion of a local isometry we can define an equivalence between states as follows.
% %
% \vspace{10pt}
% \hrule
% %
% \begin{definition}\emph{(equivalence up to local isometry)}\\[5pt]
% We say that the state $\physstate_{\rA\rB}$ is equivalent to the state $\refstate_{\rA'\rB'}$ \emph{up to a local isometry} if there exists a local isometry
% \begin{align}
% \iso_{\rA}\tp\iso_{\rB}:\mathcal{H}_{\rA}\tp\mathcal{H}_{\rB}\rightarrow\mathcal{H}_{\rA'\bar{\rA}}\tp\mathcal{H}_{\rB'\bar{\rB}}
% \end{align}
% such that
% %
% \begin{align*}
%     \iso_\rA\tp\iso_\rB[\physstate_{\rA\rB}]=\refstate_{\rA'\rB'}\tp\junk_{\bar{\rA}\bar{\rB}}
% \end{align*}
% for some state $\junk_{\bar{\rA}\bar{\rB}}$.
% \end{definition}
% \hrule
% \vspace{10pt}
% %
% \noindent Thus if $\physstate$ is equal to $\refstate$ up to a local isometry, there is a local operation that rotates $\physstate$ to a state that contains $\refstate$ shared between Alice and Bob. 
%Notice that the states $U\tp V\refstate$ and $\refstate\tp\junk$ are both equal to $\refstate$ up to a local isometry.
In the following sections we will use $\iso[\ket{\psi}]$ to denote the action of an isometry on the pure state $\ket{\psi}$ and use $\iso[\rho]$ to denote the corresponding transformation on a density matrix. As we will see, the notion of a local isometry will be central to our definitions of self-testing.

\subsection{Self-testing of states}
We are now ready to define what it means to self-test a quantum state.\\[5pt]
\hrule
\begin{definition}\label{st_states_pure}\emph{(self-testing of pure states)}\\[5pt]
The correlations $p(a,b \vert x,y)$ self-test the state $\refstate_{\rA'\rB'}$ if for any state $\physstatedm_{\rA\rB}$ compatible with $p(a,b \vert x,y)$ (for some choice of local measurements) and for any purification $\physstate_{\rA\rB\rP}$ of $\physstatedm_{\rA\rB}$ there exists a local isometry
\begin{equation}
\iso_\rA\tp\iso_\rB:\mathcal{H}_{\rA}\tp\mathcal{H}_{\rB}\rightarrow\mathcal{H}_{\rA'\bar{\rA}}\tp\mathcal{H}_{\rB'\bar{\rB}}  \nonumber 
\end{equation}
such that
\begin{align}
    \iso_\rA\tp\iso_\rB\tp\openone_{\rP}[\physstate_{\rA\rB\rP}]=\refstate_{\rA'\rB'}\otimes\junk_{\bar{\rA}\bar{\rB}\rP} \label{st_pure} 
\end{align}
for some state  $\junk_{\bar{\rA}\bar{\rB}\rP}$. 
\end{definition}
\hrule
\vspace{10pt}

\noindent The above can be understood via the idea of \emph{extraction}. If we trace out the purification space in \eqref{st_pure} then we have 
\begin{align}
    \iso_\rA\tp\iso_\rB\left[\physstatedm_{\rA\rB}\right]=\refstateproj_{\rA'\rB'}\otimes\junkdm_{\bar{\rA}\bar{\rB}},
\end{align}
where $\junkdm_{\bar{\rA}\bar{\rB}}=\tr_\rP\,\proj{\xi}$. Thus, if one self-tests the state $\refstate_{\rA'\rB'}$, then there necessarily exists a local channel (given by the isometry) that allows one to extract $\refstate_{\rA'\rB'}$ from $\physstatedm_{\rA\rB}$ into the ancilla space. The state $\junk_{\bar{\rA}\bar{\rB}\rP}$ in \eqref{st_pure}, which contains everything else from $\physstate_{\rA\rB}$ after the reference state has been extracted, is called a \emph{junk state}. We note that it is not necessary that one actually perform the isometry in the laboratory; all that is needed is a mathematical proof that such as procedure exists in principle. In section \ref{sec:example} we will see how this is possible with an explicit example.  %This is because the isometries used to prove self-testing statements are typically functions of the physical measurement operators, which themselves are unknown (see section \ref{sec:example}) given the device-independent scenario.

Definition \ref{st_states_pure} is the definition that is most commonly used in self-testing works, although the identity channel on the purification space is often left implicit. Note that it is important that the isometry does not act on the purification space since, for example, the purification of a mixed separable state would result in an entangled state, which would give the devices access to entanglement for free. Finally, the above definitions can be straightforwardly generalised to define self-testing of multipartite states; one simply uses an isometry that is local with respect to the subsystems of the multipartite state to be self-tested.

\subsection{Self-testing of measurements}
The correlations that are used to self-test the reference state often allow one to identify the measurements that were performed. Since our physical measurements are projective, this definition will apply to self-testing of projective measurements only. The idea is to prove that under the action of the isometry, the physical measurements map to the reference measurements acting on the reference state. More specifically,
\vspace{10pt}
\hrule
\begin{definition}\label{def_MST2}(\emph{self-testing of states and measurements)}\\[5pt]
The correlations $p(a,b \vert x,y)$ self-test the state and measurements $\refstate_{\rA'\rB'},\{\M'_{a|x}\},\{\N'_{b|y}\}$ if for any state and measurements $\physstatedm_{\rA\rB},\{\M_{a\vert x}\}, \{\N_{b\vert y}\}$ compatible with $p(a,b \vert x,y)$ and for any purification $\physstate_{\rA\rB\rP}$ of $\physstatedm_{\rA\rB}$ there exists a local isometry $\iso_\rA\tp\iso_\rB$ such that
\begin{align}\nonumber
%    &\iso\otimes \openone_{\rP}[\physstate^{\rA\rB\rP}]=\refstate^{\rA'\rB'}\otimes\junk^{\rA\rB\rP} \quad \text{and} \\[4pt] \label{measST2}
\begin{multlined}
    \iso_\rA\tp\iso_\rB\otimes \openone_{\rP}\left[\M_{a\vert x}\tp \N_{b\vert y}\tp\openone_\rP\physstate_{\rA\rB\rP}\right] \\ =\left(\M'_{a\vert x}\tp \N'_{b\vert y}\refstate_{\rA'\rB'}\right)\otimes\junk_{\bar{\rA}\bar{\rB}\rP}
\end{multlined}
\end{align}
for all $a,x,b,y$, for some state $\junk_{\bar{\rA}\bar{\rB}\rP}$. 
\end{definition}
\hrule
\vspace{10pt}
\noindent If we wish to make a statement solely involving the measurements, we will only be able to say something about the part of the measurement that acts on the support of the physical state. Some recent works (see e.g.\ \cite{Jed1}) adopt this approach and use a definition of measurement self-testing in the following spirit (for example, for Alice's measurements).
\vspace{10pt}
\hrule
\begin{definition}\label{def_MST2a}(\emph{self-testing of measurements)}\\[3pt]
The correlations $p(a,b \vert x,y)$ self-test the measurements $\{\M'_{a|x}\}$ for Alice if for any measurements $\{\M_{a\vert x}\}$ and state $\physstatedm_{\rA\rB}$ compatible with $p(a,b \vert x,y)$ there exists a unitary operator $U$ such that
\begin{align}\nonumber
    U_\rA[\tilde{\M}_{a\vert x}]U^{\dagger}_\rA={\M'}_{a\vert x}\otimes\openone
\end{align}
for all $a,x$, where $\tilde{\M}_{a\vert x}=\Pi\,\M_{a\vert x}\,\Pi$ and $\Pi$ is the projector onto the support of $\physstatedm_{\rA}=\tr_{\rB}\,\physstatedm_{\rA\rB}$.
\end{definition}
\hrule
\vspace{10pt}
\subsection{Self-testing via a Bell inequality and the geometry of the set of quantum correlations}\label{st_geometry}
It is known that only those correlations that are extremal points of the quantum set of correlations and are achievable with finite dimensional quantum systems can be used to self-test both a state and measurements \cite{Goh2018}. Such points can often be witnessed by the maximum violation of some Bell inequality over the set of quantum correlations. As a result, one often does not need the full set of probabilities  $p(a,b\vert x,y)$ in order to prove self-testing statements; the maximum quantum violation of a Bell inequality may already imply the existence of the desired isometry. One can thus consider self-testing relative to a Bell inequality by replacing the observation of the correlations by the value of a Bell inequality $\mathcal{I}[p(a,b\vert x,y)]$ in the previous definitions. Many of the well known Bell inequalities, such as CHSH and CGLMP have been used to this effect (see section \ref{sec:bipartite} for such results).

In this light one might ask if the maximal violation of every nontrivial Bell inequality, \emph{i.e.} one which can be violated in quantum theory, is also a self-test of some entangled state. Or even more generally, do all extremal points of the set of quantum correlations self-test some state? These questions are examined in \cite{Goh2018} where it was shown that the relation between self-testing, maximisers of non-trivial Bell inequalities and the boundary of the quantum set is not as simple as one might hope for.  %In particular, there are nontrivial Bell inequalities in the simplest scenario, which do not self-test any quantum state in the sense of definition \ref{st_states_pure}. This is related to the flatness of some of the boundary of the set of quantum correlations. 

We also note here that bipartite mixed state correlations can always be reproduced by a pure state of the \emph{same} dimension \cite{Sikora}. This implies that self-testing of bipartite mixed states following the same spirit as definition \ref{st_states_pure} above is impossible. Since the isometry preserves the purity of the input, applying the isometry to the pure state that gives the same correlations cannot result in the desired mixed state in tensor product with a junk state.

\subsection{Robust self-testing}\label{robust-intro} 
One encounters two problems when trying to prove self-testing statements as defined above from experimental data: (i) the experiment will inevitably contain some level of noise that will dampen the correlations, and (ii) the precise values of $p(a,b\vert x,y)$ will be uncertain due to a finite sample size. In practice, this will mean that proving perfect self-testing of states and measurements is impossible. Point (ii) can be addressed using tools of statistical inference \cite{iid1}. Point (i) can be tackled by proving approximate self-testing statements, and is known as robust self-testing. In short, the aim of robust self-testing is to prove that if the correlations are sufficiently close to the ideal correlations, then the state and measurements must be close (in some well-defined sense) to the desired ones. 

We will focus on two notions of `closeness' that are frequently used in the literature. Our first definition is as follows. Imagine we have identified an isometry that allows us to prove a self-testing statement as in definition \ref{st_states_pure}. If the correlations are close to the ideal then one would expect that the two vectors appearing on either side of \eqref{st_pure} be approximately equal up to some vector norm. This leads to the following.
\vspace{10pt}
\hrule
\begin{definition}\label{robdef2}\emph{(robust self-testing of states, vector norm)}\\[5pt]
The correlations $p(a,b\vert x,y)$ self-test the state $\refstate_{\rA'\rB'}$ with distance $\delta$ in the vector norm $\vert\vert \cdot \vert\vert$ if for any state $\physstatedm_{\rA\rB}$ compatible with $p(a,b \vert x,y)$ and for any purification $\physstate_{\rA\rB\rP}$ of $\physstatedm_{\rA\rB}$ there exists a local isometry $\iso=\iso_\rA\tp\iso_\rB$ 
such that
\begin{align}
    \vert\vert\iso\otimes\openone_{\rP}[\physstate_{\rA\rB\rP}]-\refstate_{\rA'\rB'}\otimes\junk_{\bar{\rA}\bar{\rB}\rP}\vert\vert\leq \delta \nonumber
\end{align}
for some state $\junk_{\bar{\rA}\bar{\rB}\rP}$. 
\end{definition}
\hrule
\vspace{10pt}
\noindent This definition was used as the first definition of robust self-testing.

Our second definition follows the intuition that if in the ideal case one can extract the reference state $\refstate$, then in the noisy case one should be able to extract something close to $\refstate$. Here, it is usually easiest to adopt the fidelity, defined as $F(\ket{\psi},\rho)=\bra{\psi}\rho\ket{\psi}$ as the notion of closeness. First, define $\rho^{\text{\tiny{EXT}}}$ as the extracted state of the ancillas after the application of the isometry, that is,
\begin{align}
   \rho^{\text{\tiny{EXT}}}_{\rA'\rB'} =\tr_{\bar{\rA}\bar{\rB}}\iso [\physstatedm_{\rA\rB}].
\end{align}
We then have the following definition.
\vspace{10pt}
\hrule
\begin{definition}\label{robdef1}\emph{(robust self-testing of states, fidelity)}\\[5pt]
The correlations $p(a,b\vert x,y)$ self-test the state $\refstate_{\rA'\rB'}$ with fidelity $f$ if for any state $\physstatedm_{\rA\rB}$ compatible with $p(a,b\vert x,y)$ there exists a local isometry $\iso=\iso_\rA\tp\iso_\rB$ such that 
\begin{align}
    F\left(\refstate_{\rA'\rB'}, \rho^{\text{\tiny{EXT}}}_{\rA'\rB'}\right)\geq f.
\end{align}
\end{definition}
\hrule
\vspace{10pt}
%\noindent 
Ideally, one would like replace the fidelity in the above definition by the trace distance $T(\rho,\sigma)=\frac{1}{2}\|\rho - \sigma\|_1=\frac{1}{2}\tr[\sqrt{(\rho-\sigma)^\dagger(\rho - \sigma)}]$, since the trace distance is both a metric (unlike the fidelity) and relates directly to the probability of distinguishing the two states. The fact that the fidelity is much more commonly used is because its linearity in the state $\physstatedm$ makes bounds generally much easier to compute. One can nevertheless prove an upper bound to $T(\rho^{\text{\tiny{EXT}}},\refstateproj)$ from a bound on the fidelity using the relation $T \leq \sqrt{1-F}$ \cite{bible}. We point the reader to \cite{Bardyn2009} where a useful discussion about appropriate figures of merit for robust self-testing can be found. 

Note that a local isometry can always prepare any pure product state of the ancillas for free by simply ignoring the physical state and applying the necessary unitaries on the ancilla space. Hence, the best bound achievable via this strategy defines a \emph{trivial bound} that can always be achieved. As an example, consider the task of self-testing the state $\ket{\psi} = \cos\theta\ket{00} + \sin\theta\ket{11}$ for some $\theta \in (0,\pi/4)$. This state has fidelity $\cos^2(\theta)$ and euclidean distance $\sqrt{2-2\cos\theta}$ to the state $\ket{00}$. Thus a self-tested fidelity or distance is interesting only if it surpasses the corresponding bound. Taking definition \ref{robdef1}, this trivial fidelity is equal to the square of the largest Schmidt coefficient of the state.

\subsubsection{Extractability relative to a Bell inequality}\label{subsec:extract}
As with ideal self-testing statements, it is most common to consider robust self-testing relative to a Bell inequality $\mathcal{I}$. For example, taking definition \ref{robdef1} as the figure of merit, one aims to find a function $f(\beta)$ that gives a lower bound on the fidelity as a function of the Bell inequality violation $\mathcal{I}(p(a,b\vert x,y))=\beta$. This can be linked to the notion of \emph{extractability} of the physical state with respect to the reference state for the Bell inequality $\mathcal{I}$. Note that any CPTP map can be realised by  performing an isometry and discarding some degrees of freedom \cite{stinespring_1955}. Thus the map $\tr_{\bar{\rA}\bar{\rB}}\iso [\physstatedm_{\rA\rB}]$ is equivalent to a general local CPTP map $\Lambda_\rA\tp\Lambda_\rB:\mathcal{H}_\rA\tp\mathcal{H}_\rB\rightarrow\mathcal{H}_{\rA'}\tp\mathcal{H}_{\rB'}$ applied to $\physstatedm_{\rA\rB}$. Given a physical state $\physstatedm$, the extractability $\Xi$ is the maximum fidelity of $\Lambda_\rA\tp\Lambda_\rB [\physstatedm]$ and $\refstate$ over all CPTP maps:
\begin{equation}
    \Xi(\physstatedm \rightarrow \refstate) = \max_{\Lambda_\rA,\Lambda_\rB}F(\Lambda_\rA\otimes\Lambda_\rB[\physstatedm],\refstate).
\end{equation}
To get optimal robust self-testing statements for a given inequality $\mathcal{I}$ one therefore needs to minimise the extractability over all states compatible with $\mathcal{I}=\beta$ for all values of $\beta$. This leads to the extractability-violation trade-off function
\begin{equation}\label{tradeof1}
    \mathcal{Q}_{\psi,\Bb_{\mathcal{I}}} = \inf_{\physstatedm|\tr[\Bb_{\mathcal{I}}\physstatedm] = \beta}\Xi(\physstatedm \rightarrow \refstate).
\end{equation}
Finding the extractability-violation trade-off function for a given Bell inequality is a difficult task since it involves a minimisation of the fidelity over all compatible states and a maximisation over all possible CPTP maps. Moreover, the optimal CPTP map (or equivalently, isometry) generally depends on the observed violation. More commonly, one fixes a single isometry for all violations and minimises the fidelity only, leading to a sub-optimal curve.  

Finally, one can use similar ideas to the above to define the robust self-testing of measurements. We do not give any definition here, but point the reader to section \ref{robmeas} for work in this direction. 

\subsection{Generalisations and alternative definitions}\label{sec:generalisations}
\subsubsection{The issue of complex conjugation}\label{cc}
 When self-testing quantum states in the bipartite scenario, it is sufficient to consider real reference states only, i.e.\ states such that  $\refstate=\refstate^*$, where $*$ denotes complex conjugation with respect to a fixed basis. This follows since all pure states are local unitary equivalent to a real state via the Schmidt decomposition \cite{Preskill_1998}. A similar argument for measurements however is not possible. As a result, definition \ref{def_MST2} suffers from a serious drawback; it can only be used to self-test sets of measurements that are invariant under the complex conjugation of all measurement operators. To see this note that since $p(ab\vert xy)=(p(ab\vert xy))^*$ then (assuming a real state $\refstate$)
\begin{align}\nonumber
    p(ab\vert xy)=&\tr\left[\refstateproj \;\M'_{a\vert x}\otimes \N'_{b\vert y}\right]\\=&\tr\left[\refstateproj \;(\M'_{a\vert x})^*\otimes (\N'_{b\vert y})^*\right]. \label{conjmeas}
\end{align}
Thus any correlations obtained using $\{\refstate,\M'_{a\vert x},\N'_{b\vert y}\}$ can also be obtained using $\{\refstate,(\M'_{a\vert x})^*,(\N'_{b\vert y})^*\}$. These two realisations are generally not equivalent under local unitary operations. In this case, one cannot self-test the set $\{\refstate,\M'_{a\vert x},\N'_{b\vert y}\}$ using definition \ref{def_MST2} since there is always another realisation $\{\refstate,(\M'_{a\vert x})^*,(\N'_{b\vert y})^*\}$ that is not related to the first via a local isometry but results in the same correlations. 

A straightforward solution to this problem first proposed in \cite{McKague2011} is to generalise the definition of measurement self-testing so that one self-tests the measurements $\{\M'_{a\vert x},\N'_{b\vert y}\}$ if one can show that on the support of $\refstate$, the physical measurements act as some unknown convex combination of $\{\M'_{a\vert x},\N'_{b\vert y}\}$ and $\{(\M'_{a\vert x})^*,(\N'_{b\vert y})^*\}$. This is in line with the general spirit of self-testing in which one aims to certify the measurements up to all the intrinsic limitations of the device-independent scenario. See appendix \ref{App:complexdef} for a possible definition along these lines and section \ref{Pauli} for an example of such a self-test. 

In principle, there may be more state and measurement transformations other than complex conjugation that do not affect the observed probabilities. Determining this set is still an open problem. While in the case of qubit bipartite systems one can aim at self-testing states and measurements up to local isometries and complex conjugations, it is unclear if more transformations may be present when considering higher dimensional systems or multipartite scenarios.

\subsubsection{Self-testing via simulation}\label{simulation}
Another recent approach presented in \cite{Jed2} is to adopt the philosophy that self-testing a reference state or measurements should imply that the physical state or measurements be capable to \emph{simulate} the reference state or measurements. For states, this translates to finding a local quantum channel $\Lambda_\rA\otimes\Lambda_\rB$ that maps the physical state to the reference state, thus allowing the simulation of any measurement on the reference state by first applying the channel followed by the desired measurement. Note that this definition is equivalent to definition \ref{st_states_pure} since via Stinespring's dilation theorem \cite{choi_1975,stinespring_1955} any local channel can be realised by first applying a local isometry then tracing out any irrelevant degrees of freedom. 

For measurements, one considers unital channels, i.e. quantum channels that preserve the identity (and thus map sets of measurements to sets of measurements). The idea is then (say, for Alice) that if one can find a unital channel such that $\Lambda[\M_{a\vert x}]=\M'_{a\vert x}$, then one can simulate the reference measurement  $\M'_{a\vert x}$ on any state by first applying the dual quantum channel $\Lambda^\dagger$ on the state followed by the physical measurement $\M_{a\vert x}$. Since
\begin{align}
    \tr\left[\Lambda^\dagger[\varrho]\M_{a\vert x}\right]=\tr\left[\varrho \Lambda[\M_{a\vert x}]\right]=\tr\left[\varrho \M'_{a\vert x}\right]
\end{align}
one recovers the same statistics as making the reference measurement on any state $\varrho$. This approach was used in \cite{MO} to self-test the Bell state measurement (see section \ref{sec:StEntMeas}) .

\subsubsection{Measurement self-testing based on commutation}\label{def:meascom}
Yet another approach to measurement self-testing focuses on certifying that the physical measurements satisfy some desired commutation relation on the support of the physical state. This can be advantageous, as commutation relations are often the only relevant features that one is interested in, and since they are invariant under isometry maps the approach can lead to simpler proofs of self-testing. Furthermore, in the case of perfect statistics, certifying a particular commutation relation may be enough to prove full self-testing statements of the form of definition \ref{def_MST2}. This approach has been used to prove measurement self-testing statements for anti-commuting qubit observables \cite{Jed1} and sets of mutually unbiased bases in dimension 3 \cite{maxmaxmax}. It is very close in spirit to one of the earliest self-testing statements given in \cite{Popescu1992}.  For further discussion on this technique, see section \ref{comm}.

%\section{for Box: SOS decompositions and Bell inequalities}

\begin{boxfigure}[label=box:bell]{Bell Nonlocality and the CHSH inequality}
Bell nonlocality is a counter-intuitive property of quantum correlations discovered by John Bell in 1964 \cite{bell}. The correlations $p(a,b\vert x,y)$ are called \emph{local} if they can be reproduced by  shared classical information. To formalise this, we represent the shared information by a classical random variable $\Lambda\sim \pi(\lambda)$. Averaging over this information, the possible correlations that Alice and Bob can achieve is given by (see (a) below for the corresponding classical causal network)
\begin{align}\label{localdecomp}
    p(a,b\vert x,y)=\int_{\Lambda}\pi(\lambda)p_{A}(a\vert x,\lambda)p_{B}(b\vert y,\lambda) \text{d}\lambda. 
\end{align}
%Note that Alice's probability to output $a$, $p_{A}(a\vert x\lambda)$, depends only on the local information $x$ and $\lambda$ that is available to her (and similarly for Bob). 
Notice that local measurements on any separable state $\rho=\int_\Lambda\text{d}\lambda\pi(\lambda) \sigma_{\lambda}^\rA\tp\sigma_{\lambda}^{\rB}$ lead to correlations of the above form, and so \eqref{localdecomp} is precisely those correlations that can be achieved using separable states.  If we collect all of the probabilities into a single vector $\vec{p}=(p(00\vert 00),p(01\vert 00),\cdots)$ then the set of local correlations forms a convex polytope, the facets of which are called \emph{Bell inequalities} (see (b), below). Remarkably, if Alice and Bob share an entangled quantum system, they may produce correlations that are \emph{nonlocal}, i.e.\ which violate a Bell inequality and therefore cannot be written in form \eqref{localdecomp}. An important Bell inequality, called the CHSH Bell inequality \cite{chsh}, already exists in the simplest scenario in which Alice and Bob have two inputs ($x,y=0,1$) and two outcomes ($a,b=\pm1)$. It is given by
\begin{align}
    \beta_{\tx{CHSH}}=\La \A_0\B_0\Ra + \La \A_1\B_0\Ra + \La \A_0\B_1\Ra - \La \A_1\B_1\Ra \leq 2,
\end{align}
where $\La \A_x \B_y \Ra=\sum_{a,b}a\cdot b \, p(ab\vert xy)$ denotes the \emph{correlator} for the inputs $x$, $y$. Measurements on the maximally entangled state $\maxent=[\ket{00}+\ket{11}]/\sqrt{2}$ can violate this inequality up to $\beta_{\tx{CHSH}}=2\sqrt{2}$. For a comprehensive review on Bell nonlocality,  see \cite{bellreview}.
\begin{center}
    \includegraphics[width=\textwidth]{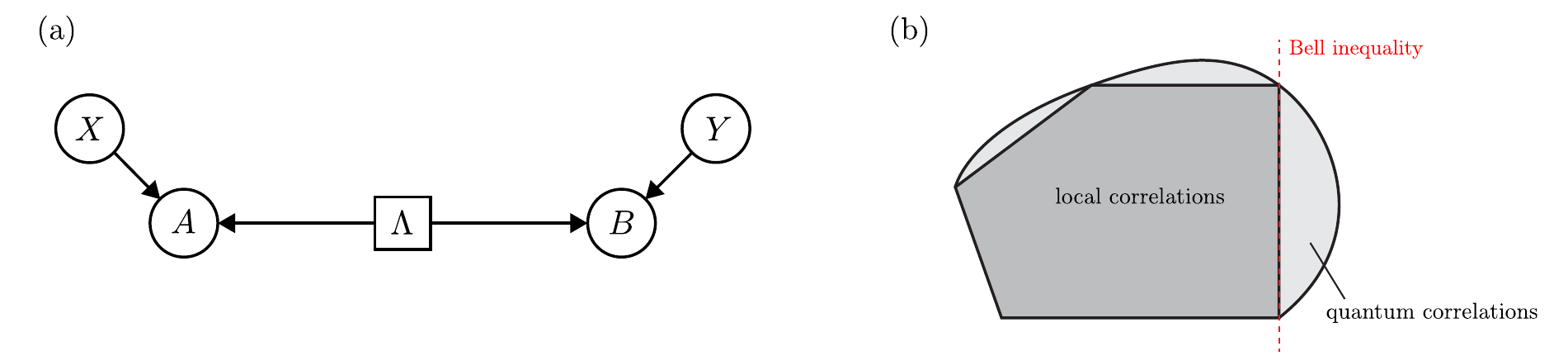}
\end{center}
\end{boxfigure}

\section{A first example}
\label{sec:example}
The maximally entangled state of two-qubits
\begin{equation}
\maxent = \frac{1}{\sqrt{2}}(\ket{00}+ \ket{11})
\end{equation}
is the quantum state which is most emblematic of the significance entanglement has in quantum theory and is used in a wide class of information processing protocols \cite{teleportation,sdc,ekert}. In this section we show how to prove formal self-testing statements for this state and locally anti-commuting observables. Many of the techniques used to self-test more complex states and measurements can be understood as a generalisation of those presented here. We work in a simple scenario in which Alice and Bob each have two inputs ($x,y=0,1)$ and two outputs ($a,b=\pm1)$. We chose the convention of  having $\pm1$ valued outcomes since it will be convenient to work with the observables
\begin{align}\label{obsdefine}
    \A_{x}=\M_{+\vert x}-\M_{-\vert x}\; ;\quad \B_{y}=\N_{+\vert y}-\N_{-\vert y},
\end{align}
where $\M_{a\vert x}$, $\N_{b\vert y}$ are the physical measurements operators in \eqref{pab}. Note that since the physical measurement operators are projective (see section \ref{sec:scenario}), the operators $\A_{x}$, $\B_{y}$ are by construction Hermitian and unitary. We thus have
\begin{align}\label{obs_properties}
\A_{x}^\dagger=\A_{x}\;;\quad \A_{x}^2=\openone\;;\quad \B_{y}^\dagger=\B_{y}\;;\quad \B_{y}^2=\openone.
\end{align}
Following definition \ref{st_states_pure} we work with a purification $\physstate_{\rA\rB\rP}$ of the physical state where the measurements act trivially on $\mathcal{H}_{\rP}$. In the following we do not explicitly write the identity on the purification space, e.g.\ $\A_x$ should be understood as $\A_x\tp\openone_\rP$.

The central object we will use for self-testing is the CHSH Bell inequality \cite{chsh} (see box \ref{box:bell} for a summary of Bell nonlocality)
\begin{equation}\label{chsh}
\beta_{\text{CHSH}} = \La \A_0\B_0\Ra + \La \A_1\B_0\Ra + \La \A_0\B_1\Ra - \La \A_1\B_1\Ra \leq 2. \nonumber
\end{equation}
Performing certain local measurements on the state $\maxent$ leads to a violation of $\beta_{CHSH}=2\sqrt{2}$.
More specifically, both Alice and Bob use anticommuting measurement observables to achieve this violation. Alice measures $\A_0=\sx$ and $\A_1=\sz$, while Bob measures $\B_0=(\sz+\sx)/\sqrt{2}$ and $\B_1=(\sx-\sz)/\sqrt{2}$. The reverse statement, that the violation $2\sqrt{2}$ can only be achieved by measurements applied on $\maxent$, represented the first self-testing statement. Early proofs of this statement can be found in \cite{Summers1987}, \cite{Popescu1992}, \cite{BMR} and \cite{Tsirelson1993}, a decade and a half before the term self-testing was coined. 

In the following we show how to self-test $\maxent$ from correlations achieving the maximal violation of the CHSH inequality. The central step in the proof will be to show that Alice and Bob's local observables anticommute on the support of their shared state, i.e.\ $\{\A_0,\A_1\}\physstate=\{\B_0,\B_1\}\physstate=0$. We present two methods to achieve this; (i) a geometrical argument for anticommutativity of Bob's observables (section \ref{geo}), and (ii) an algebraic argument (section \ref{alg}). Once this is achieved, the anticommuting observables can be used to build the required local isometry that is needed to prove a formal self-testing statement (section \ref{swap}). 

\subsection{Geometrical proof of anticommutativity}\label{geo}

In this section we give a simple geometric proof that the correlations maximally violating the CHSH inequality can be achieved only by measuring locally anticommuting observables. This will require knowing all correlations, not only the observation that $\beta_{CHSH}=2\sqrt{2}$. %The proof relies on knowing all CHSH correlations instead of the violation of the CHSH inequality. 
The ideal correlations achieving this violation are
\begin{equation}\label{CHSHcorr}
\begin{split}
\physstatebra \A_0\B_0 \physstate = \frac{1}{\sqrt{2}}, &\quad \physstatebra \A_0\B_1 \physstate = \frac{1}{\sqrt{2}}, \\
\physstatebra \A_1\B_0 \physstate = \frac{1}{\sqrt{2}}, &\quad \physstatebra \A_1\B_1 \physstate = -\frac{1}{\sqrt{2}}.
\end{split} 
\end{equation}
Let us define vectors
\small
\begin{align*}
&\vec{a}_0 \equiv \frac{1}{\sqrt{2}}(\A_0+\A_1)\physstate, &&\quad \vec{a}_1 \equiv \frac{1}{\sqrt{2}}(\A_0-\A_1)\physstate,\nonumber \\
&\vec{b}_0 \equiv \B_0\physstate, &&\quad \vec{b}_1 \equiv \B_1\physstate. \label{vec_def}
\end{align*}
\normalsize
Equations \eqref{CHSHcorr} imply the following inner product values:
\begin{equation}\label{scProd}
\vec{a}_0\cdot {\vec{b}_0}^\dagger = 1, \qquad\qquad  \vec{a}_1\cdot {\vec{b}_1}^\dagger = 1.
\end{equation}
The Cauchy-Bunyakovski-Schwarz inequality $\vec{a}\cdot {\vec{b}}^\dagger \leq |\vec{a}||\vec{b}|$ implies 
\begin{equation*}
  |\vec{a}_i||\vec{b}_i| \geq 1 \qquad \textrm{for} \quad i = 0,1, 
\end{equation*}
where $|\vec{a}_i| = \sqrt{\vec{a}_i\cdot{\vec{a}_i}^\dagger}$. Since operators $\A_i$ and $\B_j$ are unitary, vectors $\vec{b}_0$ and $\vec{b}_1$ have unit norm, which implies 
\begin{equation}\label{norms}
|\vec{a}_i| \geq 1, \qquad \textrm{for} \quad i = 0,1.
\end{equation}
The norms of the vectors $\vec{a}_0$ and $\vec{a}_1$ satisfy
\begin{equation}
|\vec{a}_0|^2 + |\vec{a}_1|^2 = 2
\end{equation}
by construction, which together with \eqref{norms} implies $|\vec{a}_0| = |\vec{a}_1| = 1$. Since eqs. \eqref{scProd} represent the saturation of the Cauchy-Bunyakovski-Schwarz inequality, vectors $\vec{a}_i$ and $\vec{b}_i$ for $i = 0,1$ must be parallel, i.e. $\vec{b}_i = \vec{a}_i$. This implies 
\begin{align}\label{acRel}
&\{\B_0,\B_1\}\physstate = (\B_0\B_1+\B_1\B_0)\physstate\\
=& \frac{(\A_0-\A_1)\B_0 + (\A_0+\A_1)\B_1}{\sqrt{2}}\physstate\nonumber\\
=& \frac{(\A_0-\A_1)(\A_0+\A_1) + (\A_0+\A_1)(\A_0-\A_1)}{{2}}\physstate \nonumber\\
=& 0,\nonumber
\end{align}
and thus $\B_0$ and $\B_1$ anti-commute on the support of $\physstate$. Note that since the correlations are symmetric, the same result holds for Alice's observables. 

\begin{boxfigure}[label=box:sos]{SOS decompositions}
To every Bell inequality $\mathcal{I}=\sum_{a,b,x,y}w_{ab}^{xy}p(a,b\vert x,y)$ corresponds a Bell operator 
\begin{equation}
    \mathcal{B} = \sum_{a,b,x,y}w_{ab}^{xy}\,\M_{a\vert x}\otimes\N_{b\vert y}
\end{equation}
such that the violation is obtained as $\beta = \tr\left[\mathcal{B}\rho\right]$. If the maximal violation achievable by using quantum resources (\emph{i.e.} the quantum bound) is $\beta_Q$ the \textit{shifted Bell operator} is defined as $ \beta_Q\openone - \mathcal{B}$. Every shifted Bell operator is by construction positive semidefinite since $\bra{\psi}\mathcal{B}\ket{\psi}\leq \beta_{Q}$ for all $\ket{\psi}$. Imagine the shifted Bell operator admits a decomposition
\begin{equation}\label{boxSOS}
    \beta_Q\openone - \mathcal{B} = \sum_{\lambda}P_{\lambda}^\dagger P_\lambda,
\end{equation}
where each $P_\lambda$ is a polynomial in the operators $\M_{a\vert x}$ and $\N_{b\vert y}$. The decomposition \eqref{boxSOS} is called a \emph{sum of squares (SOS) decomposition} of the shifted Bell operator. If the polynomials are of degree at most $n$ in either $\M_{a\vert x}$ or $\N_{a\vert x}$ we say the SOS decomposition is of $n$-th degree.  

SOS decompositions for Bell inequalities are typically hard to find. One can use numerical methods to find SOS decompositions of various degrees via the NPA hierarchy \cite{npa1,npa2} (in particular, see \cite{pna} for a link to SOS decompositions).

SOS decompositions allow one to extract potentially useful information about the physical state $\physstate$ and measurements used to achieve the maximal violation of the corresponding Bell inequality. From \eqref{boxSOS} we have
\begin{align}\nonumber
 \physstatebra   \beta_Q\openone - \mathcal{B} \physstate = 0 \qquad &\Rightarrow \quad \sum_\lambda \physstatebra P_{\lambda}^\dagger P_\lambda \physstate = 0 \quad \Rightarrow \qquad \sum_\lambda \|P_\lambda\physstate\|^2 = 0\\ \label{sosimplie}
 &\Rightarrow \qquad P_\lambda\physstate = 0 \qquad \forall \lambda
\end{align}
Since $P_\lambda$ is a function of the operators used to obtain the maximal violation, the relations of the form  $\{P_\lambda\physstate = 0\}_\lambda$ often represent nontrivial statements about the strategy used to maximally violate the Bell inequality under consideration. 
Additionally, if a non-maximal violation $\beta_Q - \epsilon$ is observed the approximate relations analogous to  \eqref{sosimplie} can be obtained:
\begin{align}\label{sosimpliesapprox}
 \physstatebra   \beta_Q\openone - \mathcal{B} \physstate = \epsilon \qquad &\Rightarrow \qquad \| P_\lambda\physstate\| \leq \sqrt{\epsilon} \qquad \forall \lambda
\end{align}
These relations are often significant for proving robust self-testing statements.
\end{boxfigure}

\iffalse
\begin{figure*}[t]
\centering
\includegraphics[scale=1.1]{newcircuit.pdf}
\caption{Swap gate isometry used to self-test the maximally entangled state of two qubits. After the application of the circuit, the maximally entangled state is extracted from $\physstate$ to the ancilla qubits.\label{fig:example}}
\end{figure*}
\fi
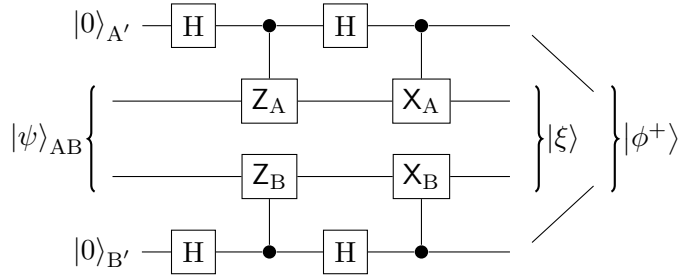
\begin{figure*}
  \centerline{
    \begin{tikzpicture}
    %
    % `operator' will only be used by Hadamard (H) gates here.
    % `phase' is used for controlled phase gates (dots).
    % `surround' is used for the background box.
    \tikzstyle{operator} = [draw,fill=white,minimum size=1.5em] 
    \tikzstyle{phase} = [fill,shape=circle,minimum size=5pt,inner sep=0pt]
   % \tikzstyle{surround} = [fill=blue!10,thick,draw=black,rounded corners=2mm]
   %
   % Qubits
   \node at (0.3,0) (q1) {$\ket{0}_{\rA'}$ };
   %\node at (0,-1) (q11) {$\ket{+}^{\rC_i''}$ };
   \node at (0.3,-1) (q2) {};
   \node at (0.3,-2) (q3) {};
   \node at (0.3,-3) (q4) { $\ket{0}_{\rB'}$};
   \draw[decorate,decoration={brace, mirror},thick] (0.2,-0.8) to 	node[midway,left] (bracket) {$\ket{\psi}_{\rA\rB}$} (0.2,-2.2);
    % Column 1
    %\node[phase] (phase11) at (1.5,0) {} edge [-] (q1);
    %\node[operator] (phase12) at (1.5,-1) {${\tilde{\X}}_{1,A}$} edge [-] (q2);
    %\draw[-] (phase11) -- (phase12);
    %\node[phase] (phase13) at (1.5,-3) {} edge [-] (q4);
    %\node[operator] (phase14) at (1.5,-2) {${\tilde{\X}}_{1,B}$} edge [-] (q3);
    %\draw[-] (phase13) -- (phase14);
    % Column 2
    \node[operator] (op12) at (1.5,0) {H} edge [-] (q1);
    \node[operator] (op22) at (1.5,-3) {H} edge [-] (q4);
    %
 % Column 3
    \node[phase] (phase21) at (2.5,0) {} edge [-] (op12);
    \node[operator] (phase22) at (2.5,-1) {${{\Zz}}_{\rA}$} edge [-] (q2);
    \draw[-] (phase21) -- (phase22);
    \node[phase] (phase23) at (2.5,-3) {} edge [-] (op22);
    \node[operator] (phase24) at (2.5,-2) {${{\Zz}}_{\rB}$} edge [-] (q3);
    \draw[-] (phase23) -- (phase24);
     % Column 4
    \node[operator] (op14) at (3.5,0) {H} edge [-] (phase21);
    \node[operator] (op24) at (3.5,-3) {H} edge [-] (phase23);
    %  
 % Column 5
    \node[phase] (phase31) at (4.5,0) {} edge [-] (op14);
    \node[operator] (phase32) at (4.5,-1) {${{\Xx}}_{\rA}$} edge [-] (phase22);
    \draw[-] (phase31) -- (phase32);
    \node[phase] (phase33) at (4.5,-3) {} edge [-] (op24);
    \node[operator] (phase34) at (4.5,-2) {${{\Xx}}_{\rB}$} edge [-] (phase24);
    \draw[-] (phase33) -- (phase34);
    \node (end1) at (5.8,0) {} edge [-] (phase31);
    %\node (end5) at (9,-1) {} edge [-] (op81); 
    \node (end2) at (5.8,-1) {} edge [-] (phase32);
    \node (end3) at (5.8,-2) {} edge [-] (phase34);
    %\node (end6) at (9,-4) {} edge [-] (op82);
    \node (end4) at (5.8,-3) {} edge [-] (phase33);
    %
    % Bracket
    \draw[decorate,decoration={brace},thick] (6,-0.8) to 	node[midway,right] (bracket) {$\ket{\xi}$} (6,-2.2);
    \node at (5.82,0) (eend1) {};
    \node at (6.9,-1) (eend11) {};
    \draw[-] (eend1) -- (eend11);
    \node at (5.82,-3) (eend4) {};
    \node at (6.9,-2) (eend44) {};
    \draw[-] (eend4) -- (eend44);
    \draw[decorate,decoration={brace},thick] (7,-0.8) to 	node[right] (bracket) {$\maxent$} 	(7,-2.2);
    %
    % Background Box
  %  \begin{pgfonlayer}{background} 
   % \node[surround] (background) [fit = (q1) (op31) (bracket)] {};
  %  \end{pgfonlayer}
    %
    \end{tikzpicture}
  }
  \caption{
   The partial swap gate isometry used to self-test the maximally entangled state of two qubits. H is the Hadamard gate. After the application of the circuit, the maximally entangled state is extracted from $\physstate$ to the ancilla qubits. \label{fig:example}
  }
\end{figure*}

\subsection{Algebraic proof of anticommutativity}\label{alg}
In principle, it is not easy to find correlations which self-test some state and measurements. %It is clear that the correlations must be extremal. 
Natural candidates, however, are correlations that maximally violate a particular Bell inequality. Moreover, the structure of the Bell inequality can be useful for proving self-testing statements, especially in cases when simple geometric considerations are not possible. Here we show how one can deduce an  anticommutation relation for Bob's observables from the observation $\beta_{\tx{CHSH}}=2\sqrt{2}$. As a starting point we take the SOS decomposition of the shifted CHSH Bell operator (see box \ref{box:sos} for a summary of SOS decompositions):
\small
\begin{multline}\label{sos}
2\sqrt{2}\openone - \mathcal{B}_{\tx{CHSH}} = \\ \quad\frac{1}{\sqrt{2}}\left[\left(\frac{\A_0+\A_1}{\sqrt{2}} - \B_0\right)^2 + \left(\frac{\A_0-\A_1}{\sqrt{2}} - \B_1\right)^2\right],
\end{multline}
\normalsize
which follows from the properties \eqref{obs_properties}. For any state $\ket{\psi}$ leading to $\beta_{\tx{CHSH}}=2\sqrt{2}$, i.e. $\bra{\psi}\mathcal{B}_{\tx{CHSH}}\ket{\psi} = 2\sqrt{2}$ we thus have 
\begin{equation}\label{a=b}
 \frac{\A_0 \pm \A_1}{\sqrt{2}}\ket{\psi} = \B_{0/1}\ket{\psi},
\end{equation}
as explained in equation \eqref{sosimplie} in box \ref{box:sos}. With these relations we can prove that $\B_0$ and $\B_1$ anticommute in the same way we did in \eqref{acRel}.

%Define inequality, SOS decomposition, provide SOS
\subsection{Swap gate}\label{swap}

We now prove a formal self-testing statement for the state $\maxent$ in the form of definition \ref{st_states_pure}. This will require proving the existence of an isometry $\iso$  mapping the physical state $\physstate$ to our reference state $\refstate=\maxent$. In the majority of self-testing proofs the isometry is explicitly constructed and in most cases it takes the form of the partial Swap gate given in figure \ref{fig:example}. 

The main idea behind this particular isometry is as follows. In the case that the physical state is a two-qubit state and the operators are ${\Zz}_{\rA}=\sz^{\rA}$, ${\Xx}_{\rA}=\sx^{\rA}$, ${\Zz}_{\rB}=\sz^{\rB}$ and ${\Xx}_{\rB}=\sx^{\rB}$, the action of the circuit is to swap the physical state with the state $\ket{00}$ of the registers $\rA$ and $\rB$. Of course, given the device-independent scenario we cannot assume that the physical state is a two-qubit state or any particular form of the operators. However, from sections \ref{geo} and \ref{alg}, we know that like $\sz$, $\sx$, the operators $\A_0$, $\A_1$ and $\B_0$, $\B_1$ anti-commute on the support of the state. The idea is then to use these operators to create new operators ${\Zz}_{\rA}$, ${\Xx}_{\rA}$, ${\Zz}_{\rB}$, ${\Xx}_{\rB}$ which act in an analogous way to $\sz$, $\sx$ on $\physstate$. Since we expect our physical state to be $\maxent$ (up to a local isometry), the hope is that by using these operators in the place of $\sz$ and $\sx$ one will still be able to extract $\maxent$ into the ancilla space. Indeed, this is the case. More precisely, we choose 
\begin{align}
&{\Zz}_{\rA} = \frac{1}{\sqrt{2}}(\A_0 + \A_1), \quad &&\Xx_{\rA} = \frac{1}{\sqrt{2}}(\A_0 - \A_1), \nonumber \\ &\Zz_{\rB} = \B_0, \quad &&\Xx_{\rB} = \B_1. \label{swap_ops}
\end{align}
Note that we have 
\begin{equation}\label{zaxa}
    \{\Zz_{\rA},\Xx_{\rA}\} = 0
\end{equation}
by construction and 
\begin{equation}\label{xbzb}
    \{\Zz_{\rB},\Xx_{\rB}\}\physstate = 0
\end{equation}
from \eqref{acRel}. Furthermore from \eqref{a=b} we have
 \begin{align}\label{zzxx}
 \Zz_{\rA}\physstate = \Zz_{\rB}\physstate, \quad \Xx_{\rA}\physstate = \Xx_{\rB}\physstate.
 \end{align}

In order for $\iso$ to be a valid isometry,  $\Zz_{\rA}$, $\Xx_{\rA}$, $\Zz_{\rB}$ and $\Xx_{\rB}$ must be unitary. This is automatically the case for $\Zz_{\rB}$ and $\Xx_{\rB}$ (see \eqref{obs_properties}), however it is not necessarily the case for $\Zz_{\rA}$ and $\Xx_{\rA}$. To deal with this problem, we need to \emph{regularise} these operators so that they are unitary. Formally, to regularise a Hermitian operator $Z$, one changes all zero eigenvalues of $Z$ to $1$, resulting in a new Hermitian operator $Z^*$. The regularised operator is then obtained by normalising the eigenvalues of $Z^*$, i.e. $\hat{Z}=\vert Z^* \vert^{-1} Z^*$. Note that $\hat{Z}$ is unitary by construction. One can often show that the regularised operators act in same way as the original operators on the physical state, i.e. $\hat{Z}\physstate=Z\physstate$. This is indeed the case for our example (see appendix \ref{regularisation}). From hereon we therefore take $\Zz_{\rA}$ and $\Xx_{\rA}$ to be the regularised unitary operators and continue to use the substitutions \eqref{swap_ops} without problem.

After a straightforward calculation one finds that the output of the isometry is
\small
\begin{multline*}
    \iso[\physstate] = \frac{1}{4}\bigg[\ket{00}\otimes (\openone+ \Zz_{\rA})(\openone+\Zz_{\rB})\physstate \\
    + \ket{01}\otimes (\openone+ \Zz_{\rA})\Xx_{\rB}(\openone-\Zz_{\rB})\physstate\\
    + \ket{10}\otimes \Xx_{\rA}(\openone- \Zz_{\rA})(\openone+\Zz_{\rB})\physstate\\
    + \ket{11}\otimes \Xx_{\rA}(\openone- \Zz_{\rA})\Xx_{\rB}(\openone-\Zz_{\rB})\physstate\bigg],
\end{multline*}
\normalsize
or in more compact form
\begin{equation}\label{swoutput}
        \iso[\physstate] = \sum_{i,j \in \{0,1\}}\ket{ij}_{\rA'\rB'}\otimes\hat{f}_{ij}\physstate_{\rA\rB\rP} ,
\end{equation}
where 
\begin{align*}\label{fij}
\hat{f}_{ij} = \frac{1}{4}\Xx_{\rA}^i(\openone+(-1)^i\Zz_{\rA}) \Xx_{\rB}^j(\openone+(-1)^j\Zz_{\rB}).
\end{align*}
Using  \eqref{zzxx} expressions of the form $(\openone\pm \Zz_{\rA})(\openone\mp\Zz_{\rB})\physstate$ are automatically equal to zero, setting $\hat{f}_{01}\physstate = \hat{f}_{10}\physstate = 0$. The expression $\hat{f}_{11}\physstate$ can be simplified in the following manner:
\begin{align*}
    \hat{f}_{11}\physstate &= \frac{1}{4}\Xx_{\rA}(\openone- \Zz_{\rA})\Xx_{\rB}(\openone-\Zz_{\rB})\physstate\\
    &= \frac{1}{4}(\openone+ \Zz_{\rA})\Xx_{\rA}(\openone+\Zz_{\rB})\Xx_{\rB}\physstate\\
    &= \frac{1}{4}(\openone+ \Zz_{\rA})(\openone+\Zz_{\rB})\physstate\\
    &= \hat{f}_{00}\physstate
\end{align*}
The second line is obtained by using anticommutativity relations \eqref{zaxa} and \eqref{xbzb}, while \eqref{zzxx} and the unitarity of $\Xx_\rB$ was used to obtain the third line. 
Finally we see that the output of the Swap isometry is
\begin{equation}\label{fin}
\Phi[\physstate_{\rA\rB\rP}] = \maxent_{\rA'\rB'}\otimes \ket{\xi}_{\rA\rB\rP},
\end{equation}
where $\ket{\xi} = \sqrt{2}\hat{f}_{00}\physstate$. 
Note that $\ket{\xi}$ is necessarily normalised since the circuit of figure \ref{fig:example} is unitary. We have thus self-tested the state $\maxent$ in the sense of definition \ref{st_states_pure}. Although we have worked with a purification of the physical state, the isometry does not act on the purification space, as needed from definition \ref{st_states_pure}. This is because $\iso$ is constructed from the measurement operators themselves, which by assumption act only on the local Hilbert spaces of Alice and Bob and therefore not on the purification space of $\physstate$.

%Before moving to self-testing of measurements few comments are in order. It is noted that the SWAP gate is a valid isometry if all the operators $\tilde{Z}_{\rA/\rB}$ and $\tilde{X}_{\rA/\rB}$ are unitary. While in our example $\tilde{Z}_{\rB}$ and $\tilde{X}_{\rB}$ are unitary by choice, since they are dichotomic measurement observables, the unitarity of the operators $\tilde{X}_{\rA}$ and $\tilde{Z}_{\rA}$ on the support of the state $\physstate$ has to be proven. This can be proven in a simple manner by utilizing eqs. \eqref{zz} and \eqref{xx}. Proving the unitarity of such observables is more complicated when the partial SWAP gate is used for robust self-testing. In such cases the technique of regularization is used, which will be discussed in more details in Sec. \ref{ni}.\\

\subsubsection{Partial vs full Swap gates}
The partial Swap gate was used in self-testing protocols for the first time in \cite{McKague2012} and in a large number of self-testing proofs since then. The full Swap gate differs from the partial one in that it contains another controlled gate before the first Hadamard is applied to the ancillary qubit. This controlled gate can be omitted if the ancilla is initiated in the state $\ket{0}$. In order to get better robust self-testing protocols it might be useful that Alice and Bob each have a local pair of maximally entangled ancillas. In this case the full Swap gate has to be used (see \cite{McKagueDQC}). The generalisation of the Swap gate useful for self-testing states of local dimension larger than two is introduced in \cite{Yang}. For more details on different types of Swap gates used for self-testing see Appendix \ref{sec:differentswaps}.

\subsection{Self-testing of measurements}\label{exmeas}

The measurements Alice and Bob use to maximally violate the CHSH inequality can also be self-tested via the Swap isometry. Here we explicitly show how to self-test Bob's measurement observable $\B_0$. For that purpose we check how the partial Swap gate transforms the state $\B_0\physstate$, which can also be written as $\Zz_{\rB}\physstate$:
\begin{align*}
        \iso[\B_0\physstate] &= \sum_{i,j \in \{0,1\}}\ket{ij}_{\rA'\rB'}\hat{f}_{ij}\B_0\physstate\\
        &= \sum_{i,j \in \{0,1\}}\ket{ij}_{\rA'\rB'}\hat{g}_{ij}\physstate,
\end{align*}
where 
\small
\begin{align*}
\hat{g}_{ij} &= \frac{1}{4}\Xx_{\rA}^i(\openone+(-1)^i\Zz_{\rA})\otimes \Xx_{\rB}^j(\openone+(-1)^j\Zz_{\rB})\Zz_{\rB}\\
%&= (-1)^j\frac{1}{4}\tilde{X}_{\rA}^i(\mathbb{1}+(-1)^i\tilde{Z}_{\rA})\otimes \tilde{X}_{\rB}^j(\mathbb{1}+(-1)^j\tilde{Z}_{\rB})\tilde{Z}_{\rB}\\
&= \frac{1}{4}\Xx_{\rA}^i(\openone+(-1)^i\Zz_{\rA})\otimes \Xx_{\rB}^j((-1)^j\openone+\Zz_{\rB})\\
&= (-1)^j\hat{f}_{ij}.
\end{align*}
\normalsize
This relation implies $\hat{g}_{01} = \hat{g}_{10} = 0$ and $\hat{g}_{11} = -\hat{g}_{00}$. Thus the output of the Swap isometry will be
\begin{equation}\label{njegos}
\Phi(\B_0\physstate) =  \left(\openone\otimes\sz\maxent_{\rA'\rB'}\right)\otimes \ket{\xi}_{\rA\rB\rP},
\end{equation}
\textit{i.e.} the measurement observable acts on the support of $\ket{\psi}$ as $\sz$. A similar method can be used to self-test all other measurement observables used for the maximal CHSH  violation. Note that \eqref{njegos}  implies a self-test of the measurement operators. Since from \eqref{obsdefine} one has $\N_{b\vert 0}=(\openone+b\B_0)/2$ it follows by linearity of $\iso$ that 
\begin{equation}
\Phi(\N_{b\vert 0}\physstate) =  \left(\openone\otimes\frac{\openone+b\sz}{2}\maxent_{\rA'\rB'}\right)\otimes \ket{\xi}.
\end{equation}
Combining this and the previous section, we can prove a full state and measurement self-testing statement as follows. This concludes the introductory sections of the review.  
\newpage\hrule
\vspace{10pt}
\noindent \emph{Self-testing statement for the CHSH inequality}\\[5pt]
\noindent Let  $\{\physstate_{\rA\rB\rP},\A_0, \A_1,\B_0,\B_1 \}$ be the state and the $\pm1$ valued observables maximally violating the CHSH inequality. Then there exists a local isometry $\iso$ such that 
\begin{align*}
    \Phi(\physstate) &= \maxent\otimes \ket{\xi},\\[5pt]
     \Phi(\A_0\physstate) &= (\frac{\sx+\sz}{\sqrt{2}}\otimes\openone \maxent)\otimes \ket{\xi}\\
     \Phi(\A_1\physstate) &= (\frac{-\sx+\sz}{\sqrt{2}}\otimes\openone \maxent)\otimes \ket{\xi},\\
     \Phi(\B_0\physstate) &=  \left(\openone\otimes\sz\maxent\right)\otimes \ket{\xi},\\
     \Phi(\B_1\physstate) &=  \left(\openone\otimes\sx\maxent\right)\otimes \ket{\xi}.
\end{align*}
\emph{for some state $\ket{\xi}$.}
\vspace{10pt}
\hrule

\section{Self-testing of bipartite states}\label{sec:bipartite}

In this section we give an overview of the existing results in the self-testing of bipartite quantum states. All of the results are for the self-testing of pure states, since mixed states cannot be self-tested (see section \ref{st_geometry}). In \ref{sec:twoqubit} we present the known results from self-testing qubit states, focusing first on the large literature dedicated to the maximally entangled pair of qubits. In \ref{sec:qudit} we move to self-testing of bipartite states of a higher local dimension. Finally, in \ref{parallel} we review the results and methods to self-test many copies of the maximally entangled pair of qubits. 

\subsection{Self-testing of two-qubit states}\label{sec:twoqubit}

\subsubsection{The maximally entangled pair of qubits}\label{singlet}
The fact that the maximal violation of the CHSH inequality can be obtained only by using the maximally entangled pair of qubits or a mixture of maximally entangled qubit states corresponding to different degrees of freedom was reported already in \cite{Summers1987,Popescu1992,BMR,Tsirelson1993}. An alternative method to self-test the maximally entangled pair of qubits is presented in \cite{Mayers2004}, today mostly known as the Mayers-Yao self-test. While \cite{Summers1987,Popescu1992,BMR,Tsirelson1993} can be considered as the avant-garde self-testing papers, \cite{Mayers2004} stands out as the founding work which defined self-testing as a protocol `on its own' and pointed out its importance.  It is worth mentioning that Mayers and Yao made a similar statement already in \cite{Mayers98}, where they called the reference correlations `self-checking'. In the Mayers-Yao protocol, Alice and Bob both measure three observables, $\sz$, $\sx$ and $(\sz+\sx)/\sqrt{2}$. The proof is geometric in spirit and the isometry the  authors use does the same job as the Swap gate, but the authors do not make the connection to the idea of applying a swap unitary. The self-test was made robust in \cite{Magniez}. A simplified proof of the Mayers-Yao self-test, in which Alice makes the same measurements, while Bob measures only $\sz$ and $\sx$ appeared in the supplementary material of \cite{McKague2014}.
 
 The concept of robustness and relevant figures of merit when self-testing the maximally entangled pair of qubits were introduced in \cite{Bardyn2009}, alongside with some explicitly calculated robustness bounds. The first completely device-independent robust self-test of the maximally entangled pair of qubits, both CHSH and Mayers-Yao based, appeared in \cite{McKague2012}. Further inequivalent proofs for self-testing the maximally entangled pair of qubits were reported in \cite{Miller2013}, where the authors gave a condition for a given binary XOR game to be a robust self-test, and in \cite{SASA}, where the chained Bell inequalities were used to self-test the maximally entangled state and an arbitrary number of real measurements. An improvement of the robustness bounds were provided numerically in \cite{PhysRevLett.113.040401} and \cite{PhysRevA.91.022115}, and analytically in \cite{Jed1}, which is currently the best self-test of the maximally entangled pair of qubits in terms of robustness. An important contribution to the self-testing of the maximally entangled pair of qubits is \cite{1367-2630-18-2-025021}, which characterises all the correlations that self-test the state using two dichotomic measurements per party. The robustness of these self-tests was estimated in \cite{Li:19}.
  
  All the results presented so far used only real measurements. The self-testing of maximally entangled pairs of qubits using $\sy$ observables was introduced in \cite{McKague2011} based on the chained Mayers-Yao conditions, also in \cite{toni} and \cite{PhysRevA.96.032119}  based on the elegant Bell inequality \cite{Gisin}, and in \cite{Jed1} based on the extended version of CHSH introduced in \cite{Slofstra} (more on the issue of self-testing complex measurements will follow in section \ref{Pauli}).

\subsubsection{Self-testing of partially entangled states}\label{partially}

All pure entangled states of two qubits admit a Schmidt decomposition
\begin{equation}\label{Sch}
\ket{\psi_{\theta}} = \cos(\theta)\ket{00} + \sin(\theta)\ket{11}\quad\quad \theta \in (0,\pi/4].
\end{equation}
Such states are known as partially entangled pairs of qubits, and they maximally violate the tilted CHSH inequalities 
\cite{amp}:
\small
\begin{equation*}\label{tilted}
\alpha \La A_0 \Ra + \La A_0B_0\Ra + \La A_0B_1\Ra + \La A_1B_0\Ra - \La A_1B_1\Ra \leq 2 + \alpha.
\end{equation*}
\normalsize
The maximal quantum violation $\sqrt{8+2\alpha^2}$ is achieved by the corresponding partially entangled state \eqref{Sch} for $\tan 2\theta = \sqrt{2\alpha^{-2}-1/2}$. To achieve the maximal violation Alice measures $A_0 = \sz$ and $A_1 = \sx$ while Bob measures $B_0 = \cos\mu\,\sz+\sin\mu\,\sx$ and $B_1 = \cos\mu\,\sz-\sin\mu\,\sx$, with $\tan\,\mu = \sin\,2\theta$. The proof that the maximal violation of the tilted CHSH inequality self-tests the corresponding partially entangled pair of qubits appeared in \cite{Yang}. It relied on an SOS decomposition of the shifted Bell operator, but the proof appeared to have an error which made the self-testing proof invalid. The work \cite{Bamps} introduced a systematic way to find SOS decompositions for arbitrary shifted Bell operators. A whole family of SOS decompositions corresponding to the tilted CHSH Bell operator is introduced which was used to show that every tilted CHSH inequality self-tests the corresponding partially entangled pair of qubits. Improved robustness bounds for self-testing partially entangled pairs of qubits through violation of the tilted CHSH inequalities were presented in \cite{tim}.
Two different Bell inequalities, inequivalent to the tilted CHSH inequality and useful for self-testing the partially entangled pairs of qubits appeared in \cite{Flavio} and \cite{wagnerPOVM}. 

The nonlocal character of partially entangled pairs of qubits can be assessed through the Hardy test \cite{Hardy1,Hardy2}. In \cite{Rabelo} it is proven that Hardy test can be used as a robust self-test for the following states
\begin{equation*}
    \ket{\psi_{\varphi}} = \alpha(\ket{01}+\ket{10}) + e^{i\varphi}\sqrt{1-2\alpha^2}\ket{11},
\end{equation*}
where $a = \sqrt{(3-\sqrt{5})/2}$ and $\varphi$ is a free parameter. 

A recent contribution \cite{erik} presents a self-test for any partially entangled pair of qubits and all three Pauli measurements (up to complex conjugation) on Alice's side. Bob needs to apply six measurements. The self-test is proven from the value of three Bell inequalities; two maximally violated tilted CHSH inequalities and one non-maximally violated CHSH inequality. 

\subsection{Self-testing of qudit states}\label{sec:qudit}
The self-testing of bipartite entangled states of higher local dimension (qudits) is more complicated task than the self-testing of qubit states. The good understanding of the qubit case has inspired the use of methods that we call `subspace methods' in which different two-qubit subspaces of the state are self-tested until enough information is gained to self-test the full state. In subsection \ref{subspace} we review this approach, before focusing on more genuinely d-dimensional methods in subsections \ref{genuine} and \ref{groupt}. Some states of local dimension $2^n$ can be seen as a tensor product of $n$ qubit states. In such cases the so-called parallel self-testing is often used, described in section \ref{parallel}. %Self-testing of qudit states and measurements which non-trivially act on the whole Hilbert space is not yet well understood, especially in prime dimensions. The few scattered results are presented in Sec. \ref{genuine}

\subsubsection{Subspace methods}\label{subspace}

Self-testing of maximally entangled states of any dimension is discussed for the first time in \cite{Yang}. The isometry for self-testing introduced there is a  high-dimensional generalisation of the Swap gate. The authors provided a set of correlations which self-test the maximally entangled state of two qudits 
\begin{equation}
\ket{\Phi_d^{\tx{+}}} = \frac{1}{\sqrt{d}}\sum_{i = 0}^{d-1}\ket{ii}.
\end{equation}
One party performs three measurements and the other four. The idea is to self-test separately maximally entangled subnormalised sub-states $\ket{\psi_{0,1}} = \ket{00} + \ket{11}$, $\ket{\psi_{2,3}}=\ket{22} + \ket{33}$, $\cdots$, $\ket{\psi_{d-2,d-1}}=\ket{d-2,d-2} + \ket{d-1,d-1}$. This can be done if all of the substates maximally violate the CHSH inequality (although in \cite{Yang} the authors used a different correlation to test the substates). For this, both parties apply the measurements which are direct sums of the ideal CHSH measurements. This step is not enough, since the mixed state $1/d\sum_{i = 0}^{d-2}\ket{\psi_{i,i+1}}\bra{\psi_{i,i+1}}$ could also pass the test. Another necessary step is self-testing of the substates $\ket{\psi_{d-1,0}}$, $\ket{\psi_{1,2}}$ and so on. It is clear that the mixed state given above cannot provide correlations necessary for this step, where the two parties use again measurements which are the direct sum of the ideal CHSH measurements in a shifted basis, i.e.\ they now self-test the states $\ket{\psi_{1,2}},\ket{\psi_{3,4}}, \cdots, \ket{\psi_{d-1,0}}$. The direct sum of $\sigma_z$ measurements are the same in both bases, thus one party applies three measurements in total while the other applies four. 

An arbitrary pure bipartite state admits the Schmidt decomposition
\begin{equation}\label{sch}
\ket{\psi} = \sum_{i = 0}^{d-1}\lambda_{i}\ket{ii}.
\end{equation}
  The generalisation of the above explained method to self-testing states of the form \eqref{sch} is given in \cite{Coladangelo2017}. In the first step the sub-states $\ket{\psi_{i,i+1}} = \lambda_{i}\ket{ii} + \lambda_{i+1}\ket{i+1,i+1}$ for $i = 0,\cdots, d-2$ are self-tested via the maximal violation of the tilted CHSH inequalities. The second step self-tests the shifted states  $\ket{\psi_{i,i+1}} = \lambda_{i}\ket{ii} + \lambda_{i+1}\ket{i+1,i+1}$ for $i = 1,\cdots, d-1$. This result completed the problem of self-testing all bipartite pure states. The Bell inequalities corresponding to this type of the self-test for maximally entangled states are described in \cite{Coladangelo2018}.

\subsubsection{Self-testing from qudit correlations}\label{genuine}
The method for self-testing all pure bipartite entangled states presented in the previous section relied on self-testing two-qubit sub-states. The measurements used in the self-test were also block-diagonal, where all blocks were either $2\times 2$ or $1\times 1$. It is surprisingly difficult to prove self-testing statements about high-dimensional states without resorting to such methods. In this section we outline a few protocols for self-testing qudit states that use genuinely qudit measurements. 

The first such results were proven in \cite{PhysRevA.91.022115}, \cite{PhysRevLett.113.040401} and \cite{abeille} where two-qutrit states were self-tested by using the numerical Swap method (for details see section \ref{sm}). In \cite{PhysRevA.91.022115} and \cite{PhysRevLett.113.040401} the maximal violation of the CGLMP inequality \cite{cglmp} was used to self-test the partially entangled state of two qutrits:
\begin{align}\label{cglmpstate}
    \ket{\psi}=\frac{1}{\sqrt{2+\gamma^2}}\left(\ket{00}+\gamma\ket{11}+\ket{22}\right)
\end{align}
where $\gamma=(\sqrt{11}-\sqrt{3})/2$, and in \cite{abeille} the SATWAP Bell inequality is introduced and used to self-test the maximally entangled pair of qutrits.

An important contribution in this direction is the analytic self-test presented in \cite{maxmaxmax}. The maximally entangled pair of qutrits is self-tested through the maximal violation of a generalised CHSH inequality. These inequalities, introduced in \cite{maxmaxmax} can be seen as a special class of Buhrman-Massar inequalities \cite{Buhrman}, represent good candidates for self-testing maximally entangled states in any prime dimension $d$. Alice and Bob, both have $d$ inputs, and the measurements necessary for the maximal violation are mutually unbiased bases. For higher dimensions, the SOS-decomposition of the shifted Bell operator is provided, but the self-testing statement is still lacking. In fact, for $d = 5$ and $d = 7$, it is proven that the maximal violation can be achieved by using inequivalent quantum realisations, however all of them involve the maximally entangled state in dimensions $5$ and $7$, respectively. 

Another contribution to self-testing maximally entangled states of qudits in the context of nonlocal games is given in \cite{Mancinska}. There, the author considers a specific type of nonlocal games, the so-called pseudo-telepathy weak projection games. A nonlocal game is called pseudo-telepathy game if it can be won with probability equal to one by using quantum finite dimensional strategy, but cannot be won by using classical strategies \cite{telepathy}. Weak projection games belong to a sub-class of pseudo-telepathy games and \cite{Mancinska} shows that every such game can be used to self-test maximally entangled states in finite dimensions.

\subsubsection{Group theoretic tools}\label{groupt}

Self-testing properties of non-local games were elaborately explored in \cite{Slofstra} and \cite{Coladangelo2017c}. The common method for both works is the `algebraisation' of the winning strategies in nonlocal games. The idea of relating representations of a Clifford algebra to the optimal strategies to win the CHSH game was used already in \cite{Summers1987} and \cite{Tsirelson1987}.  In \cite{Slofstra}, to each XOR game $\mathcal{G}$ is associated a $C^*$ algebra $\mathcal{A}$, such that optimal strategies to win $\mathcal{G}$ correspond to  representations of $\mathcal{A}$. Furthermore, there is a relation between near-optimal strategies and approximate representations. Using these techniques a self-testing statement for high-dimensional maximally entangled states via a generalisation of the CHSH game is implicitly given in \cite{Slofstra}. 

 In \cite{Coladangelo2017c}, the authors study self-testing properties of a class of pseudo-telepathy games, known as linear-constraint system games, of which the magic square and magic pentagram games are two popular examples \cite{MSGPer,MSG1}. In these games, the players are asked for assignments to a subset of variables in a system of linear equations, and they win the game if they return consistent and valid assignments. The authors extend the representation theoretic framework of \cite{CleveMittal}, \cite{CleveLiuSlofstra} and \cite{slofstra16}  and obtain a generic self-testing result for linear-constraint system games of a certain kind. They apply this result to obtain a self-testing protocol for a tensor product of $n$ maximally entangled pairs of qubits. The self-testing condition is the perfect score in either the magic square game or the magic pentagram game.
\iffalse
In \cite{Coladangelo2017c} the main result is the self-testing protocol of a tensor product of $n$ maximally entangled pairs of qubits. The self-testing condition is the perfect score in either the magic square game or the magic pentagram game. Since both games can be won with the probability equal to one they belong to the class of pseudo-telepathy games. The players' answers must satisfy one of the linear constrains corresponding to the game, so the games also belong to the class of linear constraint games.
\fi
It is proven in \cite{CleveMittal} that perfect strategy for every linear-constraint system game which is also a pseudo-telepathy game must involve a maximally entangled state. On the other side in \cite{CleveLiuSlofstra} it is shown that a solution group can be associated to every linear-constraint system game. Moreover, the operators used in the winning strategy must satisfy certain algebraic relations determined by the solution group. In \cite{Coladangelo2017c} the authors use these results and by exploiting algebraic properties of the solution group corresponding to the magic square and magic pentagram games prove the self-testing statement for a tensor product of maximally entangled pairs of qubits. The self-test is also proven to be robust.

\subsection{Self-testing \textit{n} maximally entangled pairs of qubits}\label{parallel}
In this section we outline methods and results for self-testing $n$ copies of the maximally entangled state of two qubits (which itself is a maximally entangled state of dimension $2^n$). Here, there are two main approaches; \emph{sequential self-testing} and \emph{parallel self-testing}. 
%In the standard self-testing procedure each entangled particle is measured by a different uncharacterized measurement device (black box).  In this view, it seems remarkable to self-test many entangled bipartite states shared between only two non-communicating parties. The typical problem in this scenario is self-testing of a tensor product of $n$ EPR pairs, shared by two parties. Depending on the structure of the self-test there are two main approaches to this problem. The first approach corresponds to the so-called \textit{sequential} self-testing.

\subsubsection{Sequential self-testing}
The first result relating to the self-testing of $n$ maximally entangled pairs of qubits (here also called EPR pairs) appeared in \cite{ruv}. In this scheme, in each round of the experiment the devices receive inputs $(x_i,y_i)$ for $i=1,\cdots,n$, labelling the measurement bases for $i$-th maximally entangled pair. The inputs are given to the devices sequentially: first the inputs $(x_1,y_1)$ are given and the outputs $(a_1,b_1)$ are returned; then the inputs $(x_2,y_2)$ are given and the outputs $(a_2,b_2)$ are returned. This process is continued until  the $n$-th pair of outcomes is collected and it is characterised by the following transcript
\small
\begin{eqnarray*}
&a_1,b_1\quad \textrm{given}\quad &x_1,y_1,\\
&a_2,b_2 \quad \textrm{given}\quad &a_1,x_1,x_2,b_1,y_1,y_2, \\ &\cdots, \\
&a_n,b_n \quad \textrm{given}\quad &a_1,x_1,\cdots, a_{n-1}, x_{n-1},x_n, b_1,y_1,\\&\quad &\cdots, b_{n-1}, y_{n-1},y_n
\end{eqnarray*}
\normalsize
There is no assumption that in each round the source emits the same state and the measurement strategies in the rounds may depend on the inputs and outputs in all previous rounds. In \cite{ruv} the authors prove that if the parties win CHSH game in $\omega^*n$ rounds, where $\omega^*$ is the optimal probability to win the CHSH game, there is isometry mapping the state the parties shared at the beginning of the procedure to the tensor product of $n$ EPR pairs. The result is stated in its robust form: if the parties win CHSH game in $(1-\epsilon)\omega^*n$ rounds, then at the beginning of any randomly chosen block of $m = n^{\Omega(1)}$ rounds the state of the parties can be mapped to a state $f(\epsilon)$-close to the tensor product of $m$ EPR pairs. The drawback of the work is a very low robustness, \textit{i.e.} as $\epsilon$ increases the number of rounds necessary to extract a state $f(\epsilon)$-close to $m$ EPR pairs grows very fast.  \\
% \begin{figure}
%     \centering
%     \includegraphics[width=\columnwidth]{SequentialST.pdf}
%     \caption{Sequential self-testing \joe{more beauty}}
%     \label{seqSt}
% \end{figure}

\subsubsection{Parallel self-testing}
\begin{figure}
    \centering
    \includegraphics[width=0.7\columnwidth]{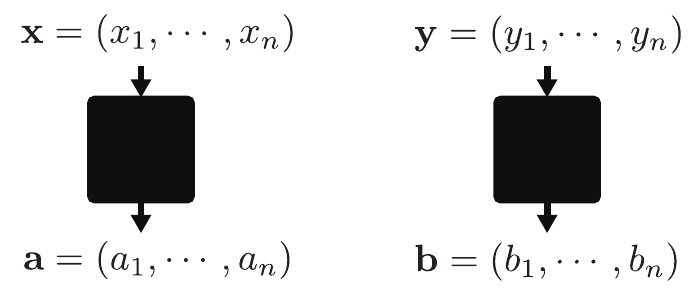}
    \caption{Scenario for parallel self-testing. Alice and Bob both receive a list of $n$ inputs and provide $n$ outputs, which correspond to measurements made on $n$ independent copies of a physical state. The aim is to prove that the statistics self-test $n$ independent copies of the reference state \label{fig:parallel}}
    \label{parSt}
\end{figure}

A more popular approach for self-testing $n$ EPR pairs is parallel self-testing (see figure \ref{fig:parallel}). Here, the inputs are not given sequentially but all at the same time, \textit{i.e.} the devices receive input vectors $\vec{x} = (x_1,\cdots , x_n)$, $\vec{y} = (y_1,\cdots , y_n)$ and return outputs vectors $\vec{a} = (a_1,\cdots , a_n)$, $\vec{b} = (b_1,\cdots , b_n)$. In principle this makes it more difficult to prove self-testing statements than in the sequential scenario, since one assumes less structure on how the outcomes are generated. %Crucially, dishonest parties have significantly more power in the parallel scenario because they have more information. 
To self-test a single maximally entangled pair we saw in section \ref{sec:example} that it is enough to identify a pair of anticommuting observables. In the case of $n$ pairs, one has not only to find $n$ pairs of anticommuting obervables, but also to show that observables from different pairs mutually commute. An important feature of parallel self-tests is their robustness.  A parallel self-test is  robust if any strategy producing correlations that are $\epsilon$-close to the ideal ones must use a state which is $f(\epsilon,n)$-close to $\refstate^{\otimes n}$, where $f(\epsilon,n)$ is a monotonically increasing function in $\epsilon$. How quickly the function  $f(\epsilon,n)$ increases with $\epsilon$ and $n$ determines how good robustness is.

%\footnote{In some self-testing papers one can find the notion of the \emph{strictly parallel} self-test (see \cite{McKague2016}). This notion appears when the self-test for $n$ EPR pairs can be seen as a set of sub-tests for individual EPR pairs, where each sub-test is completely independent of the other sub-tests. This can happen if and only if all possible questions are asked. \textcolor{red}{JD:unclear}}

%Self-tests which are not strictly parallel require fewer questions and consequently less randomness is consumed. %In the remainder of this section we do not make a distinction between strictly parallel self-tests and those that are not strictly parallel.

%
The first parallel self-test was proven for $2$ EPR pairs in \cite{Wu2016}. The work gives two different self-tests: one based on the optimal success in a double use of the CHSH game and the other on the optimal success in the Magic Square game \cite{MSG1,MSGPer}. The result was subsequently generalised for arbitrary $n$: via parallel repetition of the CHSH game in \cite{Coladangelo2017a} and \cite{McKague2017a} and via parallel repetition of the Magic Square Game in \cite{Coladangelo2017a} and \cite{Coudron2016}. The usefulness of the Magic Pentagram game \cite{MSG1} for self-testing a tensor product of three EPR pairs was proven in \cite{2058-9565-3-1-015002}. Self-testing of $n$ EPR pairs via parallel repetition of the Mayers-Yao self-test is given in \cite{McKague2016}.  We also note that the self-test \cite{Coladangelo2017c} discussed in the previous section belongs to this class of parallel self-tests.

In recent years, several works appeared which also managed to self-test $n$ EPR pairs shared by two parties but not by parallel repetition of any single self-test of a single pair. The first of such results was presented in \cite{Ostrev2016}, based on the XOR games introduced in \cite{Slofstra}. Later, self-testing of $n$ EPR pairs with measurements performed on few of them in each round was the subject of \cite{Chao2016}. This self-test is nondestructive: not all entanglement is consumed in the self-test, but can be used for eventual later protocols. 

A combination of self-testing based on nonlocal games with the quantum version of the linearity test from \cite{Blum}, named Pauli braiding test \cite{Natarajan} led to the first self-test of $n$ EPR pairs in which robustness does not get worse if the number of EPR pairs tested increases. Another parallel self-test keeping this desirable property is presented in \cite{Natarajan2018}. The test can be seen as a quantum version of the classical plane-vs-point
test for multivariate low-degree polynomials \cite{Raz:1997:SEL:258533.258641}.

Finally, while all self-tests presented in  this section certify $n$ EPR pairs and tensor products of the observables $\sx$ and $\sz$, it is possible to extend them to involve certification of $\sy$ also. This was first done in \cite{Coladangelo2017b} and later in \cite{BSCA}, up to the uncertainty of the complex conjugation of the full $n$-qubit measurement operators as explained in section \ref{sec:generalisations}. 

%but in parallel self-testing when certifying a tensor product of several $\sy$ measurements it can be at least proven that the transposition is performed on all of them or on none (for more details see Sec. \ref{Pauli} ).

\iffalse
\begin{tabular}{ |c|c|c|c|c| } 
 \hline
 $\quad$ & Robustness &  Total number of inputs & Inputs size & Outputs size \\
  \hline\hline
   \cite{Chao2016} & $O(n^{\frac{5}{2}}\epsilon^{\frac{1}{2}})$ &  $n(n-1)$ & $2$ $n$-its & $1$ bit\\
 \hline
   \cite{Coladangelo2017a} &
 $O(n^{\frac{3}{2}}\epsilon^{\frac{1}{2}})$ &  $2^n$ & $n$ bits & $n$ bits\\
 \hline
  $\qquad$ \cite{Coladangelo2017c}$\qquad$ & $O(n^{5}\epsilon^{\frac{1}{2}})$ &  $3^{\frac{n}{2}}$ & $\frac{n}{2}$ trits & $n$ bits\\
 \hline
   \cite{Coladangelo2017b} & $O(\epsilon^{\frac{1}{4}})$ &  $3^n + n!$ & $2n$ bits & $2$ bits \\
 \hline
  \cite{Coudron2016} & $O(n^2\epsilon^{\frac{1}{2}})$ &  $3^{\frac{n}{2}}$ & $\frac{n}{2}$ trits & $n$ bits \\
 \hline
 \cite{McKague2016} & $O((n^2\epsilon^{\frac{1}{2}}+ n\epsilon^{\frac{1}{4}})^{\frac{1}{2}})$  & $3^n$ & $n$ trits & $n$ bits \\
 \hline
 \cite{McKague2017a} & $O(n^{\frac{9}{8}}\epsilon^{\frac{1}{8}})$ &  $2^n$ & $n$ bits &  $n$ bits\\
 \hline
 \cite{Natarajan} & $\mathrm{poly}(\epsilon)$  &  $O(n)$ & $2n$ bits & $2$ bits \\
 \hline
 \cite{Natarajan2018} &  $\mathrm{poly}(\epsilon)$ &  $X$ & $O(\mathrm{log}~{n})$ & $\mathrm{poly}(\mathrm{log~log}~n)$\\
 \hline
  \cite{Ostrev2016} & $\epsilon = O(n^{-5})$ &  $n(n-1)$ & $2$ $n$-its & $1$ bit\\
 \hline
 \end{tabular}
\fi

%
%
%\begin{widetext}
%\centering
\begin{table*}[t]
\centering
\begin{threeparttable}
\begin{tabular}{ |c|c|c|c| } 
 \hline
 $\quad$ & Robustness &  Inputs size (in bits) & Outputs size (in bits) \\
  \hline\hline
   \cite{BSCA} & $\qquad\mathrm{poly}(n,\epsilon)\qquad$ &  $O({n})$ & $n$\\
 \hline
   \cite{Chao2016} & $\qquad\mathrm{poly}(n,\epsilon)\qquad$ &  $O(\mathrm{log}~{n})$ & $1$\\
 \hline
   \cite{Coladangelo2017a} &
 $\qquad\mathrm{poly}(n,\epsilon)\qquad$ &  $O(n)$  & $n$ \\
 \hline
  $\qquad$ \cite{Coladangelo2017c} $\qquad$ & $\qquad\mathrm{poly}(n,\epsilon)\qquad$ &  $O(n)$ &  $n$ \\
 \hline
   \cite{Coladangelo2017b} & $\qquad\mathrm{poly}(\epsilon)\qquad$ &  $O(n~\mathrm{log}~n)$  & $2$ \\
 \hline
  \cite{Coudron2016} & $\qquad\mathrm{poly}(n,\epsilon)\qquad$ &  $O(n)$ &  $n$ \\
 \hline
 \cite{McKague2016} & $\qquad\mathrm{poly}(n,\epsilon)\qquad$  & $O(n)$ & $n$ \\
 \hline
 \cite{McKague2017a} & $\qquad\mathrm{poly}(n,\epsilon)\qquad$ &  $O(n)$  &  $n$\\
 \hline
 \cite{Natarajan} & $\qquad\mathrm{poly}(\epsilon)\qquad$  &  $O(n)$ & $2$  \\
 \hline
 \cite{Natarajan2018} &  $\qquad\mathrm{poly}(\epsilon)\qquad$ &  $O(\mathrm{poly}(\mathrm{log}~{n}))$ &  $\mathrm{poly}(\mathrm{log~log}~n)$\\
 \hline
  \cite{Ostrev2016} & $\qquad\mathrm{poly}(n,\epsilon)\qquad$ &  $O(\mathrm{log}~{n})$  & $1$ \\
 \hline
 \end{tabular}
 \caption{Comparative properties of different self-tests of $n$ EPR pairs. The most important aspect of a self-test of $n$ EPR pairs when it comes to practical usefulness is its robustness to noise (or rigidity). The other relevant property is its complexity, in terms of the size of the inputs. The size of the outputs is also a relevant factor, especially in possible applications for randomness expansion. For now, the self-testing protocol presented in \cite{Natarajan2018} 
 has the best properties in terms of the total number of inputs (polynomial) and robustness bounds (independent on $n$). The papers use different distance measures, but all the bounds given here are in terms of the Euclidean distance using definition \ref{robdef2}. The work \cite{Coladangelo2017b} self-tests $\sy$ measurements on each EPR pairs, and the number of inputs increases in order to deal with the issue of complex conjugation (see \ref{cc}.) If one omits self-testing of $\sy$ from the protocol the number of inputs is $O(n)$.} 
\label{tabPST}
\end{threeparttable}
\end{table*}
%\end{widetext}

\subsubsection{Overlapping qubits}

A standard parallel self-test of $n$ EPR pairs proves the existence of $n$ pairs of anticommuting observables, where any two observables belonging to different anticommuting pairs necessarily commute. As explained in \cite{overlap} each anticommuting pair of observables defines a qubit. Hence the dimension of the underlying Hilbert space in this case must be at least $2^n$. The main contribution  of \cite{overlap} is the estimation of the dimension of the underlying Hilbert space if observables from different anticommuting pairs do not commute exactly, which might happen when the self-testing conditions are approximately satisfied. This leads to the concept of `overlapping qubits', which, depending on the amount of the overlap can be 'packed' in the Hilbert space whose dimension grows polynomially with $n$.

\section{Self-testing of multipartite states}\label{sec:multipartite}
All bipartite pure states admit a Schmidt decomposition, which simplifies the characterisation of bipartite entanglement and the self-testing of bipartite pure states. Multipartite states do not admit such a simple characterisation,  although some generalisations of the Schmidt decomposition exist in the entanglement literature \cite{aajt,BKraus}. While all bipartite pure entangled states can be self-tested, when it comes to self-testing of multipartite states, only some partial results exist. Furthermore, from a `loop-hole free' perspective, multipartite self-testing is considerably more demanding than bipartite self-testing, since it requires space-like separation between multiple measurement devices.

In this chapter we identify four main methods for self-testing multipartite entangled states: self-testing of graph states based on the structure of their stabilizer operators (section \ref{stab}); tailoring Bell inequalities to self-test specific states (section \ref{tailoring}); reductions to bipartite methods (section \ref{reduction}); parallel self-testing of multipartite states (section \ref{parallelMulti}); and self-testing from marginal information only (section \ref{marginal}). 
% \begin{itemize}
% \item exploiting stabilizer structure, which is the subject of Sec. \ref{stab}.
% \item Bell inequalities based, described in Sec. \ref{tailoring},
% \item reduction to bipartite methods, outlined in Sec. \ref{reduction}
% \item parallel self-testing, described in Sec. \ref{parallelMulti},
% \item self-testing using only marginal information, concluding this chapter in Sec. \ref{marginal}.
% \end{itemize}

\subsection{Self-testing of graph states from stabilizer operators}\label{stab}
The first  multipartite states to be self-tested were graph states \cite{McKague2014}. Formally, given a graph $G$ defined by a set of vertices $V=\{1,\cdots,N\}$ and a set of edges $E$ (pairs of connected vertices of $V$), the graph state corresponding to $G$ is given by
\begin{align}
    \ket{G}=\prod_{(i,j)\in E}\mathsf{CZ}_{i,j}\ket{+}^{\tp N},
\end{align}
where $\mathsf{CZ}_{i,j}$ is the controlled-$\sz$ two-qubit unitary $\mathsf{CZ}=\tx{diag}(1,1,1,-1)$ acting on qubits $i$ and $j$. Equivalently, $\ket{G}$ can be defined as the unique state that is stabilized by (i.e.\ is a $+1$ eigenstate of) a set of $N$ local stabiliser operators $\sigma_x^{i}\otimes_{j \in n(i)}\sigma_z^j$, where $n(i)$ is the neighbourhood of vertex $i$; the set of vertices connected to $i$ on $G$.

A self-testing protocol for any graph state corresponding to a connected graph is provided in \cite{McKague2014}. Note that graph states corresponding to graphs that are not connected must be separable with respect to at least one bipartition. %In McKague's self-testing protocol, a graph state to be self-tested is shared by the number of parties equal to the size of the corresponding graph, \textit{i.e.} each particle is held by a different party.
The reference measurements needed for the self-testing are given by the stabilizer operators themselves aided by a few measurements generalising those from Mayers-Yao self-test. More specifically, for an arbitrary graph state, one party has to measure three observables: $\sz$, $\sx$ and $(\sx + \sz)/\sqrt{2}$, while all the other parties measure only $\sx$ and $\sz$.  The self-test is robust to small imperfections and the isometry is the multipartite generalisation of the Swap gate.

% A graph is connected if for any two vertices $v_i$ and $v_j$ there exist a sequence of adjacent vertices starting with $v_i$ and ending with $v_j$. A graph state $\ket{\psi_G}$ corresponding to the graph $G$ is obtained by preparing $N$ qubits in the state $\ket{+} = (\ket{0}+\ket{1})/\sqrt{2}$ and applying a controlled-$Z$ gate on each two qubits connected with an edge. The stabilizer group is generated by $N$ local operators $\sigma_x^{i}\otimes_{j \in n(i)}\sigma_z^j$, where $n(i)$ is the set of vertices connected with $i$~\footnote{The state $\ket{\psi}$ is stabilized by an operator $A$ if $A\ket{\psi} = \ket{\psi}$.}.  
The approach from \cite{Flavio} can be also placed in the following subsection, but since it is intrinsically related to stabilizers we discuss it in this group. %An n-qubit stabilizer state is the simultaneous eigenvector of an Abelian subgroup of the $n$-fold Pauli group which does not include $-\openone^{\otimes n}$.
Starting from any graph state, the authors introduce a method to construct a Bell inequality that is maximally violated by the corresponding state. Moreover, the derived Bell inequality can be used to self-test the state. Each party measures an anti-commuting pair of observables from the real plane of the Bloch sphere. Beyond graph states, the method can be used to self-test the so-called partially entangled GHZ states $\cos{\theta}\ket{0}^{\otimes n} + \sin{\theta}\ket{1}^{\otimes n}$ for any $n \geq 2$. 

\subsection{Tailoring Bell inequalities}\label{tailoring}
In \cite{Pal} the authors introduce a method to build permutationally invariant Bell inequalities with two measurement settings per party useful for self-testing multipartite states. The method is tailored for a specific state $\refstate$ and the measurements leading to the maximal violation are chosen from the real plane of the Bloch sphere. A linear program can be used to find a Bell operator, whose eigenstate is $\refstate$ and maximises the ratio of the quantum and classical bound. The derived Bell inequality is just a suitable candidate for self-testing, which further must be checked by utilising the numerical Swap method technique (see section \ref{sm}). Since the self-testing proof relies on the Swap method, it becomes too costly when the number of parties becomes larger than four. Examples of the successful implementation of this method involve the tripartite $W$ state, the tripartite and four-partite GHZ state and the four-qubit linear cluster state.  

Another method for developing Bell inequalities, tailored for self-testing of multipartite qubit states is described in \cite{Sekatski}. As in \cite{Pal}, all parties can perform two different measurements and the constructed Bell inequalities are suitable candidates  for self-testing applications. The starting point for choosing a Bell operator is not the permutational invariance, but the structure of the stabilizers of the state. The method can be applied to multipartite states that are not graph states, in which case these stabilizer operators will not all be tensor products of Pauli operators. %Note that any state has at least one (potentially nonlocal) stabilizer. The stabilizer states mentioned in the previous section are stabilized by a tensor product of Pauli operators. 
Of all the Bell operators mimicking the structure of the stabilizers, constructed from the arbitrary two real measurements per party, the optimal candidate is the one whose maximum eigenvalue is the local maximum with respect to the small perturbation of the local measurement directions. The robust self-test is then checked by using semidefinite programming to find the lower bound to the fidelity of the state providing the maximal violation and $\refstate$ (see section \ref{oi}). As example the authors apply the method to self-test a family of four qubit states $CU_{\phi}\maxent\otimes\maxent$, where $CU_{\phi} = \ketbra{0}{0}\otimes \openone + \ketbra{1}{1}\otimes \exp[-i\phi\sx]$. Since the self-testing is proven by employing numerical methods, it becomes infeasible when the number of parties increase. 

\subsection{Reductions to bipartite methods}\label{reduction}
Self-testing protocols for multipartite states can be constructed by reusing self-testing protocols for bipartite states. The idea is as follows: when $n-2$ parties perform appropriate projective measurements, they might collapse the state of the remaining two parties into some pure bipartite entangled state, which in principle can be self-tested by using methods from section \ref{sec:bipartite}. By repeating the process of projecting and self-testing for different pairs of parties one might expect to gather enough information to self-test the whole multipartite state. An important restriction is that the measurement used by any party in the projecting part must be some of the measurement the same party uses in the self-testing part of the protocol. 

The idea was first used in \cite{Wstate} to self-test $W$-state $\ket{W} = (\ket{001} + \ket{010} + \ket{100})/\sqrt{3}$. Whenever one of the parties performs the measurement in the computational basis and obtains outcome $+1$ the state of the remaining two parties becomes maximally entangled $\sim (\ket{01} + \ket{10})$. This state can be self-tested by maximally violating the CHSH inequality, for example. The authors of \cite{Wstate} show that by repeating the above process twice for different parties measuring in the computational basis, the whole state can be self-tested using the Swap isometry. They also show that a similar method, based on self-testing of partially entangled two-qubit states, can be used to self-test states of the form $\ket{W_{\gamma}} = (\ket{001} + \ket{010} + \gamma\ket{100})/\sqrt{2+\gamma^2}$. The method was generalised in \cite{Ivan} to prove self-testing of all permutationally invariant qubit Dicke states, all qubit graph states, and all multipartite states of any local dimension admitting the Schmidt decomposition $\ket{\psi_{\mathbf{\lambda}}} = \sum_{i = 0}^{d-1}\lambda_i\ket{i,i,\cdots, i}$, representing the first self-test of a high-dimensional multipartite state. Self-testing of $W$-states for any number of parties was also proven in \cite{Wu2017}, and self-testing of all Dicke state was proven in \cite{Fadel2017}. 

The self-testing of graph states whose underlying graph is a triangular lattice is shown in \cite{Hayashi}. The whole graph is shared by three parties and if one party measures its qubits in the $\sz$ basis it prepares maximally entangled pairs of qubits for the remaining two parties, which are in \cite{Hayashi} self-tested through the Mayers-Yao criterion. \\

\subsection{Parallel self-testing of multipartite states}\label{parallelMulti}
In section \ref{parallel} we saw many ways to self-test $n$ EPR pairs by using parallel repetition of CHSH or Magic Square game. Up to date, the only parallel self-test of some multipartite state is shown in \cite{Diagram}. The authors use diagramatic proofs based on categorical quantum mechanics \cite{category}, to prove that parallel repetition of the GHZ game robustly self-tests $n$ copies of the GHZ state. %The robustness bound is $O(n^4\sqrt{\epsilon})$.

\subsection{Self-testing using only marginal information}\label{marginal}
Almost all the protocols for self-testing multipartite states presented so far require measuring full-body correlators, that is, they depend on correlations between all parties. This quickly becomes a practical problem since measuring such correlations is typically experimentally very challenging. The possibility of self-testing by measuring only few-body correlators is the subject of \cite{marginal}. The authors use the numerical Swap method (see section \ref{sm}) to self-test the tripartite  $W$-state,  a class of $W$-like states $(\ket{001} + \ket{010} + \gamma\ket{100})/\sqrt{2+\gamma^2}$ and the states maximally violating Bell inequalities defined in \cite{Jordi} by using only two-body correlators. The four-partite $W$-state is also self-tested using three-body correlators.

\begin{figure*}
\centering
\includegraphics[width=1.0\textwidth]{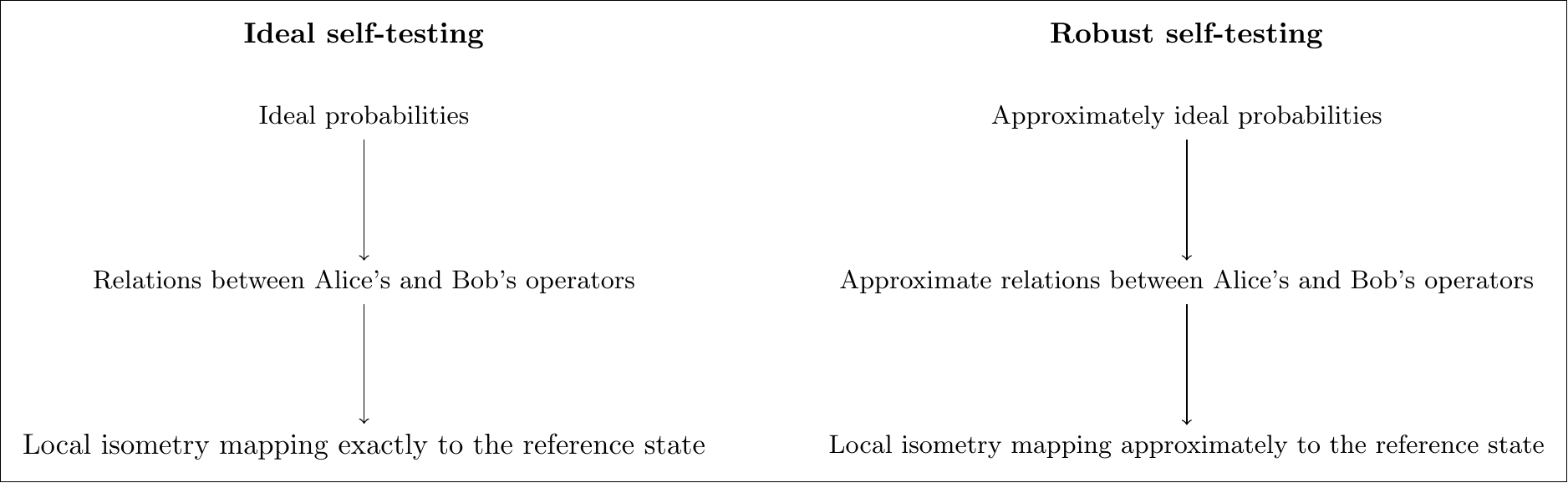}
\caption{The program for robust self-testing. The aim is to show that approximately satisfied self-testing conditions imply the existance of the local isometry which approximately maps the physical state to the reference one. \label{fig:robustnessArrows} }
\end{figure*}

\section{Robust self-testing of states}\label{sec:robust}
It is impossible to meet exactly the conditions for ideal self-testing. On one hand, experimental noise and imperfections undermine hope to reproduce exactly the reference correlations (\textit{i.e.} those obtained by performing reference measurements on the reference quantum state). On the other hand, even if all noise contributions are eliminated, one must work with a finite sample size and so the precise probabilities cannot be known, but only estimated up to some statistical confidence level. In order to make self-testing protocols practically meaningful, it is therefore crucial to make them robust to deviations from the ideal case. For possible definitions of robust self-testing see Defs. \ref{robdef2} and \ref{robdef1}. 

The first self-testing protocol to be made robust was the Mayers-Yao self-test of maximally entangled pair of qubits \cite{Magniez}. Robust self-testing through the CHSH inequality was explored in \cite{Bardyn2009} and simple robust self-testing protocols based on both the Mayers-Yao and the CHSH criterion were presented in \cite{McKague2012}. The techniques presented therein remained the main tool for making self-testing protocols robust in the majority of later contributions. As more is known about ideal self-testing the focus of the research is shifting towards finding better techniques for assessing robustness. Arguably, it remains a principal challenge in the field.

In this section we review the main contributions to robust self-testing and the techniques predominantly used in the literature. In section \ref{sec:robustmethods} we identify and explain five approaches:
\begin{itemize}
\item An approach based on the vector norm inequalities (Section \ref{ni});
\item Methods relying on the use of Jordan's lemma (Section \ref{JL});
\item  An approach based on the operator inequalities (Section \ref{oi});
\item The numerical Swap method (Section \ref{sm});
\item An algebraic method (Section \ref{algRob}).
\end{itemize}
Finally, in Section \ref{rnt} we discuss recent progress on noise-tolerant self-testing of a tensor product of many EPR pairs. 

\subsection{Robust self-testing methods}\label{sec:robustmethods}
\subsubsection{Norm inequalities method}\label{ni}
The bulk of self-testing protocols start from the observed probabilities $p(a,b|x,y) = \bra{\psi}\M_{a|x}\otimes \N_{b|y}\ket{\psi}$ or the maximal violation of some Bell inequality and deduce equations of the type (see e.g.\ \eqref{xbzb} and \eqref{zzxx} from section \ref{sec:example}) 
%
% \begin{equation}\label{steq}
%     f_A(\{\M_{a|x}\}_{a,x}) \otimes \openone \physstate = \openone\otimes f_B(\{\N_{b|y}\}_{b,y})\physstate,
% \end{equation}
\begin{equation}\label{steq}
    f(\{\M_{a|x}\},\{\N_{b|y}\}) \physstate=0,
\end{equation}
where $f$ is some polynomial function  in the measurement operators. Such relations can be drawn from either geometrical arguments (like in \cite{Mayers2004}), algebraic identities (like in \cite{McKague2014}) or SOS decompositions (like in \cite{Bamps}). The relations \eqref{steq} are a necessary step in proving that the appropriate isometry (the Swap gate in most cases) maps the physical state to the reference one,
\begin{equation}\label{isosteqId}
 \Phi(\physstate) = \refstate \otimes \ket{\xi}.  
\end{equation}
Although the self-testing proof requires that the correlations be ideal, one could hope to follow the same proof in which the exact relations are exchanged with approximate ones, leading to a noise-dependent bound on the self-tested fidelity. More precisely, when the observed probabilities are within $\varepsilon$ distance from the ideal ones (or the violation of the Bell inequality is $\varepsilon$-far from the maximal value), analogously to \eqref{steq}, the aim is to find approximate relations:
\begin{equation}\label{rsteq}
    \|f(\{\M_{a|x}\},\{\N_{b|y}\}) \physstate\| \leq g_f(\varepsilon),
\end{equation}
where $\| \cdot \|$ is a vector vector norm (usually taken to be Euclidean)  and $g_f(\epsilon)$ are some increasing functions for which $g_f(0)=0$.
One can then often guarantee that the appropriate isometry $\iso$ satisfies
\begin{equation}\label{isosteq}
 \|\Phi(\physstate) - \refstate \otimes \junk \| \leq g_\iso(\varepsilon),
\end{equation}
where $g_\iso$ is obtained by propagating the uncertainties \eqref{rsteq} through the isometry circuit. 
%
% Probability $p(a,b|x,y)$ can be seen as an inner product between vectors $\M_{a|x}\otimes\openone\physstate$ and $\openone\otimes \N_{b|y}\physstate$. This allows to
% \fi

Relations of the type \eqref{rsteq} are usually obtained via various vector norm inequalities, such as Cauchy-Bunyakovski-Schwarz, triangle or H\"{o}lder inequalities. These techniques were first used in \cite{McKague2012} and later in \cite{McKague2014,Wstate}. If a self-testing proof relies on the maximal violation of a Bell inequality, the relations \eqref{steq} can be conveniently obtained from the SOS decomposition of the shifted Bell operator (see box \ref{box:sos} and equation \eqref{sosimpliesapprox} therein). The usefulness of SOS decompositions for robust self-testing was first noted in \cite{Yang} and later used in e.g.\ \cite{Bamps,SASA}.

Techniques based on vector norm inequalities are useful in making self-testing protocols robust, but the robustness bounds are typically not very good due to large constants appearing in them. The asymptotic behaviour of the function $g_{\iso}$ for different self-testing protocols based on this method is given in Table \ref{table2}. 

\begin{table}
\begin{tabular}{ |c|c| } 
 \hline
  & \small{Asymptotic behaviour of} $g_{\iso}$  \\
  \hline\hline
 \cite{McKague2012}  & $O(\epsilon^{\frac{1}{4}})$ \\
 \hline
 \cite{McKague2014} & $O(\epsilon^{\frac{1}{4}})$\\
 \hline
 \cite{Bamps} & $O(\epsilon^{\frac{1}{2}})$\\
 \hline
  \cite{SASA} & $O(\epsilon^{\frac{1}{2}})$\\
 \hline
  \cite{Wstate} & $O(\epsilon^{\frac{1}{4}})$\\
 \hline
\end{tabular}
\caption{\label{table2}Comparative properties of different robust self-tests based on vector norm inequalities.}
\end{table}

\subsubsection{Utilising Jordan's lemma}\label{JL}

One of the main difficulties in the device-independent description of quantum experiments is related to the inability to fix the dimension of the underlying Hilbert space, which prevents the parameterisation of the measurements and states used in the experiment. This difficulty stays the prime hurdle towards calculating robust self-testing bounds. A very useful theoretical asset enabling a solution in scenarios where each party has two dichotomic measurements is the Jordan lemma \cite{Pironio09} (see lemma 2 therein). It allows to effectively reduce an arbitrary-dimensional experiment to the one in which the local subsystems are qubit systems. %involving only two qubits. 

For the purposes of robust self-testing the Jordan lemma was first time used in \cite{Bardyn2009} to obtain robust self-testing of the maximally entangled pair of qubits through violation of the CHSH inequality. Later, it was used in \cite{Sekatski} for the robust self-testing of a arbitrary multipartite states using the Bell inequalities introduced therein and described in section \ref{tailoring} of this review. For simplicity, here we give a short description of the method to the bipartite scenario, while keeping in mind that, as described in \cite{Sekatski}, it can straightforwardly be applied to the multipartite case.

The Jordan lemma states that given two Hermitian matrices of finite or countably infinite dimension and with eigenvalues $\pm1$, there exists a unitary transformation that simultaneously block diagonalises them, where each block is of size at most $2\times2$. Consider a self-testing protocol in which Alice and Bob each have a pair of $\pm 1$ valued observales $\A_x$, $x=0,1$ for Alice and $\B_y$, $y=0,1$ for Bob. It follows there is a choice of local basis in which these observables take the block structure described above. One can further assume that each of the blocks is of size $2\times 2$, since a one-dimensional block is equivalent to a two-dimensional block where the state has support only on one of these dimensions. One can then apply additional unitary rotations to each of the blocks so that they take real values only. Given this structure, one can paramaterise the observables as follows
\begin{equation}\label{JLobs}
 \begin{split}
    \A_x = \bigoplus_i \A_i &= \bigoplus_i \cos\alpha_i\sx + (-1)^x \sin\alpha_i\sz,\\
    \B_y = \bigoplus_j \B_j &= \bigoplus_j \cos\beta_j\sx + (-1)^y \sin\beta_j\sz
 \end{split}
\end{equation}
 This parameterisation covers all possibilities: $\alpha_i=0$ implies that the observables commute in that block, whereas $\alpha_i=\pi/4$ implies anticommutation in that block.
 Consequently, the Bell operator can be written  as $\mathcal{B} = \oplus_{ij}\mathcal{B}(\rA_i,\rB_j)$. Following such parametrisation the Bell violation can be written as 
\begin{equation}
\beta = \sum_{ij}p_{ij}\tr[\mathcal{B}(\rA_i,\rB_j)\physstatedm_{ij}]
\end{equation}
where $p_{ij}\physstatedm_{ij}$ are projections of the physical state $\physstatedm$  onto the blocks of Alice's and Bob's observables. Each block can then be treated separately to achieve an expression of the form
\begin{equation}\label{cond}
    F(\Lambda^i_\rA\otimes\Lambda^j_\rB(\physstatedm_{ij}),\refstateproj) \geq f(\beta).
\end{equation}
 In \cite{Sekatski} it is proven that if $f$ is a convex function of $\beta$ there exist maps $\Lambda_\rA$ and $\Lambda_\rB$ such that the fidelity between $\Lambda_\rA\otimes\Lambda_\rB(\physstatedm)$ and $\refstateproj$ given the violation $\beta$ is lower bounded by $f(\beta)$. In \cite{Bardyn2009} a similar convexity argument is used to obtain the final bound.

The remaining challenge is to obtain relations of the form \eqref{cond}. In \cite{Bardyn2009} the problem is solved analytically and the isometry used is just the one that rotates the blocks of the observables to obtain the form given in \eqref{JLobs}.  The work \cite{Sekatski} provides a general recipe: \eqref{cond} can be solved by using a nonlinear optimisation with one variable per party. %The optimization can be done for an arbitrary channels $\Lambda_A$ and $\Lambda_B$. For examples they solved the isometry is the $\alpha$-dependent dephasing channel described in the next section. 

\subsubsection{Operator inequalities method}\label{oi}
An analytic approach to robust self-testing, introduced in \cite{Jed2} currently gives the best robustness bounds for the self-testing of two-qubit states. It is suited for self-testing protocols based on a Bell inequality violation. 
\iffalse
The figure of merit used  for robust self-testing is extractability $\Xi$ of the reference state $\psi$ from the physical state $\rho'$, which is fidelity between these two states optimised over all local quantum channels ($\Lambda_A, \Lambda_B$) acting  on $\rho'$:
%
\begin{equation*}
    \Xi(\rho' \rightarrow \psi) = \max_{\Lambda_A,\Lambda_B}F(\Lambda_A\otimes\Lambda_B(\rho'),\psi)
\end{equation*}
%
To quantify closeness of the physical state providing Bell violation $\beta \leq \beta_Q$ with the reference state providing the maximal violation $\beta_Q$ one needs to minimise extractability $\Xi$ over all states compatible with the violation $\beta$. This leads to the extractability-violation trade of function
%
\begin{equation*}
    \mathcal{Q}_{\psi,\Bb} = \inf_{\rho'|\tr[\Bb\rho'] = \beta}\Xi(\rho' \rightarrow \psi).
\end{equation*}
%
$\mathcal{Q}_{\psi,\Bb}$ is infimum of extractability over all states providing the violation $\beta$.
\fi
The method uses the notion of extraction (see section \ref{subsec:extract}) and works by proving an operator inequality of the form 
\begin{equation}\label{op-in}
    K \geq s\Bb_{\mathcal{I}} + \mu \openone,
\end{equation}
for all Bell operators $\Bb_{\mathcal{I}}$ for the Bell inequality $\mathcal{I}$ in question, where $K = \Lambda_\rA^\dagger\otimes\Lambda_\rB^\dagger\left(\ketbra{\psi'}{\psi'}\right)$ and $\Lambda^\dagger$ is the dual channel of $\Lambda$ with respect to the Hilbert-Schmidt inner product. This allows one to make linear robust self-testing statements, that is, to prove the existence of real parameters $s$ and $\mu$ such that the extractability-violation trade-off defined in \eqref{tradeof1} satisfies
\begin{equation}\label{tradeof}
    \mathcal{Q}_{\psi,\Bb_{\mathcal{I}}}(\beta) \geq s\beta + \mu.
\end{equation}
%
%Proving the operator inequality \eqref{op-in} for all Bell operators $\Bb$ directly implies \eqref{tradeof} and therefore a more direct self-testing statement
One thus has
\begin{equation}\label{jedfidelity}
    F(\Lambda_\rA\otimes\Lambda_\rB(\physstatedm),\refstate) \geq s\beta + \mu
\end{equation}
for all states $\physstatedm$ achieving violation greater that $\beta$. 

\begin{figure}
%\centering
    \begin{tikzpicture}[scale=0.95]
      \begin{axis}[
       xlabel = CHSH violation,
    ylabel = {Fidelity lower bound},
        xmin=2,
        xmax=2.828427,
        xtick={2,2.2,2.4,2.6,2.8},
        ymin=0.5,
        ymax=1,
        ytick={0.5,0.6,0.7,0.8,0.9,1},
	axis background/.style={fill=white} , set layers, cell picture=true,
	legend pos = north west
        ]
        \addplot [color=orange, draw=orange, thick, mark=none]
        table[row sep=crcr]{
2.828427 1\\
2.105823 0.5\\};
\addlegendentry{\cite{Jed2}}
        \addplot[color=magenta, draw=magenta, very thick, dashed, mark=none]
        table[row sep=crcr]{
2.828427 1\\
2.414214	0.5\\
        }; \addlegendentry{\cite{Bardyn2009}}
           \addplot[color=blue, draw=blue, very thick, dotted, mark=none]
        table[row sep=crcr]{
2.828427 1\\
2.35	0.4814\\};
        \addlegendentry{\cite{PhysRevA.91.022115}}
      \end{axis}
    \end{tikzpicture}
\caption{Lower bounds on the self-tested fidelity with the maximally entangled pair of qubits as a function of the observed violation of the CHSH inequality for three methods. A trivial lower bound on the fidelity is $0.5$, achievable with the separable state $\ket{00}$. Finding the optimal curve remains as an open question. The impossibility to obtain a fidelity higher than $0.5$ for every CHSH violation $>2$ is proven in \cite{tim}. The proof is constructive: there exists a state $\rho$ providing the CHSH violation of  $\approx 2.0014$, nevertheless there is no local channel $\Lambda$ such that fidelity between $\Lambda(\rho)$ and $\Phi^+$ is higher than $0.5$.}
\label{fig:CompCHSH}
\end{figure}
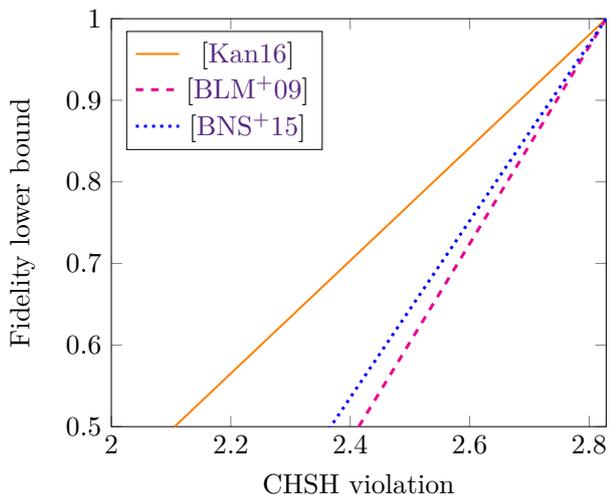

In principle it is a difficult task to prove the operator inequality \eqref{op-in} for all Bell operators regardless of the dimension.  In \cite{Jed2} Jordan's lemma is exploited to derive the current best robustness bounds for self-testing the maximially entangled state of two qubits. The method uses the CHSH inequality. The local channel $\Lambda_{\rA}\otimes\Lambda_{\rB}$ appearing in \eqref{jedfidelity} is as follows. First, local unitary transformations are applied to Alice and Bob's subsystems so that via the Jordan lemma, their local observables take a block diagonal form as in \eqref{JLobs}.
\iffalse
Then, for Alice's observable corresponding to input $x$, each of the blocks takes the form 
%
\begin{align}
\A_x=\cos \alpha \,\sx + (-1)^x \sin\alpha \,\sz.
\end{align}
%
This parametrisation covers all possibilities ;$\alpha=0$ implies that the observables commute in that block, whereas $\alpha=\pi/4$ implies anticommutation in that block. 
\fi
Then, for each block, one applies the $\alpha$-dependent dephasing channel
\begin{align}
   \Lambda_\alpha[\rho]=\frac{1+g(\alpha)}{2}\rho+\frac{1-g(x)}{2}\Gamma(\alpha)\rho\Gamma(\alpha) . \nonumber
\end{align}
Here $g(\alpha)=(1+\sqrt{2})(\sin \alpha +\cos \alpha -1)$ and 
\begin{align}
    \Gamma(\alpha)=\begin{cases} \sx \quad \alpha\in[0,\pi/4]\\
    \sz \quad \alpha \in (\pi/4,\pi/2]. \end{cases}
\end{align}
Bob's channel $\Lambda_{\rB}$ is defined analogously. This choice is shown to imply the lower bound \eqref{jedfidelity} to the fidelity with $s=(2+\sqrt{2})/8$ and $\mu=-(1+2\sqrt{2})/4$ (see figure \ref{fig:CompCHSH} for a plot). 

In \cite{Jed2} inequality \eqref{op-in} is also proven for Mermin inequalities in order to self-test the tripartite GHZ state. Moreover, the fidelity lower bound for the Mermin inequality is proven to be optimal in the sense that for any violation there always exists a state achieving that violation with the self-tested fidelity to the reference state. The method has also been used for robust self-testing of partially entangled pairs of qubits \cite{tim} and to assess the performance of different self-tests of a maximally entangled pair of qubits \cite{Li:19}. 

\begin{boxfigure}[label=box:npa]{The NPA Hierarchy}
%
% \begin{wrapfigure}{r}{0.35\textwidth}
%   \begin{center}
%     \includegraphics[width=0.35\textwidth]{fig_npa.pdf}
%   \end{center}
% \end{wrapfigure}
%
In its most general form, the NPA hierarchy \cite{npa1,npa2,pna} is a method to tackle optimisation problems involving polynomials of non-commuting variables, and as a result is suited to certain optimisation problems in quantum theory. Define optimisation problems of the form 
\begin{align}\label{npaprob}
    \min_{\ket{\psi},\M_{a\vert x},\N_{b\vert y}} \bra{\psi}\mathcal{P}(\{\M_{a\vert x},\N_{b\vert y}\})\ket{\psi}  \quad\text{subject to}\quad\quad \bra{\psi}\mathcal{F}_i(\{\M_{a\vert x},\N_{b\vert y}\})\ket{\psi}\geq 0 \;\;\forall i\;,
\end{align}
where $\mathcal{P}$ and $\mathcal{F}_i$ are polynomials in the measurement operators $\{\M_{a\vert x},\N_{b\vert y}\}_{a,x,b,y}$ and where the dimension of the state and measurement can be potentially infinite. The NPA hierarchy is a sequence of convex optimisation problems that provide increasingly better lower bounds to the optimal solution of the above by relaxing the problem to a minimisation over a larger set. Each of these relaxations can be solved via a corresponding semi-definite program \cite{boyd}. Many problems in quantum information can be cast in the above form, particularly in the device-independent setting where the state and measurements are unknown. \\

The NPA hierarchy works as follows. Consider a generic state and measurement operators $\{\ket{\psi},\{\M_{a\vert x}\},\{\N_{b\vert y}\}\}$. Then, define sets $\mathcal{S}_k$ (each corresponding to a level of the hierarchy) comprised of the identity operator and all (non-commuting) products of operators $\M_{a\vert x},\N_{b\vert y}$ up to degree $k$; e.g.\ $S_1=\{\openone\}\cup_{a,x}\{\M_{a\vert x}\}\cup_{b,y}\{\N_{b\vert y}\}$, $S_{k+1}=S_k\cup_{i,j} \{S_k^{(i)}S_1^{(j)}\}$, where $S_k^{(i)}$ is the $i^{\text{th}}$ element of $S_k$. Define the moment matrix of order $k$, $\Gamma^{k}$, by $\Gamma^{k}_{i,j}=\bra{\psi}S_{k}^{(i)\dagger}S_{k}^{(j)}\ket{\psi}$. For any state and measurements $\{\ket{\psi},\{\M_{a\vert x}\},\{\N_{b\vert y}\}\}$, the matrix $\Gamma^k$ is Hermitian positive semidefinite and satisfies some linear constraints given by the orthogonality conditions of the measurement operators. One can thus tackle optimisation problems of the form \eqref{npaprob} by minimising the corresponding elements of the matrix $\Gamma_k$, under linear constraints on $\Gamma_k$ and $\Gamma_k\succeq  0$. Such a problem is an instance of a semidefinite program which, via duality theorems, can be run to obtain certified lower bounds to the optimal solution. 
\end{boxfigure}

\subsubsection{Numerical Swap method}\label{sm}
The analytic techniques presented in the previous two subsections are only useful for either small amounts of noise (norm inequalities method), or (for now) solvable in simple cases, mostly when each party applies two binary measurements (operator inequalities method). For self-testing protocols which cannot be made robust with analytic methods, one can resort to a numerical method called the Swap method, introduced in \cite{PhysRevA.91.022115,PhysRevLett.113.040401}. While its applicability is still limited to simpler protocols due to computational resource requirements, it is responsible for the majority of  practically relevant robust self-testing bounds.

% Let us explain in a nutshel the idea of the SWAP method. Given a violation of a Bell inequality  $\beta$ and the isometry $\iso$, the aim of robust self-testing is to bound the distance between the reference state and $\iso[\rho']$ for all $\rho'$ compatible with the violation $\beta$. The choice of the isometry, the distance measure and the method are different in different methods.  

The Swap method uses the Swap gate isometry (see section \ref{swap}) and makes use of the fidelity of the extracted state as a figure of merit, as in definition \ref{robdef1}. To get a lower bound on the fidelity, one needs to minimise the fidelity between the state of the output registers of the Swap gate and the reference state, given that the input state to the Swap gate provides the violation $\beta$. For example, for two-qubit states we have seen in section \ref{sec:example}, equation \eqref{swoutput} that this fidelity is given by 
\begin{equation*}\label{swapfidelity}
    F(\physstatedm_{swap},\refstate) = \sum_{i,j,k,l \in \{0,1\}}c_{ij}^{kl}\physstatebra\hat{f}_{kl}^\dagger\hat{f}_{ij}\physstate,
\end{equation*}
where $\physstatedm_{swap} = \tr_{\rA\rB}\left[\iso[\physstatedm]\right]$, and $c_{ij}^{kl} = \langle kl \refstateproj ij\rangle$. Note that from the definition of $f_{ij}$, the above is equal to $\physstatebra\mathcal{P}(\{\A_a^x,\B_{b}^y\})\physstate$, where $\mathcal{P}$ is a polynomial in the measurement operators. As a result, lower bounds to the minimum fidelity (subject to a Bell inequality violation) can be found numerically via  a corresponding semi-definite program defined by the NPA hierarchy; see box \ref{box:npa} for more details. 

The first applications of the method are given in \cite{PhysRevLett.113.040401,PhysRevA.91.022115}, which involve the following self-testing results: the self-test of the singlet state and Bob's measurements from the CHSH inequality, the self-test of the singlet state from the Mayers-Yao criterion, the self-test of partially entangled pair of qubits from the tilted CHSH inequality, the self-test of a pure two qutrit state maximally violating CGLMP \cite{cglmp} inequality and the self-test of entangling measurements. Subsequently the method has been used to devise robust self-tests of the maximally entangled pair of qutrits \cite{abeille}, the whole family of pure entangled qutrit states \cite{bristol}, the three qubit W-state \cite{Wstate,Pal}, three- and four- qubit GHZ states and the four-qubit linear cluster state \cite{Pal}, a family of tripartite pure states, including the W-state from only marginal information \cite{marginal}, and a tensor product of two singlet states \cite{Wu2016}. It has also been used to compare the performance of different types of self-tests of the singlet state, presented in \cite{1367-2630-18-2-025021}.

\subsubsection{Algebraic method}\label{algRob}
In section \ref{genuine} we discussed self-testing through the `algebraisation' of the winning strategies in nonlocal games. Let us briefly recall that the crux of the method is associating an algebraic invariant, called the solution group, to each linear constraint system (LCS) nonlocal game. The rules of the nonlocal game allow one to define an abstract group whose representations correspond to the winning quantum strategy of the game. The correspondence between the group representations and winning strategies then allows for the use of techniques from group theory to prove self-testing statements.

In \cite{Coladangelo2017c} this reasoning is taken one step further: a quantum strategy winning the generalised magic square game with high probability allows to extract an approximate representation of the solution group, or equivalently, a mapping between the group elements and unitary operators which is approximately a homomorphism. The closeness between the approximate and the exact representation is then used to make a robust self-testing statement.

The most important ingredient for constructing robust self-tests in this way is the stability theorem for approximate representations from \cite{GowHat}. It states that for any approximate $n$-dimensional representation $f$ of a finite group $G = \{g_i\}_i$ there exists an exact unitary $m$-dimensional representation $h$ such that the Hilbert-Schmidt distance between $f(g_i)$ and $h(g_i)$ is small and $m$ is close to $n$.  
The distance between these two representations is related to the score the physical strategy gained in the LCS game under consideration. The full robustness statement is obtained through the use of the van Kampen diagrams \cite{vanKampen}. 

\subsection{Robust certification of large entanglement}\label{rnt}
In this section we discuss few contributions dealing with robustly certifying large amounts of entanglement without explicitly stating any self-testing result. There are two main reasons why such results merit attention in a review like this. The first is that for many purposes they could be used instead of robust self-testing protocols and moreover the self-testing statement is implicitly present for noiseless correlations. The second reason is the possible influence they could have on future approaches to robust self-testing.

A difference between robustness and noise tolerance when it comes to self-testing a tensor product of $n$ entangled pairs $\refstate^{\otimes n}$ is emphasised in works \cite{rotemJd} and \cite{rotemYuen}. The known self-testing protocols are robust in the sense that any strategy producing correlations that are $\epsilon$-close to the ideal ones must use a state which is $f(\epsilon,n)$-close to $\refstate^{\otimes n}$, where $f(\epsilon,n)$ scales as $an^b\epsilon^c$ for appropriate constants $a$,$b$ and $c$. This is, however, not the same as noise-tolerance since noisy source producing the state $\rho^{\otimes n}$, where $\rho$ is $\epsilon$-close to $\refstate$ is not $f(\epsilon,n)$-close to $\refstate^{\otimes n}$. The fidelity of such state with $\refstate^{\otimes n}$ drops exponentially with $n$, so there is very little hope to make any non-trivial self-testing statement about such highly entangled state. Instead of self-testing $\maxent^{\otimes n}$ \cite{rotemYuen} designs a one-shot test which is able to certify states whose entanglement of formation \cite{ef} is $\Omega(n)$. This certification method is noise-tolerant in the sense that the states $\rho^{\otimes n}$ are able to pass the test with high probability. A method to bound the one-shot distillable entanglement \cite{de} of the states produced by an uncharacterised source is presented in \cite{rotemJd}. The protocol is operationally useful since not all entanglement is consumed for certification. Both these results are implicit self-tests since the maximal score in the introduced games implies that the state produced by a source must be $\maxent^{\otimes n}$.

\section{Self-testing of measurements}\label{sec:measurements}

In many cases the correlations which self-test a quantum state also self-test the applied measurements. As a result, many of the state self-testing results presented in the previous sections are accompanied by a corresponding statement for the measurements. %In all protocols based on the SWAP gate self-testing of measurements usually follows as soon as the state has been self-tested. 
%The most commonly used approach is the one defined in Def. \ref{def_MST1}. 
In this section we give an overview of such results. In section \ref{sec:MSTresults} we review the known self-testing results for various sets of measurements, and in section \ref{sec:meas:methods} we discuss the different methods that have been used to achieve these results. We end the section with an overview of robustness techniques in measurement self-testing in section \ref{robmeas}.

\subsection{Measurement self-testing results}
\label{sec:MSTresults}

\subsubsection{Qubit measurements}
\label{sec:measurements:qubits}
The simplest set of incompatible qubit measurements is given by a pair of Pauli observables $\sigma_{\tx{x}}$ and $\sigma_{\tx{z}}$. Self-testing of these measurements (together with their rotated versions $(\sigma_{\tx{x}}\pm\sigma_{\tx{z}})/\sqrt{2}$ for the other party) can be achieved through the maximum violation of the CHSH Bell inequality (see Section \ref{sec:example}) or related self-tests. Such self-testing statements can be found in \cite{McKague2012,Jed1,PhysRevA.91.022115,Wstate}. Self-testing of the set of local observables $\{\sigma_{\tx{x}},\sigma_{\tx{z}},(\sigma_{\tx{x}}\pm\sigma_{\tx{z}})/\sqrt{2}\}$ can be achieved through the so called `Mayers-Yao' self test and its generalisations \cite{Mayers2004,McKague2014,McKague2012}. A method to self-test large sets of qubit observables that are equally spaced on the equator of the Bloch sphere was given  in \cite{SASA} based on the maximum violation of the chained Bell inequalities \cite{chained1,chained2}. A protocol for self-testing an arbitrary measurement from the real plane of the Bloch sphere is given in \cite{McKagueDQC}. Self-testing of pairs of observables of the form $\cos\mu\,\sx \pm \sin\mu\,\sz$ is given in \cite{Bamps} and \cite{Jed1} through the maximal violation of the tilted or weighted CHSH inequalities \cite{amp,llp}. The first self-testing of the set of three local Pauli observables $\{\sigma_{\tx{x}},\sigma_{\tx{y}},\sigma_{\tx{z}}\}$ first appeared in \cite{McKague2011}, using the `phase kick-back' (see section \ref{Pauli}) method and a generalised definition of self-testing to deal with the issue of complex conjugation. Other examples of such self-tests can be found in \cite{PhysRevA.96.032119,Jed1,erik}.

\subsubsection{Qudit measurements}\label{sec:measurements:qudits}
Self-testing results for measurements of dimension larger than two are much less common. The only self-test of mutually unbiased bases in a prime dimension higher that 2 was given in \cite{maxmaxmax} for dimension $d = 3$. Self-testing of the Bell state measurement was first achieved analytically in \cite{MO,Basel} (see section \ref{sec:StEntMeas} for an outline of the method). Self-tests of sets of measurements in high dimension can be achieved using the same techniques as in parallel self-testing of states (section \ref{parallel}). In this way, $n$-fold tensor products of the measurements $\{\sigma_{\tx{x}},\sigma_{\tx{z}}\}$ and $\{\sigma_{\tx{x}},\sigma_{\tx{y}},\sigma_{\tx{z}}\}$ in dimension $2^n$ have been achieved \cite{BSCA,Coladangelo2017b,Wu2016,McKague2017a,Coladangelo2017a,Natarajan,Coudron2016,Diagram,2058-9565-3-1-015002,Coladangelo2017c,Natarajan2018}. %In \cite{Tavakoli2018}, a set of qutrit measurements is self-tested in the semi-device independent prepare-and-measure scenario, see section \ref{pam} for more details. \textcolor{red}{Irreducible dimension witness.}

\subsubsection{Non-projective measurements}\label{sec:StPOVM}
Although definition \ref{def_MST2} of measurement self-testing assumes that the physical measurements are projective, one can nevertheless aim to prove that on the support of the reduced state of the self-tested state they act as some desired POVM. More specifically, suppose we have self-tested the reference state $\refstate$. Since the trace is invariant under isometry maps, the correlations can be written
\small
\begin{align}
    p(a,b\vert x,y)=\tr\left[\refstateproj_{\rA'\rB'}\otimes \sigma_{\bar{\rA}\bar{\rB}} \; \M_{a\vert x} \otimes \N_{b\vert y}\right], \nonumber
\end{align}
\normalsize
where the local measurements are projective and may act on both the primed and bared spaces. Taking the trace over the barred spaces we have
\begin{align}
    p(a,b\vert x,y)=\tr\left[\refstateproj\; \tilde{\M}_{a,b\vert x,y} \right], 
\end{align}
where
\small
\begin{align}
   \tilde{\M}_{a,b\vert x,y}=\tr_{\bar{\rA}\bar{\rB}}\left[\openone_{\rA'\rB'}\tp \sigma_{\bar{\rA}\bar{\rB}} \; \M_{a\vert x} \otimes \N_{b\vert y}\right]. 
\end{align}
\normalsize
To `self-test' non-projective measurements, one aims to show that $ \tilde{\M}_{a,b\vert x,y}=\M'_{a\vert x} \otimes \N'_{b\vert y}$, where now the reference measurements can be non-projective. Essentially, one is self-testing a Stinespring dilation \cite{stinespring_1955} of the non-projective measurement. 

In this manner, a self-test of the `tetrahedral' qubit POVM first appeared in \cite{toni}, with rigorous proofs appearing later in \cite{PhysRevA.96.032119} and \cite{andersson}, and an experimental demonstration presented in \cite{Smania}. These results were proven using the method of `post-hoc' self-testing, that we describe in \ref{posthoc}. To self-test measurements which are neither projective nor rank-one POVMs \cite{wagnerPOVM} use the approach developed by the same authors for the self-testing of quantum channels, described here in section \ref{STQC}. %On a related note, self-testing of extremal qubit POVMs has also been studied in the semi-device-independent prepare-and-measure scenario (see later section \ref{pam})  

\subsection{Methods in measurement self-testing}\label{sec:meas:methods}
In this section we outline some of the methods that have been used to prove measurements self-testing statements. 

\subsubsection{Phase kickback method for self-testing complex measurements}\label{Pauli}
As mentioned in section \ref{sec:generalisations},  definition \ref{def_MST2} is not suitable for self-testing complex-valued measurement operators. Take for example the problem of self-testing $\ket{\phi^{\text{+}}}$, the maximally entangled state of dimension 2, and $\{\sigma_{\text{z}},\sigma_{\text{x}},\sigma_{\text{y}}\}$, the three Pauli observables for say Alice. In section \ref{sec:example}, we have seen how one can self-test the state $\ket{\phi^{\text{+}}}$ and $\{\sigma_{\text{x}},\sigma_{\text{z}}\}$. Here, the issue of complex conjugation is not a problem since there exists a local basis in which the measurements and state are both real. However, there is no local basis in which the observables $\{\sigma_{\text{z}},\sigma_{\text{x}},\sigma_{\text{y}}\}$ are all real. Thus, we have two distinct possibilities for Alice's measurements, $\{\sigma_{\text{z}},\sigma_{\text{x}},\sigma_{\text{y}}\}$ and $\{\sigma_{\text{z}}^*,\sigma_{\text{x}}^*,\sigma_{\text{y}}^*\}$=$\{\sigma_{\text{z}},\sigma_{\text{x}},-\sigma_{\text{y}}\}$, both of which are compatible with the observed correlations. 

A natural question to ask is, given this uncertainty, what is the strongest possible self-testing statement that one could hope to prove? This question was first tackled by \cite{McKague2011}, see also \cite{Coladangelo2017,BSCA}. The basic idea is as follows. Consider a self-testing scenario in which Alice has (at least) three measurements given by the observables $\A_0$, $\A_1$, $\A_2$. Take a known self-testing protocol for the state $\ket{\phi^{\text{+}}}$ and observables $\{\sigma_{\text{x}},\sigma_{\text{z}}\}$ for Alice. Use this self-testing protocol three times for the pairs $\{\A_0,\A_1\}$, $\{\A_0,\A_2\}$, $\{\A_1,\A_2\}$, introducing new measurements for Bob and Alice if necessary. Since this proves that each pair $\A_i,\A_j$ anti-commute, one proves that the observables $\{\A_0,\A_1,\A_2\}$ pairwise anti-commute and should essentially be $\{\sigma_{\text{z}},\sigma_{\text{x}},\sigma_{\text{y}}\}$ or $\{\sigma_{\text{z}},\sigma_{\text{x}},-\sigma_{\text{y}}\}$. More precisely, one introduces a pair or local ancillas $\ket{00}_{\rA''\rA'}$ for Alice and another pair $\ket{00}_{\rB''\rB'}$ for Bob and proves the existence of an isometry $\iso$ such that 
\begin{align}
    &\iso[\physstate]=\maxent_{\rA'\rB'}\tp\ket{\xi}\nonumber\\
    &\iso[\A_0\physstate]=(\sz\otimes\openone\maxent_{\rA'\rB'})\tp\ket{\xi}\nonumber\\
     &\iso[\A_1\physstate]=(\sx\otimes\openone\maxent_{\rA'\rB'})\tp\ket{\xi}\nonumber\\
    &\iso[\A_2\physstate]=(\sy\otimes\openone\maxent_{\rA'\rB'})\tp\sz^{\rA'}\ket{\xi}\label{yflip}
\end{align}
where the state $\ket{\xi}$ has the form 
\small
\begin{align}\label{junkconj}
    \ket{\xi}=\ket{\xi_0}_{\rA\rB}\tp\ket{00}_{\rA''\rB''}+\ket{\xi_1}_{\rA\rB}\tp\ket{11}_{\rA''\rB''}.
\end{align}
\normalsize
In \eqref{yflip} the additional $\sz$ measurement on the junk state acts as an effective `controlled conjugation' for the measurement of $\sy$ on $\maxent$, where the probability to perform the conjugation is given by $\langle \xi_1\vert\xi_1\rangle$, which remains unknown. Note that if we consider only the space $\mathcal{H}_{\rA}\otimes\mathcal{H}_{\rB}$ then the action of $\A_2$ is to perform some unknown convex combination of $\sy$ and $-\sy$, as expected. Similar statements can be proven for Bob, where the register $\rB'$ acts as a control for a possible conjugation of his measurements. Note that given the form of \eqref{junkconj}, Alice and Bob will conjugate their measurement operators in a correlated fashion, as required from \eqref{conjmeas}. The isometry, introduced in \cite{McKague2011} and later used in \cite{Hajdusek} to prove the above self-testing statement is an extension to the Swap isometry (see figure \ref{fig:example}) introduced in section \ref{sec:example}. The full isometry consists of the regular Swap isometry, followed by two extra `phase kickback' controlled unitaries; see figure \ref{fig:pkb}.

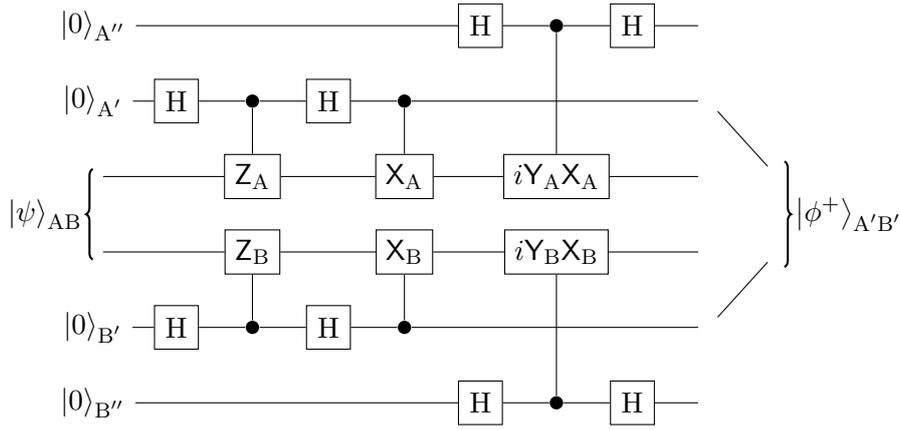
\begin{figure*}
  \centerline{
    \begin{tikzpicture}
    %
    % `operator' will only be used by Hadamard (H) gates here.
    % `phase' is used for controlled phase gates (dots).
    % `surround' is used for the background box.
    \tikzstyle{operator} = [draw,fill=white,minimum size=1.5em] 
    \tikzstyle{phase} = [fill,shape=circle,minimum size=5pt,inner sep=0pt]
   % \tikzstyle{surround} = [fill=blue!10,thick,draw=black,rounded corners=2mm]
    %
    % Qubits
    \node at (-0.1,-1) (q1) {$\ket{0}_{\rA'}$ };
    \node at (-0.1,0) (q11) {$\ket{0}_{\rA''}$ };
    \node at (-0.1,-2) (q2) { };
    \node at (-0.1,-3) (q3) { };
     \node at (-0.1,-5) (q44) {$\ket{0}_{\rB''}$ };
    \node at (-0.1,-4) (q4) {$\ket{0}_{\rB'}$ };
    % Bracket
    %\draw[decorate,decoration={brace, mirror},thick] (0,0.1) to
	%node[midway,left] (bracket) {$\ket{\Phi^+}^{A'A''}$}
	%(0,-1.1);
    \draw[decorate,decoration={brace, mirror},thick] (-0.1,-1.9) to
	node[midway,left] (bracket) {$\physstate_{\rA\rB}$}
	(-0.1,-3.1);
	%\draw[decorate,decoration={brace, mirror},thick] (0,-3.9) to
	%%node[midway,left] (bracket) {$\ket{\Phi^+}^{B'B''}$} 	(0,-5.1);
    %
    % Column 1
    \node[operator] (opH11) at (1,-1) {H} edge [-] (q1);
    \node[operator] (opH21) at (1,-4) {H} edge [-] (q4);
    %\node[operator] (op31) at (1,-2) {H} edge [-] (q3);
    %
    % Column 2
    \node[phase] (phase11) at (2,-1) {} edge [-] (opH11);
    \node[operator] (phase12) at (2,-2) {$\Zz_{\rA}$} edge [-] (q2);
    \draw[-] (phase11) -- (phase12);
    \node[phase] (phase13) at (2,-4) {} edge [-] (opH21);
    \node[operator] (phase14) at (2,-3) {${\Zz}_{\rB}$} edge [-] (q3);
    \draw[-] (phase13) -- (phase14);
    % Column 3
    \node[operator] (op12) at (3,-1) {H} edge [-] (phase11);
    \node[operator] (op22) at (3,-4) {H} edge [-] (phase13);
    %
 % Column 4
    \node[phase] (phase21) at (4,-1) {} edge [-] (op12);
    \node[operator] (phase22) at (4,-2) {$\Xx_{\rA}$} edge [-] (phase12);
    \draw[-] (phase21) -- (phase22);
    \node[phase] (phase23) at (4,-4) {} edge [-] (op22);
    \node[operator] (phase24) at (4,-3) {${\Xx}_{\rB}$} edge [-] (phase14);
    \draw[-] (phase23) -- (phase24);
 \node[operator] (opH31) at (5,0) {H} edge [-] (q11);
    \node[operator] (opH41) at (5,-5) {H} edge [-] (q44);
    %Column 5
    \node[phase] (phase31) at (6,0) {} edge [-] (opH31);
    \node[operator] (phase32) at (6,-2) {$i\Yy_{\rA}\Xx_{\rA}$} edge [-] (phase22);
    \draw[-] (phase31) -- (phase32);
    \node[operator] (phase33) at (6,-3) {$i{\Yy}_{\rB}{\Xx}_{\rB}$} edge [-] (phase24);
    \node[phase] (phase34) at (6,-5) {} edge [-] (opH41);
     \draw[-] (phase33) -- (phase34);
    % Column 6
    \node[operator] (op81) at (7,0) {H} edge [-] (phase31);
    \node[operator] (op82) at (7,-5) {H} edge [-] (phase34);
    % Column 7
    \node (end1) at (8,-1) {} edge [-] (phase21);
    \node (end5) at (8,0) {} edge [-] (op81); 
    \node (end2) at (8,-2) {} edge [-] (phase32);
    \node (end3) at (8,-3) {} edge [-] (phase33);
    \node (end6) at (8,-5) {} edge [-] (op82);
    \node (end4) at (8,-4) {} edge [-] (phase23);
    %
    % Bracket
 \node at (8,-1) (eend1) {};
    \node at (8.9,-2) (eend11) {};
    \draw[-] (eend1) -- (eend11);
    \node at (8,-4) (eend4) {};
    \node at (8.9,-3) (eend44) {};
    \draw[-] (eend4) -- (eend44);
    \draw[decorate,decoration={brace},thick] (9,-1.8) to 	node[right] (bracket) {$\ket{\phi^+}_{\rA'\rB'}$} 	(9,-3.2);
    %
    %\draw[decorate,decoration={brace},thick] (9,-0.8) to 	node[midway,right] (bracket) {$\ket{\xi}$} (9,-4.2);
    %\draw[decorate,decoration={brace},thick] (10.5,0.2) to 	node[right] (bracket) {$\ket{\Phi^{+}}$} 	(10.5,-5.2);
    %
    % Background Box
  %  \begin{pgfonlayer}{background} 
   % \node[surround] (background) [fit = (q1) (op31) (bracket)] {};
  %  \end{pgfonlayer}
    %
    \end{tikzpicture}
  }
  \caption{Swap gate with a phase kick back at the end of the circuit \label{fig:pkb}. The ${\mathsf{Y}}$ operators are constructed from the additional measurement operators in a similar fashion to \eqref{swap_ops}
  }
\end{figure*}
\iffalse
\begin{figure*}
\centering
\includegraphics[width=1.1\columnwidth]{PhaseKickBack.pdf}
\caption{Swap gate with a phase kick back at the end of the circuit \label{fig:pkb}. The $\tilde{\mathsf{Y}}$ operators are constructed from the additional measurement operators in a similar fashion to \eqref{swap_ops}.}
\end{figure*}
\fi

\subsubsection{Self-testing measurements based on commutation}\label{comm}
The majority of self-testing protocols first prove self-testing of the state, and then move to self-test the form of measurements on the support of the self-tested state. An alternative approach for binary observables was used already in \cite{Popescu1992} and revived in \cite{Jed1}. In this approach the violation of a Bell inequality is directly related to the commutation properties of measurement observables without the need to prove a statement of the form of definition \ref{def_MST2}. In the two input, two output scenario the figure of merit (for Alice's observables) is directly relatable to the maximal violation of the CHSH inequality and is given by 
\begin{align}
    t_{01} =\frac{1}{2}\tr( |[\A_0,\rA_1]|\physstatedm_{\rA}), 
\end{align}
where $\physstatedm_{\rA} = \tr_{\rB}\physstatedm_{\rA\rB}$ is Alice's reduced state of the physical state. The maximal CHSH violation implies $t_{01}=1$, which further can be used to infer anticommutativity of $\A_0$ and $\A_1$ on $\physstatedm_{\rA}$.  Beyond CHSH, it is also proven that the maximal violation of the Mermin-Ardehali-Belinskii-Klyshko inequalities \cite{mermin,ardehali,bk} implies the parties use anticommuting observables to achieve the maximal violation. 

The method is also applied to self-test a set of three mutually anti-commuting observables. Note that there exists a basis in which any two anticommuting observables $\A_0$ and $\A_1$ can be written
\begin{align}\label{ac01}
    \A_0 = \sz\otimes \openone , \qquad  \A_1 = \sx\otimes \openone.
\end{align}
If an inequality involving three observables $\A_0,\A_1,\A_2$ can be used to certify $t_{01} = t_{02} = t_{12} = 1$ then besides relation \eqref{ac01} the following relation can be extracted
\begin{equation}
    \A_2 = \sy \otimes \rA_{\text{Y}},
\end{equation}
where $\A_{\text{Y}}$ is a Hermitian $\pm 1$-eigenvalue operator. Hence, when measuring $\A_2$ the measurement result for $\rA_{\text{Y}}$ on the `junk' Hilbert spaces takes care of the complex conjugation issue in the same way as \eqref{yflip}.

Once the form \eqref{ac01} is extracted for measurement observables of all parties it is possible to also make statements about the underlying state.  Furthermore, in \cite{Jed1} this method is useful to certify observables which are not anticommuting but maximally violate weighted CHSH inequalities introduced in \cite{llp}. Going beyond binary measurements, the extension of this method is applied also in \cite{maxmaxmax} to self-test mutually unbiased bases in the dimension $d = 3$. 

\subsubsection{Post-hoc self-testing of measurements}\label{posthoc}
Once a state and sufficiently many measurements have been self-tested, further measurements can often be self-tested `for free' through a method that we call post-hoc self-testing. As an example, the maximal violation of the CHSH inequality assures that up to an isometry the shared state is $\maxent$, Alice's measurement observables are $\sz$ and $\sx$, and Bob's measurement observables $(\sz\pm\sx)/\sqrt{2}$. The self-testing protocol can be extended to certify any other real qubit observable applied by either Alice or Bob. Assume Bob uses another measurement observable $\B_2$ which we want to self-test as $\st$ on his half of the maximally entangled state. If Bob performs this measurement, we will observe the correlations $\physstatebra \A_0\B_2\physstate = \cos{\theta}$ and $ \physstatebra \A_1\B_2\physstate = \sin{\theta}$. Now, the maximal violation of the CHSH inequality implies that $\A_0\physstate$ and $\A_1\physstate$ are orthogonal. If we take these two states to be the first two states in an orthonormal basis of the full space, it must be that $\B_2\physstate = \cos\theta \A_0\physstate + \sin\theta \A_1\physstate$. Applying the isometry to this, we have
\begin{align}
    \iso[\B_2\physstate]=&\iso\left[\cos\theta \A_0\physstate + \sin\theta \A_1\physstate\right] \nonumber\\
    =&(\cos\theta \sz\tp\openone+\sin\theta\sx\tp\openone \maxent)\tp\junk \nonumber\\
    =&(\openone\tp\cos\theta \sz+\openone\tp\sin\theta\sx \maxent)\tp\junk, \nonumber
\end{align}
as required, where we have used the property $\M\otimes\openone\maxent=\openone\tp\M^T\maxent$. 

This technique can be understood from the perspective of measurement tomography. Given a set of linearly independent pure states that are tomographically complete on some subspace, one can infer the form of any measurement in this subspace from the statistics of measurement outcomes on the set of states. Given the CHSH self-test, we know that conditioned on Alice's input and output, the reduced states of Bob up to a unitary transformation are $\pi_{a\vert x}^\rB\otimes\openone_{\rB}$, where $\pi_{a\vert x}=\proj{0},\proj{1},\proj{+},\proj{-}$ depending on the value of $x,a$. These four states are informationally complete for real qubit measurements, and can thus be used to infer further such measurements for Bob when interpreted as states in a measurement tomography protocol. The technique can be applied in a similar way for higher-dimensional self-testing protocols and to the post-hoc self-testing of complex measurements. The first time such an approach was used was for self-testing real measurements applied on a graph state in \cite{McKagueDQC}, expanding the protocol for self-testing graph states from \cite{McKague2014}. This technique has also been particularly useful to self-test non-projective measurements (see \ref{sec:StPOVM}).

\subsubsection{Self-testing of entangling measurements}\label{sec:StEntMeas}
An entangling measurement is one whose measurement operators are non-separable with respect to some bipartition of the Hilbert space. Two recent works \cite{MO,Basel} have presented analytic methods to robustly self-test the Bell state measurement (BSM), the entangling measurement whose eigenvectors are the four maximally entangled Bell states $\{\ketbra{\phi^+}{\phi^+}, \ketbra{\phi^-}{\phi^-},\ketbra{\psi^+}{\psi^+},\ketbra{\psi^-}{\psi^-}\}$. Here the reference scenario is an entanglement swapping protocol: Bob possesses two particles, one maximally entangled with Alice's particle and the other with Carmela's. Bob performs the BSM on his two particles, and depending on his outcome projects the particles of Alice and Carmela onto one of the Bell states. These four Bell states can be self-tested by maximally violating the four different CHSH inequalities (mutually related by relabelling) conditioned on the outcome of Bob. The idea for self-testing is simple: if Alice and Carmela maximally violate all CHSH inequalities then it must be that Bob's measurement is the BSM. Importantly, to maximally violate each of the four different Bell inequalities, Alice and Carmela use the same measurements.

In \cite{MO,Basel},  the entanglement swapping scenario is used to self-test the BSM using the notion of measurement self-testing via simulation described in section \ref{simulation}. Here, one necessarily needs to identify some well defined local Hilbert spaces for Bob in order to define entanglement. This is achieved by assuming that there are two independent sources between Alice and Bob, and between Bob and Carmela. Entanglement is then defined with respect to the Hilbert spaces of these two sources. A self-testing protocol for entangling measurements whose eigenvectors are partially entangled pairs of qubits, or GHZ states is also presented in \cite{MO}. In \cite{Basel} the techniques for self-testing quantum channels is used to self-test the BSM. For more details on these techniques see section \ref{STQC}.

It is also worth mentioning here that in \cite{Rabelo2011}, a protocol to device-independently certify the existence of an entangling measurement is given, which was later shown to be robust in \cite{PhysRevA.91.022115} via the use of the numerical Swap method (section \ref{sm}). Both these works however only prove the existence of some entangling measurement but do not provide a self-testing statement for a particular measurement. In the context of witnessing irreducible dimension a way to certify entangling measurements is also presented in \cite{IrrDim} alongside with a figure of merit quantifying how entangled the measurement operators are.

\subsection{Robust measurement self-testing}\label{robmeas}
In this section we give an overview of the approaches to robust self-testing of measurements. As we saw in section \ref{robust-intro} there are a number of valid figures of merit one could consider when robustly self-testing a state. Given the increased complexity of measurements compared to states, the self-testing of measurements is even more diversified. In what follows we discuss a few approaches to quantify how close a physical measurement $\{\M_x\}$ is to some reference measurement $\{\M'_x\}$. 

The straightforward approach, in accordance with definition \ref{robdef2} uses the same methods as in the self-testing of states. In robust self-testing of states one defines a figure of merit that captures the closeness between the states $\iso[\physstatedm_{\rA\rB}]$ and $\refstateproj_{\rA'\rB'}\tp\junkdm_{\bar{\rA}\bar{\rB}}$. For measurement self-testing, one can simply use the same figure of merit between the sub-normalised states $\iso\left[\M_{a\vert x}\tp \openone\,\physstatedm_{\rA\rB}\, \M^\dagger_{a\vert x}\tp \openone\right]$ and $\left(\M'_{a\vert x}\tp \openone\,\refstateproj_{\rA'\rB'}\,\M^{'
\,\dagger}_{a\vert x}\tp \openone\right)\otimes\junkdm_{\bar{\rA}\bar{\rB}}$. This will mean that there will be a different value associated to each of the measurement operators; a single figure of merit can be obtained by, for example, taking the average or maximum of these values. Such an approach is used, for example, in  \cite{McKague2012,McKague2014,Bamps,SASA}.

The Swap method \cite{PhysRevLett.113.040401,PhysRevA.91.022115} can also be used to define a figure of merit for robust measurement self-testing. Taking the CHSH example, to estimate the closeness of Alice's measurements to the reference measurements, the Swap gate is applied only on her system (i.e. only Alice's local branch of the Swap gate is used) and the ancilla is initiated in one of the eigenstates $\ket{\varphi_\pm^{\A'_i}}$ of her reference observable $\A'_i$. In the ideal case, if Alice measures the reference observable $\A'_i$ after the Swap gate is applied she will deterministically obtain the outcome $\pm 1$. The probability that this measurement gives the result $+1$ is then used as a figure of merit to asses the closeness of the measurement, however it is not proven to give a distance measure. As with the estimation of the fidelity with the reference state, one can use the NPA hierarchy to lower bound this quantity with respect to the given CHSH violation. 

A different figure of merit, analogous to the notion of state extractability (see section \ref{robust-intro}), was suggested in different self-testing contexts \cite{Tavakoli2018,MO,Basel,arminPOVM,wagnerPOVM}. If $D$ is a suitably chosen distance measure on the set of measurements, one can define the distance $\mathcal{D}$ between the physical measurement $\M_{a\vert x}$ and the reference one $\M'_{a\vert x}$ as
\begin{equation}\label{disMeasMeas}
    \mathcal{D}(\M_{a\vert x},\M'_{a\vert x}) =\frac{1}{c_a}\max_{\Lambda} \sum_{a}D(\Lambda(\M_{a\vert x}),\M'_{a\vert x}),
\end{equation}
where $c_a$ is a normalisation factor and the maximisation is taken over all completely positive and unital maps $\Lambda: \mathcal{L}(\mathcal{H}_{\rA})\rightarrow \mathcal{L}(\mathcal{H}_{\rA'})$. Depending on the type of reference measurements $D$ can be chosen to be the overlap (as in \cite{MO,Tavakoli2018,arminPOVM}), Uhlmann fidelity (as in \cite{Basel,wagnerPOVM}) or any other distance measure. While the physical state $\physstatedm$ does not explicitly appear in  \eqref{disMeasMeas}, the map $\Lambda$ has to depend in some way on it. In \cite{wagnerPOVM,Basel}  measurements are self-tested through their action on the maximally entangled pair of qudits and the state appears explicitly in the distance measure.

\section{Extensions of self-testing to other scenarios}\label{sec:extensions}

In this section we cover three extensions to the standard scenario of self-testing. In section \ref{STQC} we cover works that self-test the action of a quantum gate in a device-independent manner. In section \ref{sec:semi} we focus on the so-called semi-device independent approaches. In section \ref{context} we focus on self-testing via contextuality. 

\subsection{Self-testing of quantum gates and circuits}\label{STQC}

The paradigm of self-testing can be useful in scenarios going beyond the certification of states and measurements. Anticipating usefulness in the certification of devices for quantum computing, one may ask if it is possible to certify quantum gates, \textit{i.e.} unitary transformations. First answers to this question came already in the early days of self-testing with two contributions devoted to the task of self-testing of quantum gates or quantum circuits, \cite{vandam} and \cite{Magniez}. Although \cite{vandam} has the phrase `self-testing' in its title, it does not correspond to the fully device-independent scenario; the certification relies on several assumptions such as knowledge of the dimension of the system, which shifts it to the landscape of semi-device-independent scenarios.  

The first protocol providing a recipe to self-test quantum gates acting on an arbitrary number of qubits is presented in \cite{Magniez}. Denote the physical implementation of the gates Alice and Bob use with $G_\rA$ and $G_\rB$. For the protocol to work, Alice and Bob must have access to the \emph{same} gate, that is, $G_\rA=G_\rB$. The core of the protocol is the Mayers-Yao self-test of the maximally entangled pair of qubits. To self-test a one-qubit unitary gate $G'_{\rA'}$ acting on her system, Alice has to share a maximally entangled pair of qubits with Bob. As usual, the state shared between Alice and Bob is $\physstate$ and they perform measurements $\{\M_{a|x}\}$ and $\{\N_{b|y}\}$. The aim is to show that there is a local isometry $\iso$ such that
\begin{multline*}
\iso\otimes\openone_\rP\left[{G}_{\rA}\,{\M}_{a|x}\otimes\openone\physstate_{\rA\rB\rP}\right] = \\ =  {G'}_{\rA'}\M'_{a|x}\otimes\openone\maxent_{\rA'\rB'}\otimes \ket{\xi}_{\bar{\rA}\bar{\rB}\rP}
\end{multline*}
where $\{\M'_{a|x}\}$ are the reference measurements for the Mayers-Yao self-test.  The protocol consists of three parts:
\begin{itemize}
\item The Mayers-Yao self-test on the input state $\physstate$,
\item The Mayers-Yao self-test on the output state ${G}_{\rA}\otimes {G}_{\rB}\physstate$,
\item A check that ${G}_{\rA}\otimes \openone_{\rB}\physstate_{\rA\rB}$ reproduces the statistics of ${G'}_{\rA'}\otimes \openone_{\rB'}\maxent_{\rA'\rB'}$ with respect to the Mayers-Yao measurements.
\end{itemize}
The first two steps serve for self-testing the underlying state and ensure that $G$ is a unitary gate (at this step potentially the identity gate). The third step can be seen as a tomography of $G$, since the measurements and state are already self-tested in the first two steps. Note that this means one can only self-test gates having real coefficients with respect to the self-tested measurements. The method is also extended to many-qubit gates. Each of Alice's qubit on which a gate acts is maximally entangled with another qubit of Bob, and the tensor product structure of Alice's and Bob's Hilbert spaces is assumed. The procedure repeats as in the case of one-qubit gates, with three steps involving the self-test of the input and output states which are now tensor products of many maximally entangled pairs of qubits and tomography of the corresponding multi-qubit gate. By using this method one can self-test the whole quantum circuit by self-testing each gate in sequence according to the recipe given above. This self-test is also proven to be robust.

Another self-test of quantum gates with simpler structure and significantly better robustness bounds is given in \cite{Sekatski}. It is more general than \cite{Magniez} since it provides a framework to lower bound the fidelity with an arbitrary quantum channel $\Gamma'$. For Alice to self-test the channel $\Gamma'$ she again has to share a maximally entangled pair of qubits with Bob, but now Bob does not perform any channel to his system. Let the physical implementation of the channel be denoted as $\Gamma$. The aim is to find the fidelity between the reference channel $\Gamma'$ and the physical one $\Gamma$. The protocol consists of two steps:
\begin{itemize}
\item The self-test of the input state $\physstate$. The result provides a lower bound to the input fidelity $F_i$ between $\Lambda^i_{\rA}\otimes\Lambda_{\rB}\physstate$ and $\maxent$, where $\Lambda^i_{\rA}$ and $\Lambda_{\rB}$ are the CPTP maps as explained in \ref{simulation}.

\item The appropriate self-test which finds a lower bound of the output fidelity $F^o$ between the state $\Lambda^o_{\rA}\otimes\Lambda_{\rB}\left({\Gamma}_{\rA}\otimes\openone_{\rB}\physstate_{\rA\rB}\right)$ and the reference output state ${\Gamma'}_{\rA'}\otimes\openone_{\rB'}\maxent_{\rA'\rB'}$.
\end{itemize}
The fidelity of the physical channel to the reference channel is proven to be lower bounded by $\cos\left(\arccos(F_i) + \arccos(F_o)\right)$. The main challenge is to find the appropriate self-test necessary for the second step. The paper gives the solution for unitary channels (\textit{i.e.} quantum gates), generalises the protocol for many-qubit channels and provides the explicit solution for arbitrary two-qubit controlled gate of the form $CU_{\varphi} = \ketbra{0}{0}\otimes \openone + \ketbra{1}{1}\otimes e^{-i\varphi\sx}$. Since such gates are necessary and sufficient for universal quantum computing (together with a set of single qubit gates) the toolkit represents an important contribution to the self-testing of all the building-blocks of a quantum computer. 

Aside from this, the paper \cite{Sekatski} is also valuable for two contributions independent of quantum channel self-testing. One is a method to self-test multipartite states, described in \ref{tailoring} and the other a technique useful for robust self-testing, described in \ref{oi}. Furthermore, by defining the gates in terms of their Krauss representation, the techniques from \cite{Sekatski} have been generalised in \cite{wagnerPOVM} to allow for self-testing of  measurements other than rank-one POVMs.

\subsection{Semi-device-independent scenarios}\label{sec:semi}
A number of works have investigated extensions of self-testing to so-called semi-device-independent (SDI) scenarios. In the device-independent scenario, all devices are treated as black boxes and one hence imposes minimal assumptions on the states and measurements. In the SDI scenario, some additional assumptions are added, without assuming a full characterisation of the entire set-up. As such, the SDI scenario can be seen as a weaker version of the DI scenario, intermediate  between the scenarios of full device independence and full characterisation. Moving to the SDI scenario can be advantageous for at least three reasons. First, the additional assumptions can overcome some of the mathematical difficulties of the DI scenario and make statements easier to prove and results more tolerant to noise; second, for some scenarios it may actually be necessary to move to SDI scenario in order to make any non-trivial statements (see \ref{pam}), and third, the additional assumptions may be very natural given a particular experimental set-up or level of trust in some devices. In this section we give an overview of three extensions of self-testing to the SDI scenario, namely one-sided device-independent self-testing, commonly known as the EPR steering scenario (section \ref{steering}), self-testing in prepare-and-measure scenarios (section \ref{pam}), and self-testing based on noncontextuality inequalities (section \ref{context}).

\subsubsection{One sided device-independent self-testing (EPR steering)}\label{steering}
The one-sided device-independent scenario (also commonly referred to as the EPR steering scenario), is equivalent to the standard self-testing scenario, with the additional assumption that there is one trusted party (here Bob) whose device is fully characterised, that is, his measurement operators are known. %L 
Thus, Bob is able to apply any quantum measurement and can in principle perform quantum state tomography of his half of the state. Alice, as in the self-testing scenario, receives classical input $x$ to her device and outputs classical output $a$. The subnormalised state of Bob conditioned on Alice's input $x$ and output $a$ is given by
\begin{equation}
\sigma_{a|x} = \tr_\rA\left[\M_{a|x}\otimes\openone\physstatedm_{\rA\rB}\right]
\end{equation}
The set $\{\sigma_{a|x}\}_{a,x}$ is called an assemblage. It is said that the assemblage $\{\sigma_{a|x}\}_{a,x}$ admits a local hidden state model (LHS) if it admits a decomposition
\begin{equation}\label{nosteer}
\sigma_{a|x} = \int_{\lambda}d\lambda\,q(\lambda) p_{a|x,\lambda}\rho_{\lambda}, \qquad \forall a,x,
\end{equation}
where $q(\lambda)$ is a normalised probability density and $\rho_{\lambda}$ is a normalised density operator acting on the local Hilbert space of Bob. If  the assemblage  $\{\sigma_{a|x}\}_{a,x}$ is incompatible with a LHS model one says that it demonstrates steering. The existence of a LHS model can be refuted by violation of steering inequalities, which take into account the correlations between Alice's outputs and the outputs of known measurements performed by Bob. Another way to prove that the assemblage $\{\sigma_{a|x}\}_{a,x}$ demonstrates steering is by using simple SDP optimisations \cite{wiseman},\cite{steeringreview},\cite{steeringreview2}.

The decomposition \eqref{nosteer} captures the types of assemblages that Bob can see if the two parties do not share any entanglement. Thus, a violation of \eqref{nosteer} demonstrates that the shared state must be entangled. A natural question is in which cases can we go beyond witnessing entanglement and recover the shared state $\physstatedm_{\rA\rB}$. This task was introduced in \cite{IvanMatty} and \cite{gheorghiu2017} under the name one-sided device-independent (1SDI) self-testing. In \cite{IvanMatty} the authors are mostly interested in the robustness of 1SDI self-testing and how it compares to the robustness of standard self-testing, while the authors of \cite{Gheorghiu} are also interested in the application to delegated quantum computing protocols (see Section \ref{dqc}). Two types of 1SDI self-testing are introduced: correlation based, which draws conclusions only from the violations of steering inequalities, and assemblage-based, which works with the full assemblage. An interesting conclusion is that in the case of the simplest self-test of the maximally entangled pair of qubits, the asymptotic behaviour of the self-tested fidelity as a function of noise is the same in both the 1SDI and DI scenarios. How general this statement is remains as an open question. The 1SDI scenario is also very useful for self-testing a tensor product of many EPR pairs. In the standard self-testing of such states the main difficulty is establishing a tensor product structure, while in 1SDI scenario this comes for free due to the fact that Bob's device is characterised. Numerical techniques, similar to the Swap method, for robust self-testing in the 1SDI scenario were also presented in work \cite{IvanMatty}. 1SDI self-testing of all pure two-qubit states is presented in \cite{steeringAnyTwoQubit}.

\begin{figure}
    \centering
    \includegraphics[scale=1.4]{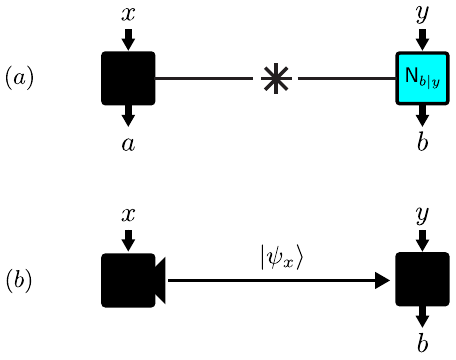}
    \caption{Semi-device-independent scenarios. (a) The one-sided device-independent scenario. One of the parties (here Bob) is assumed to have a trusted measurement device (i.e.\ his measurement operators are known). (b) The prepare-and-measure scenario. Alice sends a quantum state to Bob, conditioned on her input.}
    \label{fig:sdi}
\end{figure}

\subsubsection{Self-testing in the prepare-and-measure scenario}\label{pam}
A recent series of works have adapted the self-testing scenario to the prepare-and-measure scenario. Here, Alice sends one of a number of states $\ket{\psi_x}$, labelled by $x$, to Bob, who measures $\{\M_{b\vert y}\}$ conditioned on input $y$ and obtains outcome $b$. The statistics of the experiment are therefore given by 
\begin{align}
    p(b\vert x,y)=\tr[\proj{\psi_{x}}\M_{b\vert y}].
\end{align}
In analogy to the case of self-testing entangled states and measurements, one aims to infer from the statistics that the preparations and measurements $\{\ket{\psi_x},\M_{b\vert y}\}$ are equal to some reference set $\{\ket{\psi'_x},\M'_{b\vert y}\}$ up to some unknown isometry. 

In contrast to the Bell scenario, non-trivial statements can only be made if one places additional assumptions on the experiment. To see this, note that any statistics $p(b\vert x,y)$ can be reproduced by sending the label $x$ to Bob, i.e. with the preparations $\ket{\psi_x}=\ket{x}$, and Bob simply outputting $b$ with probability $p(b\vert x,y)$. Thus, self-testing any set of preparations that are not diagonal in the same basis is impossible without some additional assumptions. As is common in the prepare-and-measure scenario, a number of recent works \cite{Tavakoli2018,arminPOVM,marcinPOVM,Farkas2018} overcome this by assuming an upper bound on the Hilbert space dimension of the preparations and measurements. A similar assumption has also been studied in the Bell scenario \cite{Bardyn2009,gohMDI} where the state and measurements are assumed to be qubit systems, making self-testing statements significantly easier to prove.

A convenient figure of merit in prepare-and-measure scenario is the success in a game called a random access code (RAC) \cite{RAC1,RAC2,RAC3}. In a $n^d\rightarrow 1$ RAC, Alice receives $n$ dits $x=(x_1,x_2,\cdots,x_n)$ and Bob receives $y=1,2,\cdots,n$. The aim is to maximise the average probability that Bob correctly guesses the input bit $x_y$, i.e. to maximise the expression 
\begin{align}\label{RAC}
    \mathcal{A}_{n^d\rightarrow 1}=\frac{1}{2d^n}\sum_{x,y}p(b=x_y\vert x,y).
\end{align}
 In \cite{Tavakoli2018}, the authors study as a figure of merit the $2^2\rightarrow 1$ RAC. It is proven that the value $\mathcal{A}_{2^2\rightarrow 1}$ can be used to perform robust self-testing of the preparations $\{\ket{0},\ket{1},\ket{+},\ket{-}\}$ and measurements given by the qubit observables $(\sx\pm \sz)/\sqrt{2}$ under the assumption of qubit preparations and measurements, using a technique based on that in \cite{Jed2} (similar statements also appeared in \cite{erikSDI,noSR} in the context of SDI quantum key distribution protocols and dimension witnesses).  This is then generalised to a self-test of any pair of non-commuting qubit observables by adding an input-dependent bias to \eqref{RAC}. The authors also use $\mathcal{A}_{2^2\rightarrow 1}$ to self-test a non-trivial set of qutrit preparations and measurements, and implement an adaptation of the numerical Swap method (see Section \ref{sm}) to deal with the prepare-and-measure scenario. In \cite{Farkas2018}, the authors study the $2^d\rightarrow 1$ RAC. It is proven that this game provides a robust self-test of a pair of measurements that correspond to two mutually unbiased bases in dimension $d$. Further to this, the authors show how the score of the RAC can also be used to bound both the incompatibility robustness \cite{incompat_robustness} of the pair and the randomness of the measurement outputs. It is worth emphasising that self-testing claims in this scenario can be made for arbitrary dimension-bounded communication games, \emph{i.e.} it is not restricted to RACs.

Several works have investigated self-testing of non-projective measurements in the dimension-bounded prepare-and-measure scenario. In \cite{marcinPOVM} the authors start from the $3^2\rightarrow1$ RAC to develop a self-test of the extremal `tetrahedral' qubit POVM.  In \cite{arminPOVM}, a general method is given to self-test any extremal qubit POVM, given a self-test of a set of preparations with opposite Bloch vectors to the POVM on the Bloch sphere. In a similar fashion in \cite{TRR} the authors provide a self-test of $d$-dimensional SIC POVM (whenever it exists).  To prove the ideal self-testing all works use the same idea; if there is one outcome of a measurement that never occurs for a given preparation, it follows that the corresponding POVM element must be opposite to the preparation on the Bloch sphere. From this one can use a self-test of preparations to effectively tomograph the POVM measurement in a similar manner to that in Section \ref{posthoc}. In \cite{arminPOVM} the protocol is made robust  by introducing a convenient distance measure which is used in the experimental demonstration reported therein. All three works also discuss certification of the non-projective nature of a measurement, which is a weaker form of certification than robust self-testing since it does not discuss closeness to any particular POVM.

The recent work \cite{unsharpSDI} investigates the self-testing of non-projective measurements in a prepare-and-measure scenario involving a sequence of measurements, using figures of merit that are closely linked to the  $2^2\rightarrow 1$ RAC. To achieve the optimal success probability in the game, one party needs to perform a so-called L{\"u}ders instrument, which corresponds to a non-projective measurement. The authors then derive bounds on the maximal eigenvalue of the corresponding measurement operators given an observed score in the game, using a numerical approach based on an adaptation of the NPA hierarchy to the prepare-and-measure scenario.

\subsubsection{Self-testing through noncontextuality inequalities}\label{context}
 One of the important features of quantum theory is contextuality, first noticed via the Kochen-Specker theorem \cite{Ks}. In general, in any deterministic hidden variable model reproducing quantum correlations, the outcome of a measurement  $\M$ must depend on its context, \textit{i.e.} the set of compatible measurements one may perform alongside $\M$. Contextuality can be detected through the violation of noncontextuality inequalities, of which the simplest is the Klyshko-Can-Binicio{\u{g}}lu-Shumovsky (KCBS) inequality \cite{Kcbs}. It corresponds to the scenario in which five projective binary measurements $\{\M_i = (\M_{0|i},\M_{1|i})\}_{i=1}^5$ can be performed and each pair $\M_i,\M_{i+1}$ is compatible, i.e.\
 \begin{align}\label{context_compat}
     [\M_i,\M_{i+1}]=0
 \end{align}
 and exclusive \emph{i.e.} $\tr(\M_{0|i}\M_{0|i+1}) = 0$ (the labels of measurements are taken modulo $5$).  If we denote $p_i = \tr[\M_{0|i}\rho]$, where $\rho$ is the measured state the KCBS inequality reads:
\begin{equation*}
    \sum_{i=1}^5p_i \leq 2.
\end{equation*}
The inequality is satisfied in all outcome deterministic noncontextual theories, while quantum measurements achieve the value $c_q = 5\cos(\pi/5)/(1+\cos(\pi/5))$. The maximal quantum value is achieved by measuring a pure state $\rho = \ketbra{v_0}{v_0}$ and measurements $\M_{0|i} = \ketbra{v_i}{v_i}$ where
\begin{align*}
    \ket{v_0} &= (1,0,0)^T,\\
    \ket{v_i} &=  (\cos\theta,\sin\theta\sin\phi_i,\sin\theta\cos\phi_i)^T,
\end{align*}
where $\cos\theta = \cos(\pi/5)/(1+\cos(\pi/5))$ and $\phi_i = i\pi(4/5)$. 

Note that since this scenario consists of making measurements on a single quantum system, the statistics can always be simulated classically and it is thus impossible to achieve self-testing in a fully device-independent manner. In \cite{Bharti} the authors overcome this by assuming compatibility relations such as \eqref{context_compat}, thus rendering the scheme semi-device independent. Using this, they give a general scheme to self-test quantum states using noncontextuality inequalities and prove that the maximal quantum violation of the KCBS inequality robustly self-tests the above strategy. In other words, for any state $\physstatedm$ and measurements $\{\ketbra{v_i}{v_i}'\}$ which can be used to 
achieve the violation $c_q - \epsilon$ of KCBS inequality there exist a unitary $U$ such that
\begin{equation*}
    \|U\ket{v_i'} - \ket{v_i}\| \leq O(\sqrt{\epsilon}), \qquad \forall i.
\end{equation*}
The proof holds also for the generalisations of KCBS inequality given in \cite{gen1,gen2,gen3}. The crux of the proof is an equivalence between the optimal strategy for violating the KCBS inequality and the solution to a certain type of SDP optimisation known as the Lov\'{a}sz theta number of an odd cycle graph \cite{lov}.  

\section{Applications of self-testing}\label{sec:applications}
The birth of self-testing is usually associated to the Mayers-Yao paper \cite{Mayers2004} from 2004. It set the terminology and formalism, including the first usage of the term self-testing in this context and identifying local isometries as relevant transformations. A similar main result as in \cite{Mayers2004} was presented in \cite{Mayers98}, although in the context of whether untrusted sources can be pertinent for cryptographic tasks. This earlier paper used the term `self-checking' instead of `self-testing'. Moreover, \cite{Mayers98} is at the same time one of the pioneering works  in device-independent cryptography, indicating importance of self-testing for the development of device-independent protocols. Since then self-testing has been scrutinised as a task of twofold significance:
\begin{itemize}
    \item \emph{purely theoretical}, related to exploring the conditions for a probability distribution to determine a specific quantum state and/or measurements, and proving that such statements also hold approximately. This theoretical aspect was reviewed in Sections \ref{sec:example},\ref{sec:bipartite}, \ref{sec:multipartite}, \ref{sec:robust}, \ref{sec:measurements} and \ref{sec:semi}. As a result, ideas and results from self-testing can lead to progress in related theoretical areas. 
    \item \emph{practical}, relevant for creating new device-independent or semi-device-independent protocols for different tasks. In the Mayers-Yao paper \cite{Mayers2004} the authors say `\textit{We hope that it will have application in different areas of quantum information processing}'. Fifteen years later we can observe that this hope is fulfilled. 
\end{itemize}

In this section we give an overview of the applications of self-testing during the first fourteen years after the technique has been formally introduced. On the practical side, we cover the relation of self-testing with device-independent randomness generation in section \ref{dirg}, device-independent quantum cryptography in section \ref{diqkd}, and device-independent entanglement certification in section \ref{ed}. In section \ref{dqc} we describe the applications to delegated quantum computing \ref{dqc}. Finally, from the theoretical side, in section \ref{tp} we describe the influence self-testing has had in understanding the structure of the set of quantum correlations. In particular in section \ref{infinite} we highlight a link to the study of quantum correlations produced by finite vs infinite quantum systems, and in section \ref{uncertain} how self-testing has been used to prove an inequivalence between uncertainty relations and Bell nonlocality.

\subsection{Device-independent randomness generation}\label{dirg}
The probabilistic nature of quantum mechanics can be exploited for generation of random numbers. In the simplest example, measuring in the computational basis a qubit in the state $\ket{+}$  results in a perfectly random output bit. However, the certification of the random nature of bits obtained this way relies on the exact characterisation of both the quantum state and the measurement performed. The device-independent scenario offers much less stringent requirements for randomness certification, by qualitatively relating randomness with nonlocality. By treating her devices as non-communicating black boxes Alice can certify some amount of randomness by observing the violation of a Bell inequality. The first results in this direction show that the maximal violation of the Mermin  \cite{colbeck,colbeckkent} and the CHSH inequality \cite{pironio2010random} can be used in this way. For more information on certification of quantum randomness see \cite{randomnessreview} and references therein.

Here, we comment on the relation between self-testing and randomness. A self-testing protocol proves the existence of a pure entangled state and a certain set of measurements acting on it. Once this conclusion is made, certified random bits come for free since local measurements on a pure entangled state necessarily produce random outcomes. As a result, ideas from self-testing are often either implicitly present in device-independent randomness works or are explicitly used as tools to prove randomness lower bounds. 

In the pioneering works of \cite{colbeck,colbeckkent} a self-testing statement is implicitly present, where the authors prove that only an orthogonal sum of GHZ states can maximally violate the Mermin inequality. In \cite{Coudron:2014:IRE:2591796.2591873}, the sequential self-testing of $n$ EPR-pairs proven in \cite{ruv} is used as a sub-protocol for infinite randomness expansion with a constant number of devices. 

Simple symmetry-based arguments are used in \cite{dhara} to prove that the violation of some Bell inequalities can be used to certify the presence of genuine randomness. A necessary condition is that there exists a unique probability distribution maximally violating the Bell inequality. One way to prove such uniqueness is through self-testing: if the maximal violation of the Bell inequality is a self-test the maximally violating probability distribution has to be unique.  Furthermore, incomplete results from \cite{dhara}, were proven to be true by using self-testing techniques in \cite{SASA}. 

The results on self-testing properties of binary XOR games from \cite{Miller2013} were expanded in \cite{Miller:2016:RPS:2997039.2885493} and used to devise protocols for exponential randomness expansion. More recently, the authors of \cite{sublinear} directly use robust self-testing bounds for the tilted-CHSH inequality \cite{Bamps} to lower bound the randomness generated in their protocol. Self-testing techniques are also used in \cite{toni,andersson,erik} to prove that two bits of local randomness can be certified from a two-qubit entangled state.

\subsection{Device-independent quantum cryptography}\label{diqkd}

\subsubsection{Quantum key distribution}

Quantum key distribution (QKD) is the most widely studied quantum cryptographic protocol in which two parties, Alice and Bob, use quantum resources to generate a shared private key which can later be used for encryption and decryption of messages. The security of a standard QKD protocol relies on the correct characterisation of all devices, which can be difficult to achieve in practice and far from ideal from a security perspective. An alternative approach comes from device-independent quantum key distribution (DIQKD), where security is based only on the observation of the correlations, and can be proven even if the constituent devices are treated as black boxes. DIQKD is intimately related to DI randomness generation; whereas in randomness generation one aims to have random outcomes, in a DIQKD protocol one aims to have random outcomes that are also correlated between Alice and Bob (thus ensuring a shared private key). As with randomness generation, the security of DIQKD is often measured against the violation of some Bell inequality. For a concise review on the topic see \cite{QKDreview}.

An indication of a close relation between DIQKD and self-testing is their common root in the Mayers-Yao work \cite{Mayers98}. It discusses self-testing as a protocol for the first time (under the name self-checking) and recognises that it can help to use untrusted devices in cryptographic setting. \cite{Mayers98} consider the BB84 protocol \cite{bb84} in which Alice certifies an untrusted source she wants to use. The source is supposed to emit EPR pairs with Alice keeping one particle and measuring it and sending the other one to Bob. The untrusted source can be self-tested using the Mayers-Yao self-testing criterion, as explained in section \ref{sec:bipartite}. The protocol is later discussed in \cite{Mayers2002} in the context of the Ekert QKD protocol \cite{ekert}. The second Mayers-Yao paper \cite{Mayers2004}, improving the first one by characterising the measurements (and introducing the phrase `self-testing') also discussed the relation of self-testing with the BB84 protocol.

Ever since then self-testing and DIQKD have been  intertwined. A certification of some quantum resource is implicitly present in every DIQKD security proof, however in some works the relation between self-testing, as the strongest form of certification, and DIQKD was explicitly examined. The effect of the inability to self-test complex measurements on the security of cryptographic tasks has been the subject of \cite{McKague2011}. The authors prove that the  6-state QKD protocol \cite{sixstate1,sixstate2} can be secure even if the devices are untrusted, despite the issue with complex conjugation. Similarly like in the protocols for randomness expansion, the self-testing properties of XOR binary nonlocal games explored in \cite{Miller2013} were used in \cite{Miller:2016:RPS:2997039.2885493} to prove the security of certain class of QKD protocols. Finally, the concept of parallel DIQKD developed analogously to parallel self-testing was first introduced in \cite{parallelQKD} and the security proof relied on the rigidity of the magic square game \cite{Wu2016} allowing for parallel self-testing of two singlets. A simplified proof appeared in \cite{parallelQKD2}.

\subsubsection{Cryptography beyond quantum key distribution}
%QKD is by far not the only quantum cryptographic protocol whose DI version can be related to some self-testing statement. Here we mention a few primitives in which different parties do not trust each other, as opposite to QKD where they collaborate. 

\emph{Bit commitment}--- Bit commitment is a cryptographic primitive in which Alice chooses a bit $b$ that she wants to first commit, then later reveal, to Bob. The protocol should be both binding (Alice should not be able to change her choice of $b$ after the commit step) and hiding (Bob should not be able to know $b$ until Alice chooses to reveal). In a classical protocol, either Alice or Bob can cheat with probability $1$ without being caught, \emph{i.e.} either Alice can alter the bit after committing or Bob can learn it before it is revealed. Although unconditionally secure bit commitment is known to be impossible even using quantum resources \cite{impossibleQBC,lo1997}, there exist quantum protocols in which either party's probability to successfully cheat is strictly smaller than 1 \cite{QBC}. \cite{Silman2011} introduces the idea of a DI quantum bit commitment protocol in which besides not trusting each other, Alice and Bob do not trust their equipment either. The security of the protocol is based on the self-testing fact that the maximal violation of the Mermin inequality can only be achieved by measuring the GHZ state. A version of DI quantum bit  commitment based on the violation of the CHSH inequality is presented in \cite{Aharon_2016}, in which the security can be seen as consequence of the self-testing properties of the CHSH inequality.  Similarly, the violation of the CHSH inequality has been used to prove the security of DI relativistic bit commitment in  \cite{Adlam2015}, ruling out the possibility of location attacks in which devices are able to track their own space and time coordinates. None of the works explicitly relate their results to the corresponding self-testing protocol however.

\emph{Weak string erasure}--- Weak string erasure (WSE) \cite{WSE} is a primitive which can be used in two-party cryptographic protocols in which no large scale reliable quantum storage is available to the cheating party. WSE  provides a random bit string $(b_1,\cdots, b_n)$ to Alice, while sending a randomly chosen substring $(b_{i_1},\cdots,b_{i_k})$ to Bob, together with the set $(i_1,\cdots,i_k)$ specifying the location of substring bits. WSE is secure against Bob if he cannot learn much about the full string given to Alice, while it is secure against Alice if she cannot learn the location of Bob's bits. A DI version of this protocol useful for bit commitment or oblivious transfer is introduced in \cite{Kaniewski_2016} and the security is related to the self-testing properties of the CHSH inequality.

\emph{Position verification}--- Finally, we briefly mention the position verification primitive, useful in position-based cryptography in which the parties have to convince the (honest) verifiers that they are located at a particular location. Protocols for position verification that improve the security by using quantum communication have been proposed in  \cite{PV1,PV2}. The DI security of position verification against adversaries with no quantum memory is proven in \cite{Ribeiro}, and can also be traced to the self-testing properties of the CHSH inequality.

\subsection{Entanglement detection}\label{ed}
One of the most basic tasks in quantum information is that of detecting entanglement of a bipartite quantum system via local measurements on its subsystems. Device-independent entanglement detection considers this problem in the device-independent scenario, i.e.\ where all local measurement devices are treated as black boxes. Since the observation of Bell nonlocal correlations necessarily implies that the underlying state is entangled, the standard approach to DI entanglement detection involves violating a Bell inequality. However, since there exist entangled mixed states that do not violate any Bell inequality\footnote{At least in the original Bell scenario in which Alice and Bob can perform any number of non-sequential local measurements on a single copy of the state. In more complex measurement scenarios (see \cite{Cavalcanti_networks,Sen_networks,Palazuelos2012,Rabello_networks,bilocality,GHNL,HiddenNL}) it is generally unknown if such states exists.} \cite{Werner89,Barrett02,lhv_review,Bowles2016,Jevtic_steering,Bowles_GME,Hirsch2016}, this method cannot be used for all entangled states. A partial solution to this problem, allowing for the entanglement detection of all entangled states, was given in \cite{Buscemi2012} (see also \cite{MDIEW}) using the concept of a `semi-quantum game'. Here, the classical inputs in a Bell test are replaced by `quantum inputs' $\ket{\psi_x},\ket{\psi_y}$, that is, a set of known quantum states that are sent to the measurement device instead of the classical labels $x$ and $y$. This scenario is semi-device-independent since although the measurement devices are treated as black boxes, the quantum input states must be trusted. %H

In \cite{Bowles2018a,BSCA}, tools from self-testing and semi-quantum games were used to construct fully DI protocols for the entanglement detection of all entangled mixed states. The idea is as follows. If one achieves a self-test of a particular state and local measurements for Alice, then this certifies (up to a local isometry) the reduced states of Bob conditioned on a particular choice of input/output for Alice. In this way one can certify an ensemble of state preparations (conditioned on Alice's input/output) on Bob's local Hilbert space. These preparations can then be used as quantum inputs in a semi-quantum game. Since (i) the quantum inputs are now certified device-independently, (ii) the semi-quantum games scenario can be applied to all entangled states, the two can be combined to construct a fully device-independent protocol that works for all entangled states. Specifically, one needs to consider a network scenario in which the state of interest is augmented with two auxiliary bipartite states that are used to prepare the quantum inputs. Here, tools from parallel self-testing as well as the issue of complex conjugation become important for the general proof.

\subsection{Delegated quantum computing}\label{dqc}
Delegated computation is a protocol in which a party, usually called a verifier, delegates a computational task to another party, usually called a prover. The verifier aims to solve difficult computational tasks, but does not have enough computational resources. The prover, on the other side, has a very powerful computer and is able to solve any task the verifier is interested in. When one talks about delegated quantum computation (DQC) the prover possesses a quantum computer, while the verifier has either only classical computing resources or limited quantum resources but wants to solve a problem intractable for classical computing devices. For a concise review on the existing approaches in DQC see \cite{GheorghiuReview}.

 There are two desirable properties of a DQC protocol: verifiability and blindness. The protocol is said to be verifiable if the verifier can be convinced that the solution provided by the prover(s) is correct. This is non-trivial, since the verifier is unable to solve the problem. Blindness of the protocol is related to the secrecy of the computation. It is ensured when the prover(s) cannot learn anything about the computational task the verifier wants to perform. It is very difficult to construct a DQC protocol with a fully classical verifier and a single prover which is verifiable and blind. The first protocol that achieves this, under computational assumptions, is  \cite{Mahadev}. In principle, it is easier to achieve both verifiable and blind protocol with a classical verifier when there is more than one prover. In this case the provers are entangled and forbidden to communicate. Verifiability is proven if the verifier can be convinced that the two or more non-communicating provers are performing the prescribed sequence of measurements. This, of course, requires that the verifier be able to test that the provers perform measurements from a set that is universal for quantum computing. The latter is exactly a self-testing task. Thus it is no surprise, that self-testing is useful for such a delegation protocol, yet orchestrating such a computation in a verifiable fashion is a delicate task. Delegation protocols with two or more provers based on self-testing typically achieve information theoretic security.

 The first such protocol was presented by Reichardt, Unger and Vazirani (RUV) in \cite{ruv}. The protocol involves two provers sharing a tensor product of many EPR pairs and the computation model is quantum computation by teleportation \cite{GC}. The provers are able to convince the verifer that they posses $N$ EPR pairs by obtaining the optimal score in sequential playing of the CHSH game. The complexity of RUV protocol in terms of time and the number of EPR pairs needed is extremely large. The protocol was subsequently improved in \cite{Hajdusek} where each Bell pair is shared by two provers, making the number of provers increase significantly at the expense of reducing the overall complexity. 
 The advantage of parallel self-testing of $N$ EPR pairs instead of sequential was exploited  in \cite{Natarajan} and \cite{Coladangelo2017b}. The latter obtains a protocol with an almost optimal overhead (in the size of the computation) in terms or resources used, by exploiting parallel self-tests with robustness independent of the number of EPR pairs tested, discussed previously in subsection \ref{parallel}. The work \cite{Gheorghiu} shows that RUV protocol can be significantly improved if the verifier is actually quantum and wants to be convinced that they share with the prover $n$ EPR pairs. In this case the self-testing of $n$ EPR pairs through steering becomes a relevant sub-protocol.
 
Another example of self-testing incorporated into a delegated quantum computing protocol is \cite{McKagueDQC}. The DQC protocol involves many provers which share a graph state. The model of computation is measurement-based quantum computing (MBQC) \cite{mbqc1,mbqc2}. The protocol is made verifiable by using the self-testing of graph-states mostly based on \cite{McKague2014}. Based on a similar idea a significantly simplified protocol appeared in \cite{Hayashi}: since the triangular lattice graph state is universal for MBQC \cite{mhalla} the number of provers can be reduced to three and the number of necessary copies of the graph states is also considerably smaller than in \cite{McKagueDQC}. 

\subsection{Structure of the set of quantum correlations}\label{tp}
Here we highlight two ways in which self-testing as furthered our understanding of the set of quantum correlations. 

\subsubsection{Correlations from finite vs infinite dimensional quantum strategies}\label{infinite}
It is said that the correlations $\{p(a,b|x,y)\}_{a,b,x,y}$ admit a quantum strategy if there exists a state $\ket{\psi}_{\rA\rB} \in \mathcal{H}_\rA \otimes \mathcal{H}_\rB$ and projective measurements $\{\M_{a|x}\}_{a,x}, \{\N_{b|y}\}_{b,y}$ such that
\begin{equation*}
    p(a,b|x,y) = \bra{\psi}\M_{a|x}\otimes \N_{b|y}\ket{\psi}.
\end{equation*}
This strategy is also called a \emph{tensor-product strategy} due to the tensor product between Alice's and Bob's spaces. Different sets of quantum correlations arise when certain conditions are imposed on the state $\ket{\psi}$ and/or measurements $\{\M_{a|x}\}_{a,x}, \{\N_{b|y}\}_{b,y}$. The correlations obtained through a tensor-product strategy on finite-dimensional Hilbert spaces $\mathcal{H}_\rA$ and $\mathcal{H}_\rB$ constitute the set denoted as $\mathcal{C}_q$. If the Hilbert spaces $\mathcal{H}_\rA$ and $\mathcal{H}_\rB$ can also be infinite dimensional the set is denoted as $\mathcal{C}_{qs}$. The closure of the set $\mathcal{C}_q$ is called $\mathcal{C}_{qa}$. Additionally, a \emph{commuting-operator strategy} is one in which no tensor product structure is imposed but instead all measurement operators of Alice commute with all those of Bob, i.e.\,
\begin{equation*}
    p(a,b|x,y) = \bra{\psi}\M_{a|x} \N_{b|y}\ket{\psi}.
\end{equation*}
with $[\M_{a\vert x},\N_{b\vert y}]=0$. The commuting-operator model is used in algebraic quantum field theory and all such correlations are denoted by the set $\mathcal{C}_{qc}$. The set inclusion relation defines a hierarchy among these sets \cite{todorov}:
\begin{equation}\label{inclusionrelations}
    \mathcal{C}_q \subseteq \mathcal{C}_{qs} \subseteq \mathcal{C}_{qa} \subseteq \mathcal{C}_{qc}.
\end{equation}
Whether $\mathcal{C}_{qc}$ is equivalent to either $\mathcal{C}_{qs}$ or $\mathcal{C}_{qa}$ are problems known as Tsirelson's problems \cite{Tsirelson1993}. Recently, it has been proven that $\mathcal{C}_{qs} \neq  \mathcal{C}_{qc}$ \cite{slofstra16}, and also that $\mathcal{C}_{qs} \neq \mathcal{C}_{qa}$ in \cite{slofstra17}. Whether $\mathcal{C}_{qa}\subset \mathcal{C}_{qc}$ remains an open problem.

Self-testing techniques have also inspired proving a strict inclusion of $\mathcal{C}_{q}$ in $\mathcal{C}_{qs}$. The inequivalence between these two sets was proven in \cite{Coladangelo2017d} in cases when either the number of inputs or the number of outputs is infinite. The separation between  $\mathcal{C}_{q}$ and $\mathcal{C}_{qs}$ %unconditional on the size of input and output alphabets 
for finite input or output alphabets was proven by the same authors in \cite{Coladangelo2018a}. The separation is demonstrated by explicitly giving the correlations which can be obtained with infinite-dimensional quantum strategies, but not with any finite-dimensional ones. The proof is inspired by the protocol for self-testing infinite-dimensional bipartite pure states, described in section \ref{subspace}. In the finite case, the self-test of a bipartite state can be interpreted as involving one of two different protocols depending on the parity of the dimension. An infinite dimensional state does not have a defined parity and can thus be self-tested by either protocol. The authors use this fact as the theoretical basis to prove the separation. 

Following Slofstra's proof of non-closure of the quantum set of correlations \cite{slofstra17} alternative proofs appeared in \cite{Dykema2019} and \cite{Musat}. All these proofs rely on the representation theory of $C^*$-algebras. A relatively simpler proof, using embezzling entanglement \cite{embezzlement} and self-testing, is presented in \cite{Andrea19}.

\subsubsection{Uncertainty relations and Bell nonlocality}\label{uncertain}
Self-testing has also been used as a tool to understand the relationship between uncertainty relations, steering and Bell nonlocality. In \cite{Oppenheim1072}  it was shown that in the optimal quantum strategy for a large class of Bell inequalities called XOR games, the steered states of Bob after a measurement by Alice are always such that they saturate a fine-grained uncertainty relation defined by the Bell inequality itself. This was then extended to more Bell inequalties in \cite{Zhen2016}. Whether such a link was true for generic Bell inequalities remained an open question. 

In \cite{Ravi2018} (building on \cite{Zhen2016}) the authors answer this in the negative by providing specific examples of Bell inequalities for which the relationship does not hold. To achieve this,  self-testing statements for the Bell inequalities are proven, from which the form of the steered states can be determined and checked against the corresponding uncertainty relations. The self-testing statements are proven using Jordan's lemma (see section \ref{JL}) to reduce the  problem to essentially a two-qubit strategy, which simplifies the analysis significantly.

\section{Experiments}\label{sec:experiments}
The bulk of self-testing procedures are still only theoretical recipes. This is understandable since the majority of robust self-testing protocols have fidelity bounds that decrease rapidly with noise and serve only as a proof of principle.  However, in the last years there has been an increasing number of self-testing protocols robust to realistic amounts of noise (see section \ref{sec:robust}). Here we mention the few experimental realisations of such protocols.   

The biggest experimental hurdle towards fully device-independent protocols is simultaneously closing detection and locality loopholes. \cite{BaselMunich} reports  the self-testing of a  Bell state distributed over 398 meters through the violation of the CHSH inequality free of both detection and locality loopholes and furthermore free of the i.i.d. assumption. In this light it is the first fully device-independent self-testing protocol to be implemented in practice.  Entanglement between the distant atoms is generated by entangling the spin of each atom with polarisation of a single photon. The obtained fidelity is $55.54\%$ at a confidence level of $99\%$.  By applying the same theoretical tools to analyse the data obtained in the loophole-free Bell test presented in \cite{hensenSR} no fidelity higher than the trivial $50\%$ could be found.

The remainder of the experimental self-testing contributions are not based on a completely loophole-free Bell tests. 
\cite{Tan} reports a high violation of the CHSH inequality by a pair of $^9$Be$^+$ ions. The violation is used to make a self-testing statement, based on \cite{Jed2}. At the $95\%$ confidence level the pair of $^9$Be$^+$ ions has 0.958 fidelity with the maximally entangled pair of qubits. An overview of the inferred self-testing bounds from some previous works reporting CHSH violations is also presented. In \cite{expSttrip} the operator inequalities for robust self-testing from \cite{Jed2} are tested for a large number of bipartite and tripartite qubit states encoded in photon polarisation degrees of freedom.

Experimental robust self-testing of partially entangled pairs of qubits using the Swap method is presented in \cite{expSTbip}. The systems under consideration are polarisation entangled photons. Self-testing of partially entangled qubit pairs is also used to heuristically estimate the fidelity of a produced ququart state with a given reference state. Robust self-testing of partially entangled pairs of qubits encoded in photon polarisation degrees of freedom was reported in \cite{chile}. The self-testing was done through the violation of the tilted CHSH inequality and the robustness bounds were estimated by using the numerical results from \cite{tim}. In \cite{SingExp} the authors explore the certification of partially entangled pairs of photons encoded in the polarisation degree of freedom. The fidelity of the entangled pair with the corresponding partially entangled pair of qubits is estimated in two ways:  by using standard tomographic methods and self-testing. The obtained fidelities have ratio $\approx0.998$, implying that for the case of qubit states self-testing can be used to achieve almost the same conclusions as tomography. It is argued that self-testing may have an advantage over tomography even when the detection and locality loopholes are not closed since it avoids characterisation of measurements and assumptions about dimension and, in principle, requires estimating fewer average values. 

In \cite{bristol} various two-qutrit entangled states are self-tested using photons entangled in the mode degree of freedom of the waveguides in a silicon based integrated optical chip. Fidelity bounds were obtained numerically via the Swap method. The self-tested states are the maximally entangled pair of qutrits (estimated fidelity $0.799$), the state maximally violating the CGLMP inequality \cite{cglmp} (estimated fidelity $0.68$) and a state maximally violating an extension of one of the SATWAP inequalities \cite{abeille} (estimated fidelity $0.832$).

Experimental self-testing in the steering scenario is the subject of \cite{Li:19steering}. The fidelity of the underlying physical state with the GHZ state is estimated based on the violation of the Mermin's steering inequality \cite{MerminSteer}. The fidelity lower bound is estimated to be $0,7866$, while the tomographically retrieved fidelity is $0.8725\pm 0.0034$.

Finally, in \cite{BellMeasurExp} the authors report an experimental realisation of a robust self-test of a Bell state measurement based on the entanglement swapping protocols of \cite{Sekatski,MO} (see section \ref{STQC}). Photon pairs that are hyper-entangled in the spatial and polarisation degrees of freedom are used to encode the two maximally entangled pairs of qubits that are needed for the entanglement swapping protocol. 

\section{Concluding remarks and open questions}\label{sec:openquestions}
Recent years have seen an increased interest in device-independent self-testing, accompanied by the plethora of self-testing protocols and methods presented in this review. However, there are still many important unresolved questions. Without aiming to exhaust the list, we name some open questions and research directions which we believe worthy of attention. \\

\emph{Analytic methods for dimension larger than 2}---The majority of known self-testing protocols are either built to self-test multi-qubit states and measurements, or apply existing qubit protocols to the self-testing of higher dimensional systems. The self-testing of states and measurements using methods that exploit the genuine $d$ dimensional nature of quantum systems, is however still a very unexplored area. For instance, it has been known for a long time that via the numerical Swap method, the CGLMP inequality self-tests the two-qutrit state of equation \eqref{cglmpstate}. However, a corresponding analytic proof of this statement is still lacking, despite the relative simplicity of the inequality. Similarly desirable are analytic proofs for the self-testing of the maximally entangled states in dimension $d$ using the SATWAP inequalities \cite{abeille}. With respect to high dimensional measurement self-testing, one important open question is to extend the analytic self-test of a set of qutrit mutually unbiased basis measurements and the maximally entangled state of \cite{maxmaxmax} to higher dimensions. Moving beyond high dimensional systems to continuous variable systems, essentially nothing is known and there exist no protocols to self-test such states.  

\emph{Multipartite methods}---In a similar vein, techniques for self-testing general multipartite states are also needed, since current methods are only known for restricted classes such as graph states. One potential line of research in this direction would be to develop methods to self-test multi-qubit hypergraph states \cite{hypergraph}, which exhibit a richer structure than graph states but still admit a useful description in terms of Clifford group stabiliser operators. One would thus need a general method to construct Bell inequalities for such states, as was done for graph states \cite{graphBI,Flavio}. This appears more complicated for hypergraph states however since the nonlocal nature of the hypergraph stabilizer operators means they do not have an obvious interpretation as local measurement observables (although some progress has been made \cite{hypergraph_BI}). 

\emph{Identifying the set of undetectable transformations}---Part of the challenge in going beyond two qubit methods is to identify the set of local transformations defining the equivalence classes of self-testable states and measurements in higher dimensions. As we have seen, the standard definitions presented in section \ref{defs} need to be adapted in order to self-test complex valued measurements, stemming from the invariance of quantum correlations under complex conjugation of the state and measurement operators. In higher dimensions, it is still unknown whether there exist more state and measurement transformations that leave correlations invariant. If such transformations exist, an all-encompassing definition of what it means to self-test a state and measurements in general dimension is therefore still missing.

\emph{Self-testing of a state or measurements only}---In many self-testing works, a self-testing statement for the state is accompanied by an additional self-testing statement for the measurements. One interesting problem would be to find situations where the correlations allow one to identify the state, but not the measurements, even allowing for the freedom of complex conjugation of the measurements in the definition of measurement self-testing. Similarly, it would be interesting to find correlations that allow one to self-test the measurements, but not the state.

\emph{Self-testing in non-i.i.d. scenarios}---Throughout this review we have made the assumption that each round of the experiment is independent and identical to all others. It would be interesting and practically relevant to attempt to remove this assumption from self-testing protocols. One way to achieve this would be to build sequential self-testing protocols akin one in \cite{ruv} or by leveraging some recently introduced techniques \cite{BaselMunich,entaccumulation,rotem_nonIID,rotem_thesis} for such scenarios. Methods could be borrowed also from the protocols for delegated quantum computation, \textit{i.e.} \cite{Coladangelo2017b}. Since the state and measurements can now depend on settings and outcomes in previous rounds, one would also need to adapt the definition of self-testing to apply to such scenarios, as well as derive corresponding confidence bounds from finite statistics.  For example, the aim would be to show that, to high statistical confidence, the source is producing something close (by some measure) to $n$ independent copies of $\refstate$, where $n$ is the number of experimental rounds. 

\emph{Improved robustness methods}---Finally, methods to improve the robustness bounds of general self-testing protocols are much in need. In practice, the applicability of the majority of self-testing protocols is hindered by very poor tolerance to noise. Significant improvements have been achieved for simple scenarios \cite{Jed2}, however it is not clear if these methods can be extended to scenarios with more inputs and outputs due to their dependence on Jordan's lemma. Finding a good robustness bound involves a difficult maximisation over all local isometries, and as a result nearly all methods use one of the few Swap isometries that are known to give good results in the well-explored simple cases. Thus, knowing more useful isometries and understanding which work well for particular classes of states would likely lead to improved robustness bounds. 

\section*{Acknowledgements}
We are grateful to Alexia Salavrakos, Jean-Daniel Bancal, Andrea Coladangelo, Jed Kaniewski, Armin Tavakoli, Erik Woodhead, Alejandro Pozas-Kerstjens, Aleksandra Dimi\'{c}, Marc-Olivier Renou, Felix H\"{u}ber, Sébastien Designolle, Nicolas Brunner, Yeong Cherng Liang,  Nikolai Miklin and Rotem Arnon-Friedman for suggestions while preparing the manuscript. We acknowledge funding from the ERC CoG QITBOX, the Spanish MINECO (QIBEQI FIS2016-80773-P, Severo Ochoa SEV-2015-0522), Fundacio Cellex, Generalitat de Catalunya (SGR 1381 and CERCA Programme). I{\v{S}} acknowledges funding from SNSF (Starting grant DIAQ). JB acknowleges funding from the Juan de la Cierva-formación grant the AXA chair in quantum information science.  
\appendix
\section{Appendix}

\subsection{Self-testing complex measurements}\label{App:complexdef}

Here we give a possible definition of self-testing of states and complex valued measurements. 
\begin{definition}\label{def:complexST}(\emph{self-testing of states and complex measurements)}\\
We say that the correlations $p(a,b \vert x,y)$ \emph{self-test the state and measurements} $\refstate_{\rA'\rB'},\{\M'_{a|x}\},\{\N'_{b|y}\}$ if for all states and measurements $\physstatedm_{\rA\rB},\{\M_{a\vert x}\}, \{\N_{b\vert y}\}$ compatible with $p(a,b \vert x,y)$ there exists a local isometry $\iso=\iso_\rA\tp\iso_\rB$
such that for any purification $\physstate_{\rA\rB\rP}$ of $\physstatedm_{\rA\rB}$ there exists some state $\junk_{\bar{\rA}\bar{\rB}\rP}$ such that 
\begin{align}\nonumber
\begin{multlined}
    \iso\otimes\openone_\rP \left[\M_{a\vert x}\tp \N_{b\vert y}\tp\openone_\rP\physstate_{\rA\rB\rP}\right]  \\
    =\tilde{\M}_{a\vert x}\otimes \tilde{\N}_{b\vert y}\tp\openone_\rP\left(\refstate_{\rA'\rB'}\otimes\junk_{\bar{\rA}\bar{\rB}\rP}\right), \nonumber
\end{multlined}
\end{align}
for all $a,x,b,y$, and where 
\begin{align}
&\tilde{\M}_{a\vert x}=\M'_{a\vert x}\otimes \Ss_0^{\bar{\rA}} + (\M'_{a\vert x})^*\otimes \Ss_1^{\bar{\rA}} \nonumber \\
&\tilde{\N}_{b\vert y}=\N'_{b\vert y}\otimes \T_0^{\bar{\rB}} + (\N'_{b\vert y})^*\otimes \T_1^{\bar{\rB}} \nonumber\\[5pt]
&\Ss_0+\Ss_1\openone_{\bar{\rA}},\quad \T_0+\T_1 = \openone_{\bar{\rB}}, \nonumber\\
&\bra{\xi}(\Ss_0\tp \T_0 +\Ss_1 \tp \T_1)\tp\openone_\rP\ket{\xi} =1. \nonumber
\end{align}
\end{definition}
Here, the $\Ss_i$ and $\T_i$ part of the measurements are acting as effective controlled complex conjugations of the reference measurements. The final condition ensures that this conjugation is performed in a correlated fashion as implied from \eqref{conjmeas}. The probability that the conjugation is performed depends on the (unknown) junk state and is thus unknown. Note that if one traces out all but the $\bar{\rA}\bar{\rB}$ space, one obtains some convex combination of the reference and conjugated measurements acting on the reference state.  

\subsection{Regularisation trick}\label{regularisation}

In this appendix we give more details about the so-called regularisation trick. It refers to the case when one of the operators used to build the Swap gate (see figure \ref{fig:example}) is not unitary. This happens already in the case described in Chapter \ref{sec:example} where the operators $\Zz_{\rA}$ and $\Xx_{\rA}$ from equation \eqref{swap_ops} might have some zero eigenvalues. Let us focus on $\Zz_{\rA} = (\rA_0+\rA_1)/\sqrt{2}$. The first step in the regularisation procedure is to change all the zero eigenvalues of $\Zz_{\rA}$ to 1, resulting in a new operator $\Zz^*_{\rA}$. In the second step, all eigenvalues are normalized, \emph{i.e.} the new operator defined as $\hat{\Zz}_{\rA} = \Zz^*_{\rA}/|\Zz^*_{\rA}|$ is unitary by construction. However, one has to prove that $\hat{\Zz}_{\rA}$ acts on the physical state in the same way as ${\Zz}_{\rA}$. For that the following series of inequalities can be used (Note that  $\Zz^*_{\rA}$ acts on $\physstate$ in the same way as  $\Zz_{\rA}$  since it can be seen as $\Zz_{\rA} + P$ where $P$ is the projector on the kernel of $\Zz_{\rA}$):
\begin{align*}
    \|(\hat{\Zz}_{\rA} - {\Zz}_{\rA})\physstate\| 
    &= \|(\openone - \hat{\Zz}_{\rA}^\dagger{\Zz}_{\rA})\physstate\| \\
    &= \|(\openone - |{\Zz}_{\rA}|)\physstate\| \\
    &= \|(\openone - |{\Zz}_{\rA}{\Zz}_{\rB}|)\physstate\| \\
    &\leq \|(\openone - {\Zz}_{\rA}{\Zz}_{\rB})\physstate\| \\
    &= 0
\end{align*}
The first line is the consequence of the unitarity of $\hat{\Zz}_{\rA}$ and the second uses the definition of $\hat{\Zz}_{\rA}$. To get the third line we used the fact that ${\Zz}_{\rB}$ is unitary. The inequality follows from the operator inequality $A \leq |A|$. The last line stems from equation \eqref{a=b}. The key ingredient necessary for regularisation is exactly equation \eqref{a=b}. In general, the regularisation of any operator $A$ can be done if there is a unitary $U$ such that $A\tp\openone\physstate = \openone\tp U \physstate$.

\subsection{Swap isometries}\label{sec:differentswaps}
\begin{figure}
\centering
\includegraphics[width=1.0\columnwidth]{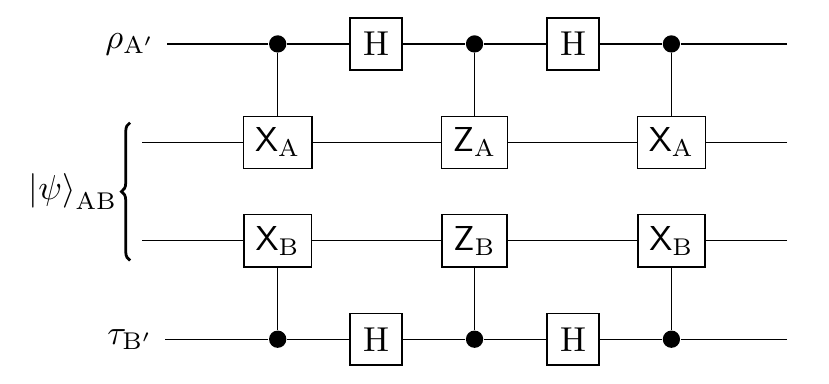}
\caption{The full Swap gate used in some robust self-testing protocols.   If the ancillas are initiated in the state $\ket{0}$ the gate reduces to the partial Swap gate, given on figure \ref{fig:example}. \label{fig:full} }
\end{figure}

In this appendix we provide further comments on the different Swap isometries used in the self-testing protocols. In section \ref{swap} we mentioned that the partial Swap gate given on figure \ref{fig:example} is appropriate only if the ancillas are initiated in the state $\ket{0}$. In the case that the ancillas are in a different state the correct isometry to use is the full Swap gate, given in figure \ref{fig:full}.

\begin{figure}
\centering
\includegraphics[width=1.0\columnwidth]{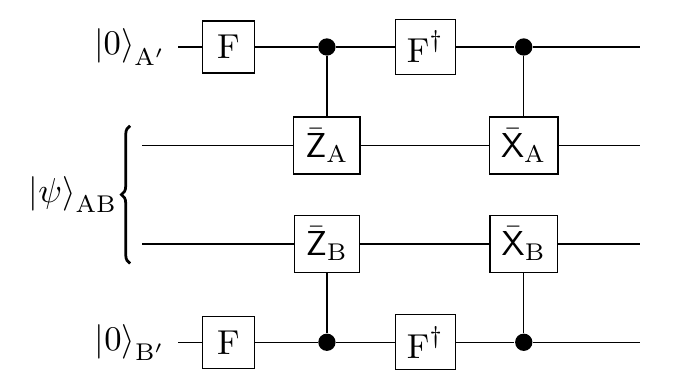}
\caption{The partial Swap gate used in some  protocols for robust self-testing of qudit entangled states. \label{fig:quditSWAP}}
\end{figure}

A generalisation of the Swap gate, given on figure \ref{fig:quditSWAP}, can be used for self-testing of bipartite qudit states $\ket{\psi} = \sum_{j = 0}^{d-1}\lambda_j\ket{jj}$, where $\lambda_j$ are positive real numbers. The gate F is the Fourier transform defined as:
\begin{equation*}
    F\ket{j} = \frac{1}{\sqrt{d}}\sum_{k=0}^{d-1}\omega^{jk}\ket{k},
\end{equation*}
where $d$ is the local dimension of the reference state, and $\omega$ is the $d$-th root of the unity. The controlled gates ${C\bar{\Zz}}$ and ${C\bar{\Xx}}$ are defined as follows\footnote{Note that $\{{\Xx}^{(j)}\}$ are $j$ different operators, while $\{\Zz^j\}$ are $j$-th powers of the operator $\Zz$.}:
\begin{align*}
    C{\bar{\Xx}}\ket{j}\ket{\psi} &= \ket{j}\bar{\Xx}^{(j)} \ket{\psi}\\
    C{\bar{\Zz}}\ket{j}\ket{\psi} &= \ket{j}\bar{\Zz}^{j} \ket{\psi}.
\end{align*}
For the gate on figure \ref{fig:quditSWAP} to work as an effective Swap gate the operators $\bar{\Xx}$ and $\bar{\Zz}$ have to satisfy certain conditions, mimicking anticommutativity from the qubit case. In \cite{Yang} the authors give the recipe: operators $\bar{\Zz}_{\rA}$ and $\bar{\Zz}_{\rB}$ have to satisfy
\begin{equation}\label{firstcondition}
 \sum_{j = 0}^{d-1} \omega^{ja}\bar{\Zz}_{\rA}^j\otimes\openone\physstate = \openone \otimes\sum_{j = 0}^{d-1} \omega^{ja}\Zz_{\rB}^j\physstate, 
\end{equation}
for all $a \in \{1,\cdots, d\}$. In addition, operators $\bar{\Xx}^{(k)}_{\rA}$ and $\bar{\Xx}^{(k)}_{\rB}$ must satisfy 
\begin{multline}\label{secondcondition}
   \lambda_0 \bar{\Xx}_{\rA}^{(k)}\otimes\sum_{j = 0}^{d-1} \omega^{jk}\bar{\Zz}_{\rB}^j\physstate \\ = \lambda_k\sum_{j = 0}^{d-1} \omega^{jd}\bar{\Zz}_{\rA}^j\otimes \left(\bar{\Xx}_{\rB}^{(k)}\right)^{\dagger}\physstate, \qquad \forall k.
\end{multline}
It can be proven that the output state of the isometry $\iso_d$ given on figure \ref{fig:quditSWAP} and built from the operators satisfying the conditions \eqref{firstcondition} and \eqref{secondcondition} has the form $\sum_{k}\lambda_k\ket{kk}_{\rA\rB}\tp\ket{\xi}_{\rA\rB}$ where $\ket{\xi}$ is some normalised state. For more details about the qudit Swap isometry see \cite{Yang,Coladangelo2017,Ivan}.

\subsection{Localising matrices in the Swap method}\label{sec:locmat}

In section \ref{sm} we presented the numerical Swap method used in robust self-testing protocols. The isometry used in the Swap method is the Swap gate and as we discussed in Appendix \ref{regularisation} when one of the operators $\Xx_\A$, $\Xx_\B$, $\Zz_\A$ or $\Zz_\B$ is defined as a sum or difference of physical measurement observables, the Swap isometry might not be unitary. Let us, for simplicity, focus the CHSH case and the operator $\Zz_\A = (\A_0+\A_1)/\sqrt{2}$ and $\Xx_\A = (\A_0-\A_1)/\sqrt{2}$ used to build the Swap isometry. In Appendix \ref{regularisation} we showed how to regularize such operators for the purposes of the ideal self-testing. 

The procedure to solve this problem when using the Swap isometry in robust self-testing protocols is introduced in \cite{PhysRevA.91.022115,PhysRevLett.113.040401}.
In the context of the Swap method one solves the problem by introducing two new operators $\rA_2$ and $\rA_3$ which are unitary and not too different from $(\A_0+\A_1)/\sqrt{2}$ and $(\A_0-\A_1)/\sqrt{2}$, respectively. That way the isometry built by defining  $\Zz_\A = \A_2$ and $\Xx_\A = \A_3$ is necessarily unitary. One way to impose the proximity of $\A_2$ and $\A_3$ to $(\A_0+\A_1)/\sqrt{2}$ and $(\A_0-\A_1)/\sqrt{2}$, respectively, is by imposing the relaxation
\begin{subequations}
\begin{gather}\label{locmatcond1}
    \A_2(\A_0+\A_1)/\sqrt{2}  \geq 0 \\ \label{locmatcond2}
    \A_3(\A_0-\A_1)/\sqrt{2}  \geq 0.
\end{gather}
\end{subequations}
This can be enforced by introducing two new moment matrices called the \emph{localizing matrices}. The condition \eqref{locmatcond1} can be imposed as a requirement that the moment matrix defined as
\begin{equation}
\Gamma^{k}_{i,j}(\A_2)=\bra{\psi}S^{(i)\dagger}\rA_2\frac{\A_0+\A_1}{\sqrt{2}}S^{(j)}\ket{\psi},
\end{equation}
where $S = \{\openone,\A_0,\A_1,\A_2\}$, is positive semi-definite. An analogous constraint can be made to enforce the condition \eqref{locmatcond2}.\\

\section{State and measurement assumptions}\label{app:assumptions}
Here we overview the assumptions made about the state and measurements in the definitions of self-testing. 
\subsection{State}
In the definitions of section \ref{defs} we use a purification $\ket{\psi}_{\rA\rB\rP}$ of the potentially mixed physical state $\rho_{\rA\rB}$. One does not assume that the state shared between Alice and Bob is pure however, since it is given by 
\begin{align}
    \rho_{\rA\rB}=\tr_\rP[\proj{\psi_{\rA\rB\rP}}].
\end{align}
Suppose we have a self-testing statement in the form of definition \ref{st_states_pure}:
\begin{align}\nonumber
    \iso_\rA\tp\iso_\rB\tp\openone_{\rP}[\physstate_{\rA\rB\rP}]=\refstate_{\rA'\rB'}\otimes\junk_{\bar{\rA}\bar{\rB}\rP}.
\end{align}
If we trace over the purification space on both sides of the above we obtain
\begin{align}\nonumber
    \iso_\rA\tp\iso_\rB[\rho_{\rA\rB}]=\refstateproj_{\rA'\rB'}\otimes\junkdm_{\bar{\rA}\bar{\rB}}.
\end{align}
where $\junkdm=\tr_{\rP}[\proj{\xi}]$. Thus a self-testing statement of the form of definition 1, implies that the same isometry maps $\physstatedm_{\rA\rB}$ to the reference state. This is possible because the isometry map acts trivially on the purification space. This follows from the fact that the isometry is constructed from the measurement operators, which by assumption themselves act non-trivially only on $\physstatedm$.

\subsection{Measurements}
We do, however, assume that the physical measurements are projective. This can lead to some confusion, since naturally one may want to repeat an analogous argument to the above that would allow one to treat the measurements as general POVM measurements, and simply make use of a Naimark dilation for mathematical convenience. We elaborate on this difficulty in doing this below.

Let us drop the projective assumption, so that the physical measurements are in general POVM measurements. Imagine we aim to self-test a set of measurements $\M'_{a\vert x}$ for Alice and the reference state $\ket{\psi'}$ in the sense of definition \ref{def_MST2}. This immediately poses a problem for the majority of self-testing proofs. For example in \eqref{sos}, one needs the physical measurements to be projective to guarantee $\A_x^2=\openone$ in order to prove anti-commutativity of the observables. In order to proceed to use the standard proofs of measurement self-testing, of which almost all assume the projective nature of the measurements, one has three options:
\begin{enumerate}
    \item Prove that even if treating the physical measurements as POVM measurements, the only physical measurements that are compatible with the observed correlations are projective measurements. 
    \item Prove a general theorem that states that if one has an isometry mapping any Naimark dilation of the physical measurements to the reference measurements, this implies a (possibly different) isometry mapping the POVM physical measurements to the reference measurements in the sense of definition \ref{def_MST2}.
    \item Argue that projective measurements are the only fundamental measurements in quantum theory. That is, POVM measurements can only be physically realised via a projective measurement on a dilated space.
\end{enumerate}

Option 1 has been done very rarely; perhaps the only example can be found in \cite{Jed2} when self-testing qubit Pauli measurements. Option 2 would be an analogue for the ability to use a purification of the state as explained above. However, such a theorem has not been proven to the best of our knowledge, and my may not be possible. We comment on option 3 at the end of this section. 

Let us focus further on option 2. Define the physical measurements for Alice as usual by $\M_{a\vert x}$, which may now be POVM. Furthermore, define a Naimark dilation of these measurements by $\tilde{\M}_{a\vert x}$. The vast majority of self-testing works prove an isometry mapping the dilated measurements $\tilde{\M}_{a\vert x}$ to the reference measurements:
\begin{align}\label{st_naimark}
    \iso\left[\tilde{\M}_{a\vert x} \tp \openone{}_{\rB} \physstate\tp\ket{\tilde{0}}\right] = \M'_{a\vert x}\tp\openone_{\rB'}\ket{\psi'}\tp\ket{\xi},
\end{align}
where we have explicitly written the ancilla state $\ket{\tilde{0}}\in\mathcal{H}_{\tilde{\rA}}$ used for the dilation, and the purification space of the state is left implicit as in section \ref{sec:example}. Now consider the physical (potentially POVM) measurements acting on the physical state. 
\begin{align}
\M_{a\vert x} \tp \openone_\rB \physstate.
\end{align}
At this point, it is tempting to define an isometry $\Omega$ that maps this measurement to it's Naimark dilation:
\begin{align}\label{projprob}
    \Omega[\M_{a\vert x} \tp \openone_\rB \physstate]=\tilde{\M}_{a\vert x} \tp \openone_{\rB'} \physstate\tp\ket{\tilde{0}},
\end{align}
One could then use the standard proof of self-testing by concatenating isometries. Note, however, that fixing an input $x$, the vectors on the right hand side of \eqref{projprob} are orthogonal for different $a$ since for $a\neq a'$
\begin{align}
    &\left(\physstatebra\tp\bra{\tilde{0}}\tilde{\M}_{a'\vert x}\tp\openone\right)\left(\tilde{\M}_{a\vert x}\tp \openone  \physstate\tp\ket{\tilde{0}}\right)\nonumber\\=&\physstatebra\tp\bra{\tilde{0}}\tilde{\M}_{a'\vert x}\tilde{\M}_{a\vert x}\tp\openone\physstate\tp\ket{\tilde{0}}=0. 
\end{align}
The corresponding inner product between the vectors inside the isometry in \eqref{projprob} is
\begin{align}
    \physstatebra\M_{a'\vert x}\M_{a\vert x}\tp\openone\physstate,
\end{align}
which is generally not equal to zero if the measurements are POVM. Thus, \eqref{projprob} is generally not valid, since isometry maps conserve the inner product between vectors. This means that one cannot simply consider a Naimark dilation `for free' since the map which takes the POVM measurements to the dilation is not an isometry.

Nevertheless, one may hope to prove that given a self-test of the dilated measurements \eqref{st_naimark}, one could recover the corresponding map for the physical measurements by somehow discarding the ancilla degrees of freedom used in the dilation, as was done for the purification space of the state. The difficulty here however is that---unlike the purification space--- the dilated space is explicitly used by the isometry, since it is constructed from the dilated measurements. 

Consider a general state $\rho$ on $\mathcal{H}_{\rA}$. For any Naimark dilation $\tilde{\M}_{a\vert x}$ of $\M_{a\vert x}$ one has 
\begin{align}
    \tr[\tilde{\M}_{a\vert x}\rho\tp\proj{\tilde{0}}]=\tr[\M_{a\vert x}\rho].
\end{align}
Tracing over the ancilla space only we have 
\begin{align}
    \tr[\openone\tp\bra{\tilde{0}}(\tilde{\M}_{a\vert x})\openone\tp\ket{\tilde{0}}\rho]=\tr[\M_{a\vert x}\rho]
\end{align}
and since this holds for all $\rho$ we must have 
\begin{align}\label{povmdef}
    \M_{a\vert x}=\openone\tp\bra{\tilde{0}}(\tilde{\M}_{a\vert x})\openone\tp\ket{\tilde{0}}.
\end{align}
Returning to \eqref{st_naimark}, we insert the identity $\openone_{\rA}\tp\sum_k\proj{\tilde{k}}_{\tilde{\rA}}$ inside the isometry, giving
\begin{align}
    &\iso\left[\big[(\openone_{\rA}\tp\sum_k\proj{\tilde{k}})\big]\tilde{\M}_{a\vert x}) \tp \openone_\rB \physstate\tp\ket{\tilde{0}}\right]\nonumber \\ &= \M'_{a\vert x}\tp\openone_{\rB'} \ket{\psi'}\tp\ket{\xi}. 
\end{align}
Taking the sum outside we find
\begin{align}
    &\sum_k\iso\left[\tilde{\M}^k_{a\vert x}\tp\openone_{\tilde{\rA}} \tp \openone_\rB \physstate\tp\ket{\tilde{k}}\right]\nonumber \\ &= \M'_{a\vert x}\tp\openone_{\rB'} \ket{\psi'}\tp\ket{\xi}. 
\end{align}
Where 
\begin{align}
    \tilde{\M}^k_{a\vert x}=\openone_{\rA}\tp\bra{\tilde{k}}(\tilde{\M}_{a\vert x})\openone_{\rA}\tp\ket{\tilde{0}}.
\end{align}
From \eqref{povmdef} we have $\tilde{\M}^0_{a\vert x}=\M_{a\vert x}$ and so
\small
\begin{align}
    &\iso\left[\M_{a\vert x} \tp\openone_{\tilde{\rA}} \tp \openone_\rB \physstate\tp\ket{\tilde{0}}\right]  \\ &= \M'_{a\vert x}\tp\openone \ket{\psi'}\tp\ket{\xi}-\sum_{k>0} \iso\left[\tilde{\M}^k_{a\vert x} \tp \openone_\rB \physstate\tp\ket{\tilde{k}}\right]. \nonumber
\end{align}
\normalsize
If the sum in the right hand side of the above is zero then we have proven that the isometry $\iso$ indeed maps the physical measurements to the reference measurements. However, it is not clear that there exists a Naimark extension such that this is always the case. Showing whether this is or is not possible would be a valuable contribution to the field. 

Finally we comment on option 3. This position can be justified on the basis that all other operations in quantum theory (CPTP maps, non-projective measurements) can be seen as the result of combining unitary evolution and projective measurement, and so there is no logical problem that arises with this stance. For some, this argument is not convincing however, since they consider POVMs to be as `real' as projective measurements, and thus deserve a place in the ontology of the theory.  

An alternative way to argue the projective-only assumption is from a information theoretic perspective. More precisely, the measurement update rule in quantum theory can be understood as a process by which an observer updates their description of a quantum state given new information. This new information is given by the outcome of a measurement, and is \emph{classical information} from the perspective of the observer (i.e. it is the classical information that the observer reads from the macroscopic degrees of freedom of the measurement device). Thus it is onto these degrees of freedom (which may have become entangled with the quantum system during the measurement procedure) that the measurement operators act. Since these degrees of freedom are perfectly distinguishable, the measurement operators are described by orthogonal projectors. If one performs another measurement, this amounts to applying a different unitary Schrodinger evolution to the state before observing the measurement procedure, thus mapping the measurement to a different projective measurement. The POVM update rule can then be understood as the effective measurement operator applied to the state that is induced by the observation of this classical information. In this sense, any measurement operator---which by definition describes the effect of obtaining classical information---is described by a projector. 

%One can counter this view by claiming that the collapse happens not at the macroscopic level of the reader of the measurement device before the measurement device could become entangled with the system, however one then runs into the problem of having a preferred Heisenberg cut, and consequently having to justify this via some as yet unknown collapse theory that is outside of standard quantum theory.  

% Let us write the Naimark dilation explicitly
% %
% \begin{align}
%     \tilde{\M}_{a\vert x} = U^\dagger[ \openone_\rA\tp\Pi_{a\vert x}] U
% \end{align}
% %
% where $\Pi_{a\vert x}$ are projectors acting on an ancilla space and $U$ acts on $\ket{\psi}\tp\ket{0}$ in the standard way as 
% %
% \begin{align}
%     U\ket{\psi}\tp\ket{0}=\sum_a \sqrt{M_{a\vert x}}\ket{\psi}\tp\ket{a}
% \end{align}

\bibliographystyle{alphaurl}
\onecolumn
\bibliography{mybibliography}

\end{document}